\documentclass[aps,prb,reprint,groupedaddress]{revtex4-1}
\usepackage[dvipdfmx]{graphicx} 
\usepackage{xcolor}

\usepackage{amsmath,amssymb} 
\usepackage{amsmath} 
\usepackage{amsfonts} 
\usepackage{bm} 

\usepackage{comment} 
\usepackage{braket} 

\DeclareMathOperator{\Tr}{Tr}

\begin{document}

\title{Purity of thermal mixed quantum states}
\author{Atsushi Iwaki}
\email{iwaki-atsushi413@g.ecc.u-tokyo.ac.jp}
\author{Chisa Hotta}
\affiliation{Department of Basic Science, University of Tokyo, Meguro-ku, Tokyo 153-8902, Japan}
\date{\today}

\newcommand{\Erase}[1]{\textcolor{red}{\sout{\textcolor{black}{#1}}}}
\newcommand{\ch}[1]{{\color{red} #1}}

\begin{abstract}
We develop a formula to evaluate the purity of a series of thermal equilibrium states 
that can be calculated in numerical experiments without knowing the exact form of the quantum state \textit{a priori}. 
Canonical typicality guarantees that there are numerous microscopically different expressions of such states, 
which we call thermal mixed quantum (TMQ) states. 
Suppose that we construct a TMQ state by a mixture of $N_\mathrm{samp}$ independent pure states. 
The weight of each pure state is given by its norm, and the partition function is given by the average of the norms. 
To qualify how efficiently the mixture is done, 
we introduce a quantum statistical quantity called ``normalized fluctuation of partition function (NFPF)". 
For smaller NFPF, the TMQ state is closer to the equally weighted mixture of pure states, 
which means higher efficiency, requiring a smaller $N_\mathrm{samp}$. 
The largest NFPF is realized in the Gibbs state with purity-0 and exponentially large $N_\mathrm{samp}$, 
while the smallest NFPF is given for thermal pure quantum state with purity-1 and $N_\mathrm{samp}=1$. 
The purity is formulated using solely the NFPF and roughly gives $N_\mathrm{samp}^{-1}$. 
Our analytical results are numerically tested and confirmed 
by the two random sampling methods built on matrix-product-state-based wave functions. 
\end{abstract}
\maketitle

\section{introduction} \label{sec:intro}
According to modern theories, 
the density matrix that represents the macroscopic state of 
a physical system in thermal equilibrium is not uniquely determined. 
By the same token, there are numerous variants of the microscopic descriptions of thermal equilibrium states 
while they are regarded as the same macroscopic state 
as far as they yield the same measurement outcomes of the local physical quantities. 
Here, we call them ``thermal mixed quantum (TMQ) states". 
\par
In a conventional statistical ensemble framework, thermal equilibrium is described by the Gibbs state, 
which is the classical mixture of exponentially large numbers of pure states. 
Let us consider purifying the Gibbs state on system $A$; 
by attaching sufficient degrees of freedom called ancilla 
or a bath to any mixed state and by entangling them with each other, 
a single pure state is realized on a whole. 
For larger degrees of classical mixtures in $A$, 
a larger entanglement with the ancilla is required to purify it. 
This means that the Gibbs state has the density matrix representation 
which is maximally entangled with the outside. 
\begin{figure}[tbp]
   \centering
   \includegraphics[width=0.86\hsize]{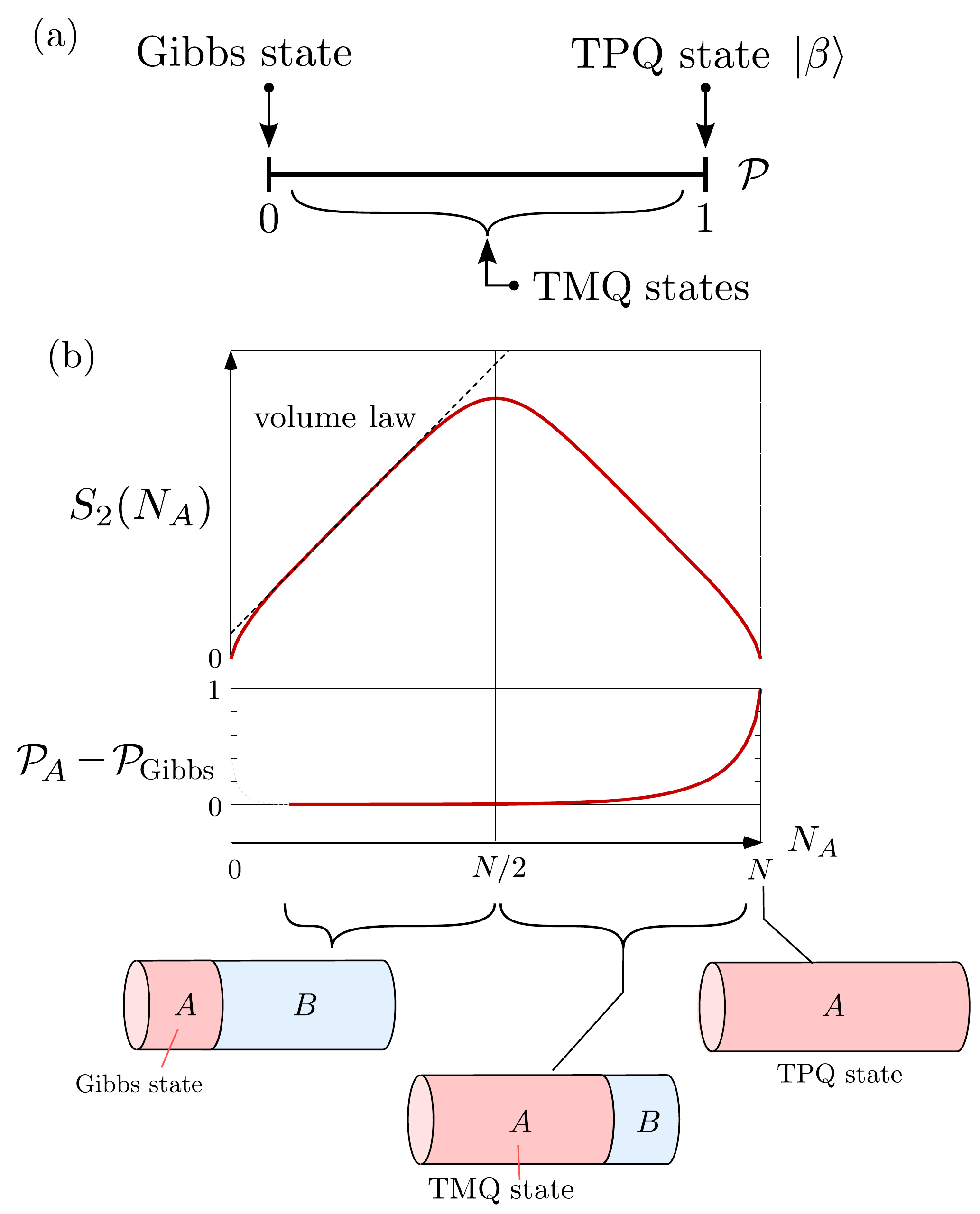}
   \caption{ 
      (a) Purity of Gibbs, TPQ, and TMQ states. 
      (b) Second R\'enyi entropy of the subsystem $A$ as a function of the subsystem size $N_A$, 
      when we divide the pure state of size $N$ into $A$ and $B$. 
      Schematic graph of the purity of thermal equilibrium states. 
      The lower panel is the purity of subsystem $A$, measured from the purity of the Gibbs state of size $N_A$. 
      Depending on $N_A$, a Gibbs, a TMQ, or a TPQ state is realized in subsystem $A$.
   }
   \label{fig:purity}
\end{figure}
\par
A single pure quantum state, on the other hand, can represent the density matrix of a thermal equilibrium without the aid of a classical mixture 
\cite{neumann1929, popescu2006, goldstein2006, sugita2006, reimann2007}, 
which is called the thermal pure quantum (TPQ) state \cite{sugiura2012, sugiura2013, hyuga2014}. 
When dividing the pure state system into small subsystem $A'$ and the rest $B'$, 
the local entanglement of the TPQ state between the two parts takes the role of the entanglement of the aforementioned Gibbs state in $A$ and the bath $B$, 
and the physical quantities measured within subsystem $A'$ match those obtained by the Gibbs state in $A$. 
It is a paraphrase of the concept called ``canonical typicality" 
\cite{neumann1929, popescu2006, goldstein2006, sugita2006, reimann2007}, 
mentioning that the density matrices are element wise equal between almost all states within an energy shell. 
\par
The macroscopically equivalent thermal equilibrium states are thus described as various microscopically different states. 
One of the measures to distinguish them is purity,
\begin{align}
\mathcal{P}=\Tr(\rho_\beta^2), 
\label{eq:purity0}
\end{align}
where $\rho_\beta$ is the density matrix of a TMQ state at temperature $\beta^{-1}$. 
The purity of the Gibbs state is the smallest among all possible choices of TMQ states 
and approaches zero in the thermodynamic limit. 
Whereas, the TPQ state has $\mathcal{P}=1$ by definition. 
Indeed, $\mathcal{P}^{-1}$ is known as an effective dimension of the quantum state \cite{linden2009, linden2010}, 
and it measures how many pure states must be mixed to describe the quantum state in question. 
As shown in Fig.\ref{fig:purity}(a), the Gibbs state and the TPQ state are the two limiting cases representing the same thermal equilibrium, and there are numerous intermediate TMQ states with $0< \mathcal{P} < 1$. 
\par
To visualize the difference between TPQ, TMQ and Gibbs states, 
we consider dividing a single TPQ state of size $N$ into two parts, $A$ and $B$, 
and focus on the quantum state realized in subsystem $A$ of size $N_A$. 
The second R\'enyi entropy of subsystem $A$ is given as 
\begin{align}
S_2(N_A)= - \log \Tr_A (\rho_A^2) = -\log \mathcal{P}_A , 
\end{align}
with $\rho_A$ being the local density matrix and $\mathcal{P}_A$ being the purity of subsystem $A$. 
Using the fact that $S_2(N_A)$ follows a Page curve \cite{page1993, garrison2018, nakagawa2018} shown in Fig.~\ref{fig:purity}(b), 
the corresponding $\mathcal{P}_A$ is derived, which is shown in Fig.~\ref{fig:purity}(b); 
when $N_A$ is sufficiently large but smaller than $N/2$, 
we find a volume law entanglement, $S_2(N_A) \propto N_A$, 
which means that the purity is bounded exponentially as $\mathcal{P}_A = e^{-\Theta(N_A)}$. 
When the subsystem $A$ is in a Gibbs state, 
the thermal entropy $s_\mathrm{th}$ is equivalent to the von Neumann entropy of subsystem $A$ as 
\begin{align}
   S_\mathrm{vN} (\rho_A) =-\Tr(\rho_A\log \rho_A)= N_A s_\mathrm{th} \quad(1 \ll N_A \ll N) 
\label{eq:sA}
\end{align}
and the volume law guarantees that each constituent of a mixed state in $A$ is minimally entangled inside $A$, 
while maximally entangled with bath $B$. 
If we take $N_A \to N$, the purity becomes $\mathcal{P}_A \to 1$ 
by definition and consistently with $S_2(N_A) \to 0$. 
\par
We may thus regard the state in $A$ in the intermediate region $N/2\lesssim N_A < N$ as 
one of the constructions of a TMQ state with $0<\mathcal{P}_A <1$. 
It is known that the entanglement entropy of a pure state is equivalent to the thermal entropy \cite{iwaki2021}. 
If the entanglement inside $A$ is not enough to cover the whole thermal entropy of the thermal state that should be realized in $A$, 
a series of quantum states in $A$ needs to entangle with its bath $B$ in order to offset the deficiency. 
\par
Naively, the purity controls the ratio of the thermodynamic entropy assigned to the internal entanglement and to the classical mixture. 
However, in exploring a state whose density matrix is unknown {\it a priori}, 
there is no clue to find the necessary and sufficient number of pure states to be mixed. 
This is because there is no way of measuring the ratio of the two contributions 
to the thermodynamic entropy in a mixed state. 
If the number of mixtures is unavailable, neither the thermal equilibrium state nor 
the purity in Eq.(\ref{eq:purity0}) using the density matrix is defined. 
Therefore we need alternative ways to evaluate the purity and to characterize the TMQ state. 
\par
In this paper, we consider a series of TMQ states generated by random sampling methods 
and obtain an analytical formula that describes the purity by a measurable quantity. 
Although the constructions differ between such stochastic TMQ state and 
the TMQ state obtained as a subsystem of a TPQ state in Fig.~\ref{fig:purity}(b), 
these two can be identified as the same, which we will discuss shortly in \S.\ref{sec:pre}B. 
Here, in stochastiacally constructing a TMQ state, the random sampling average corresponds to entangling 
the target region $A$ with $B$ in Fig.~\ref{fig:purity}(b). 
There are various stochastic finite temperature numerical solvers for quantum many-body systems. 
The quantum Monte Carlo method makes use of the Markov chain process to efficiently select a series of states, 
and the snapshots realized at each Monte Carlo step altogether form a mixed state. 
Random sampling methods approximating the finite temperature state by the mixture of matrix product states (MPS) or tensor network states are also developed \cite{garnerone2013-1, garnerone2013-2, iitaka2020, iwaki2021, goto2021}. 
The common strategy of random sampling methods is to generate a series of states based on 
independent sampling and average them to calculate physical quantities. 
The law of large numbers guarantees that the result coincides with the exact quantity for 
a large enough number of samples. 
However, the variance of a physical quantity is the only measure to judge the quality of the wave functions, 
and a necessary and sufficient number of samples are observed only empirically. 
We show that the purity can be evaluated using 
``normalized fluctuation of partition function (NFPF)", which is a key quantity we introduce in this paper. 
The NFPF is proportional to the number of samples mixed. 
\par
Previously, purity has been measured directly using Eq.(\ref{eq:purity0}) experimentally in ultra-cold atom systems for only a few numbers of atoms, 
which is used to judge whether the system keeps its isolated nature during the time evolution \cite{kaufman2016}. The bulk TMQ state we consider is far difficult to deal with both in theory and experiments 
because the Hilbert space dimensions grow exponentially with the number of degrees of freedom. 
Even in such a case, our theory enables us to calculate the purity without using Eq.(\ref{eq:purity0}). 
\par
We add some remarks that there is a necessity of obtaining a TMQ state rather than TPQ state in quantum condensed matter. 
In these systems, the TPQ state is basically obtained by operating the non-unitary imaginary time evolution to the Haar random initial state. 
In the context of quantum information, 
there has been a development to efficiently construct a unitary 2-design that 
reproduces the second moment of the finite dimension Haar random state, 
which may even amount to $N$ of a few hundred 
\cite{brown2008, dankert2009, harrow2009, diniz2011, brandao2016-1, brandao2016-2, cleve2016, nakata2017-1, nakata2017-2}. 
This indicates that the high-quality initial state for the thermal state is available. 
However, the non unitary operation to such a state is still difficult to attain, 
and there is a need to deal with a lower-purity TMQ state which can be much easily realized. 
The present framework can be applied to the studies of quantum information that considers 
a general non unitary operation. 
\par
The remainder of the paper is organized as follows. 
In \S.\ref{sec:pre}, we introduce the definitions of the related physical quantities and explain 
the overall physical implication of obtaining the form of purity. 
In \S.\ref{sec:purity_efficiency}, we develop an analytical framework for measuring NFPF and purity in the random sampling methods. 
\S.\ref{sec:demonstration} is devoted to the demonstration to verify the formula given in \S.\ref{sec:purity_efficiency} 
using two numerial methods, and we summarize our framework finally in \S.\ref{sec:summary}. 
%
%
\section{Preliminaries}
\label{sec:pre}
In this section, we introduce some basic notations 
and preliminary concepts relevant to our theory, rephrasing the context in the introduction. 
\subsection{TMQ states}
In the conventional framework of statistical mechanics, 
the density matrix operator of a Gibbs state for a given Hamiltonian $\Hat{H}$ is 
\begin{align}
   \rho_\beta^\mathrm{G} = \frac{e^{-\beta\Hat{H}}}{Z(\beta)}, 
\end{align}
where $Z(\beta) = \Tr e^{-\beta\Hat{H}}$ is the partition function at temperature $\beta^{-1}$. 
The von Neumann entropy $S_\mathrm{vN}$ evaluated using the density operator is equivalent to the 
thermodynamic entropy, 
\begin{align}
   S_\mathrm{th}(\beta) = S_\mathrm{vN}(\rho_\beta^\mathrm{G})
   = - \Tr(\rho_\beta^\mathrm{G} \log \rho_\beta^\mathrm{G}). 
\end{align}
Since the Gibbs state is a mixture of an exponentially large number of pure states, 
its purity is as small as
\begin{align}
   \mathcal{P}_\mathrm{Gibbs} &= \Tr[(\rho_\beta^\mathrm{G})^2]
   = \frac{Z(2\beta)}{Z(\beta)^2} \notag \\
   &= \exp[-Ns_\mathrm{th} (\Tilde{\beta})] \quad (\beta\le ^\exists \Tilde{\beta} \le 2\beta), 
\label{p_gibbs}
\end{align}
and approaches zero exponentially with increasing system size $N$. 
A Gibbs ensemble average of a local operator $\Hat{O}$ acting on a $D$-dimensional Hilbert space 
spanned by an orthonormal set of states $\{\ket{r}\}$ for a system of size $N$ is given by
\begin{align}
   \langle \Hat{O} \rangle_\beta
   &= \Tr (\rho_\beta^\mathrm{G} {\hat O})  \notag \\
   &= \frac{1}{Z(\beta)} \sum_{r=1}^D \bra{r} e^{-\beta\Hat{H}/2} \Hat{O} e^{-\beta\Hat{H}/2} \ket{r}, \label{o_gibbs} \\
 & Z(\beta)= \sum_{r=1}^D \bra{r} e^{-\beta\Hat{H}} \ket{r}. 
\end{align}
\par
Contrarily, the TPQ state describes the thermal equilibrium solely by itself. 
One way to construct it in a $D$-dimensional Hilbert space is to 
prepare an initial state $\ket{0}=\sum_{r=1}^D c_r \ket{r}$ 
using a randomly chosen $D$ complex numbers $\{c_r\}$ 
generated independently from the complex Gaussian distribution, 
and perform an imaginary time evolution as 
\begin{align}
   \ket{\beta} = e^{-\beta\Hat{H}/2} \ket{0}. 
\label{tpqstate}
\end{align}
The physical quantities $\langle \Hat{O} \rangle^\mathrm{TPQ}_\beta = \braket{\beta|\Hat{O}|\beta}/\braket{\beta|\beta}$ 
matches the Gibbs ensemble average within the fluctuation as 
\begin{align}
& \overline{\left(\langle \Hat{O} \rangle^\mathrm{TPQ}_\beta
    - \langle\Hat{O}\rangle_\beta \right)^2}
   \lesssim (\mathrm{const.}) \times \|\Hat{O}\|^2
   e^{-Ns_\mathrm{th}(\Tilde{\beta})}, 
\label{error:tpq}
\end{align}
where $\overline{\cdots}$ represents a random average and 
$(\mathrm{const.})$ is a constant independent of system size $N$. 
Since the density operator for the TPQ state is $\rho^\mathrm{TPQ}_\beta = \ket{\beta}\bra{\beta}/\langle\beta|\beta\rangle$, 
we find $\mathcal{P}=\Tr [(\rho^\mathrm{TPQ}_\beta)^2]=1$. 
As a consequence of typicality, 
for a local density matrix of subsystem $A$ given as 
\begin{align}
   \rho_A= \Tr_B \rho^\mathrm{TPQ}_\beta , 
\end{align}
the entanglement entropy follows a volume law in Eq.(\ref{eq:sA}). 
Since Eq.(\ref{error:tpq}) is the random fluctuation which decreases exponentially with increasing $N$, 
the TPQ state obtained in the form Eq.(\ref{tpqstate}) is basically regarded as pure. 
We use the TPQ state at size $N$ as a quantitative criterion for $\mathcal {P}=1$ 
to determine the purity of the mixed state of the same size. 
\par
In obtaining the actual form of TMQ states in a quantum many-body state of finite size 
and at finite temperature, numerical methods with some approximations are used. 
The following section focuses on the random sampling method as the most frequently used approach. 

%
\subsection{Random sampling methods} 
Although we want a standard ensemble average in classical computers, 
performing the Gibbs ensemble average in Eq.(\ref{o_gibbs}) is practically difficult, 
since $D$ grows exponentially with system size $N$. 
For a TPQ state in Eq.(\ref{tpqstate}) the full description of a $D$-dimensional quantum many-body state is limited to $N\lesssim 30$. 
Therefore one needs to approximate the state in the intermediate form of Eqs.(\ref{o_gibbs}) and (\ref{tpqstate}) by sampling over $M \ll D$ different appropriately chosen states. 
\par
Random sampling method begins by preparing $M$-independent set of states, $\{|\psi_0^{(i)}\rangle\}_{i=1}^M$. 
These states are generated from some given random distribution, 
which form an identity operator when averaged over the distribution as 
\begin{align}
   \overline{\ket{\psi_0} \bra{\psi_0}} = c \Hat{I}, 
   \label{eq:condition}
\end{align}
with a positive constant $c = \overline{\braket{\psi_0|\psi_0}} /D$. 
We obtain $\{|\psi_\beta^{(i)}\rangle\}$ from $\{|\psi_0^{(i)}\rangle\}$. 
Here, the condition for these method to work is to have 
\begin{align}
   \overline{\ket{\psi_\beta}\bra{\psi_\beta}} = c e^{-\beta\Hat{H}}. 
\label{eq:cond}
\end{align}
or equivalently,
\begin{equation}
\ket{\psi_\beta^{(i)}}=e^{-\beta \Hat H/2} \ket{\psi_0^{(i)}}. 
\end{equation}
\par
For such states, the random averages of physical quantities should coincide the Gibbs ensemble average as 
\begin{align}
   \langle \Hat{O} \rangle _\beta =
   \frac{\overline{\braket{\psi_\beta|\Hat{O}|\psi_\beta}}}{\overline{\braket{\psi_\beta|\psi_\beta}}},
\label{eq:orandomsamp}
\end{align}
and free energy as
\begin{align}
   F(\beta) = - \beta^{-1}\log \overline{\braket{\psi_\beta|\psi_\beta}} + \beta^{-1}\log c.
\end{align} 
If we sample enough large $M$, the law of large numbers gurantees that 
the sample average 
\begin{align}
   \langle \Hat{O} \rangle_{\beta, M}^\mathrm{samp}&=
\frac{\sum_{i=1}^M\braket{\psi_\beta^{(i)}|\Hat{O}|\psi_\beta^{(i)}} }
     {\sum_{j=1}^M \braket{\psi_\beta^{(j)}|\psi_\beta^{(j)}}}, 
\label{eq:samp}
\end{align}
should match Eq.(\ref{eq:orandomsamp}) with arbitrary precision. 
The density matrix representing the corresponding TMQ state is given as 
\begin{equation}
\rho(M)=\frac{ \sum_{i=1}^M  
\ket{\psi_\beta^{(i)}}\bra{\psi_\beta^{(i)}} }
{ \sum_{j=1}^M  \langle\psi_\beta^{(j)}|\psi_\beta^{(j)}\rangle}. 
\label{eq:rho_m}
\end{equation}
We show that the necessary and sufficient $M=N_\mathrm{samp}$, 
can be determined {\it not empirically} but based on the analytical formula. 
\par
We briefly mention that the random sampling method is a basic operation to construct 
the quantum thermal equilibrium state, and the TMQ state described in Eq.(\ref{eq:rho_m}) 
can be regarded as one of the examples of the TMQ state we naturally constructed 
in Fig.~\ref{fig:purity}(b) as gedankenexperiment. 
Suppose that we have a TMQ state in system $A$ of size $N_A$, 
which can be interpreted in two ways;
one is to regard this TMQ state as a subsystem of a TPQ state in a larger-size system, $N=N_A+N_B$. 
This is because it is known that any mixed state can be purified 
by attaching proper extra degrees of freedom. 
Therefore we can always find another subsystem $B$ that can form a TPQ state 
together with the numerically generated TMQ state in $A$ following Eq.(\ref{eq:rho_m}), 
although there is a facultativity in the choice and the size of $B$. 
\par
The other interpretation is to prepare some ideal TMQ state $A$ 
as a subsystem of a TPQ state ($A+B$). 
We can always perform a spectral decomposition of a TMQ state in $A$ 
to be described by $\sum_i \lambda_i|i\rangle \langle i|$, for a given basis $\{|i\rangle\}$, 
and by sampling the states following the distribution function $\{\lambda_i\}$, 
one is able to approximately construct TMQ state as a weighted 
stochastic mixture of $\{|i\rangle\}$ in a computer. 
In our random sampling method, the weight $\{ w_{i,M} \}(M\rightarrow\infty)$ 
implicitly included in Eq.(\ref{eq:rho_m}) 
and explicitly shown in Eq.(\ref{eq:weight}) corresponds to the distribution function $\{\lambda_i\}$. 
\subsection{Proper choice of sample average } 
The standard way of taking the averages over $M$ samples is often recognized as 
\begin{align}
   \langle \Tilde{O} \rangle_{\beta, M}^\mathrm{samp}&= \frac{1}{M} \sum_{i=1}^M
\frac{\braket{\psi_\beta^{(i)}|\Hat{O}|\psi_\beta^{(i)}} }
     {\braket{\psi_\beta^{(i)}|\psi_\beta^{(i)}}}, 
\label{eq:samp-wrong}
\end{align}
where we use the normalized expectation values for all $i=1 - M$ samples 
instead of evaluating the denominator and the numerator separately as in Eq.(\ref{eq:samp}). 
In such a case, the density matrix of the mixed state is given as 
\begin{align}
\Tilde{\rho}(M)=\frac{1}{M}\sum_{i=1}^M\frac{e^{-\beta
H/2}\ket{\psi_0^{(i)}}\bra{\psi_0^{(i)}}e^{-\beta H/2}}
{\bra{\psi_0^{(i)}}e^{-\beta H}\ket{\psi_0^{(i)}}}. 
\label{eq:rho_m-wrong}
\end{align}
Unfortunately, this seemingly widely accepted formulation is wrong, since its $M\rightarrow\infty$ limit 
does not extrapolate to the proper canonical ensemble average, which we show in the following. 
\par
Suppose we have a set of data 
$(x^{(i)} = \braket{\psi_\beta^{(i)}|\Hat{O}|\psi_\beta^{(i)}},
y^{(i)} = \braket{\psi_\beta^{(i)}|\psi_\beta^{(i)}} )$, $i=1\sim M$, 
where taking the random average, we find 
$x_0 = \overline{x^{(i)}} = c {\rm Tr}(\hat O e^{-\beta \Hat{H}})$ and 
$y_0 = \overline{y^{(i)}} = c {\rm Tr}(e^{-\beta \Hat{H}})$, 
and the ensemble average of operator is given by $\braket{\hat O}_\beta= x_0/y_0$. 
We now take $\overline{\sum_{i=1}^M \cdots}$, 
which is the random average of summation for $M$-samples generaged from the random distribution
(see \S.\ref{sec:variance_phys} using the same treatment). 
\par
Then, the random average of Eq.(\ref{eq:samp}) is 
evaluated up to the second moment of $\delta x = x - x_0$ and $\delta y = y - y_0$ becomes 
\begin{align}
\overline{\left(
   \frac{\sum_{i=1}^M x^{(i)}}{ \sum_{i=1}^M y^{(i)}} \right) }
=\frac{x_0}{y_0} + \frac{1}{M} \left( \frac{x_0}{y_0^3} \overline{\delta y^2} -  \frac{1}{y_0^2} \overline{\delta x\delta y} \right). 
\end{align}
As for Eq.(\ref{eq:samp-wrong}), the random average is given as
\begin{align}
   \overline{ \sum_{i=1}^M  \frac{x^{(i)}}{y^{(i)}} } 
=\frac{x_0}{y_0} -  \frac{1}{y_0^2} \overline{\delta x\delta y}. 
\end{align}
Comparing these two, we find that the $M\rightarrow\infty$ limit of the former is $x_0/y_0$, 
but for the latter, the second moment remains finite regardless of how large we take $M$. 
\par
To briefly summarize, the sample average that properly approximates the canonical ensemble average 
needs to be taken as Eq.(\ref{eq:samp}) 
using a set of data generated from the random distribution. 
Although Eq.(\ref{eq:samp-wrong}) may approximate $x/y$ with sufficient accuracy, 
it does not converge to the correct value. 
Accordingly, the TMQ state should be represented by Eq.(\ref{eq:rho_m}), 
and not by Eq.(\ref{eq:rho_m-wrong}). 
\subsection{What is the purity of the random sampling method?} 
For a generalized ensemble average, we usually prepare a set of {\it normalized} states 
$\{\ket{\phi^{(i)}}\}_{i=1}^{N_\mathrm{samp}}$ that are generated 
by {\it the physically meaningful distribution function of that ensemble}, 
and for a given $N_\mathrm{samp}$, 
the physical quantities can be calculated as 
\begin{align}
\frac{1}{N_\mathrm{samp} }\sum_{i=1}^{N_\mathrm{samp}} \bra{\phi^{(i)}} \Hat{O}_A \ket{\phi^{(i)}}, 
\label{eq:sampdist}
\end{align}
with sufficient accuracy. 
For example, the microcanonical ensemble average is chosen from a uniform distribution 
in the corresponding energy shell. 
The quantum state consisting of a classical mixture of $\{\ket{\phi^{(i)}}\}_{i=1}^{N_\mathrm{samp}}$ 
is represented by the density operator 
\begin{align}
   \rho(N_\mathrm{samp}) = \frac{1}{N_\mathrm{samp}}
   \sum_{i=1}^{N_\mathrm{samp}} \ket{\phi^{(i)}} \bra{\phi^{(i)}}. 
   \label{eq:rhosamp}
\end{align}
By assuming that the sampled states are orthogonal to each other as $\langle \phi^{(i)} | \phi^{(j)} \rangle= \delta_{ij}$, 
we can determine the purity as
\begin{align}
   \mathcal{P}_\mathrm{ens} = \Tr[\rho(N_\mathrm{samp})^2] = \frac{1}{N_\mathrm{samp}}. 
   \label{eq:purity_apprx}
\end{align}
\par
This form suggests that $\mathcal{P}^{-1}$ is a quantitative measure of 
how much the pure state is mixed in a TMQ state. 
However, the purity in Eq.(\ref{eq:purity_apprx}) does not hold in the random sampling approach 
described in the preceding section. 
This is due to the fact that the ensemble's information is contained in the imaginary time evolution; 
in Eq.(\ref{eq:samp}) the norm of sampled states 
$\braket{\psi_\beta^{(i)}|\psi_\beta^{(i)}}$ depends on $i$, 
and the numerator incorporates the weight of each sample. 
As a result, it differs from an equally weighted sample average in Eq.(\ref{eq:rhosamp}). 
In other words, although Eqs.(\ref{eq:sampdist}) and (\ref{eq:rhosamp}) 
formally resemble Eqs.(\ref{eq:samp-wrong}) and (\ref{eq:rho_m-wrong}), respectively, 
the former is physically meaningful but the latter is not. 
Whilst, Eqs.(\ref{eq:rhosamp}) and (\ref{eq:rho_m}) 
built on different distribution function, equivalently serve 
as approximate forms of $\braket{\hat O}_\beta$. 
Aside from that, the foregoing discussion is incomplete 
because we do not know the actual value of $N_{\rm samp}$ {\it a priori} nor the purity.  
\par
In numerical simulations, $N_{\rm samp}$ are typically determined to suppress the variance of physical quantities 
within a predetermined error. 
The variance depends on the types of physical quantities, 
for example, the variance of energy is substantially smaller than the variance of correlation functions.
It is also possible that $N_{\rm samp}$ is overdetermined in order to ensure accuracy. 
We want to obtain the necessary and sufficient value of $N_{\rm samp}$ 
which can serve as an effective dimension. 
Our goal is to create a formula that starts with Eq.(\ref{eq:purity0}), 
and replaces Eq.(\ref{eq:purity_apprx}) with an explicit definition of purity. 
There, the definition relies on the numerically measurable quantities 
rather than on $N_\mathrm{samp}$. 
On top of that, we discuss the necessary and sufficient sample number 
$M_e$ in the numerical calculation 
in \S.\ref{sec:sample}.

\section{Purity and efficiency of random sampling methods}
\label{sec:purity_efficiency}

In this section, we start from the physical quantity called ``efficiency" denoted as $\eta$. 
It measures the degree of uniformity of distribution of the weight of the samples. 
If all the samples equally contribute to the averages, we find $\eta=1$, 
whereas if only a few of the samples contribute, $\eta$ approaches zero. 
We derive the formula that describes $\eta$ by the normalized fluctuation of partition function (NFPF). 
This NFPF is found to be proportional to the number of samples $N_\mathrm{samp}$, 
and finally, the purity of the TMQ state is described by NFPF.  

\subsection{Efficiency}
There was previously no criterion for evaluating and comparing different types of random sampling methods. 
Recently, Goto {\it et al.} introduced the measure of efficiency
of random samplings for a general set of $\{\ket{\psi_\beta^{(i)}}\}_{i=1}^M$
\cite{goto2021}. 
When $M = N_\mathrm{samp}$,
physical quantities are measured by a sample average with sufficient accuracy, 
and a set of the state forms a TMQ state. 
The sample average of observable $\Hat{O}$ is written as 
[see also Eq.(17)]
\begin{align}
   \langle \Hat{O} \rangle_{\beta, M}^\mathrm{samp}
   = \sum_{i=1}^M \frac{\braket{\psi_\beta^{(i)}|\Hat{O}|\psi_\beta^{(i)}}}
   {\braket{\psi_\beta^{(i)}|\psi_\beta^{(i)}}}
   \frac{\braket{\psi_\beta^{(i)}|\psi_\beta^{(i)}}}
   {\sum_{j=1}^M \braket{\psi_\beta^{(j)}|\psi_\beta^{(j)}}}, 
\label{eq:osamp}
\end{align}
which takes the form of a weighted average of physical quantities 
$\braket{\psi_\beta^{(i)}|\Hat{O}|\psi_\beta^{(i)}}/\braket{\psi_\beta^{(i)}|\psi_\beta^{(i)}}$ 
with its weight given as 
\begin{align}
   w_{i, M} = \frac{\braket{\psi_\beta^{(i)}|\psi_\beta^{(i)}}}
   {\sum_{j=1}^M \braket{\psi_\beta^{(j)}|\psi_\beta^{(j)}}}.
 \label{eq:weight}
\end{align}
Using the Shanon entropy of $w_{i, M}$, 
\begin{align}
   S_M = - \sum_{i=1}^M w_{i, M} \log w_{i, M}, 
 \label{eq:sm}
\end{align}
the efficiency of random sampling methods is defined as 
\begin{align}
   \eta = \frac{e^{S_M}}{M}. 
 \label{eq:eta}
\end{align}
If the weight $\{ w_{i, M} \}_{i=1}^M$ has a uniform distribution 
we find $S_M=\log M$ and $\eta=1$, and otherwise we have $\eta<1$. 
The larger variance of $w_{i, M}$ the smaller $\eta$ becomes. 
Therefore $\eta$ gives a quantitative measure of how uniformly the samples contribute to give a 
higher efficiency in the calculation. 
\par
We illustrate the physical implications of efficiency $\eta$ using the analytical calculation, 
finally proving that it is connected to the purity of a TMQ state. 
For the sake of clarity, we introduce simplified notations of variables as 
\begin{align}
   &x^{(i)}\:= \braket{\psi_\beta^{(i)} | \Hat{O} | \psi_\beta^{(i)}} , \quad \\
   &x_0 = \overline{x} = c \Tr (e^{-\beta \Hat{H}} \Hat{O}) , \\
   &y^{(i)}\: = \braket{\psi_\beta^{(i)} | \psi_\beta^{(i)}} , \label{eq:yi}
  \quad \\
   &y_0 = \overline{y} = c Z(\beta) , 
\end{align}
where we sometimes abbreviate the superscript $(i)$ when it is not explicitly needed. 
The weight is rewritten as
\begin{align}
   w_{i, M} = \frac{y^{(i)}}{\sum_{j=1}^M y^{(j)}}.
\label{eq:wi}
\end{align}
The goal of this section is to expand $\eta$ up to the second order of 
the fluctuation of random variables given as 
\begin{align}
   \delta y = y - y_0, \quad \delta x = x - x_0. 
\end{align} 
For this purpose, 
we consider $w_{i,M}$ and take the average over $M\rightarrow \infty$ samples; 
\begin{align}
   w_{i, M}
   &= \frac{y_0 + \delta y^{(i)}}{y_0 M} \frac{1}{1+\frac{1}{y_0 M} \sum _j \delta y^{(j)}} \notag \\
   &= \frac{1}{M} \left( 1+\frac{\delta y^{(i)}}{y_0} \right)
   + \mathcal{O}\left(\frac{1}{M^2}\right). 
\end{align}
Since $\sum_i w_{i, M}=1$, we find 
\begin{align}
  & \overline{w_{i, M}} = \frac{1}{M}, \\
  & \delta w_{i, M} = w_{i, M} - \overline{w_{i, M}}
   = \frac{\delta y^{(i)}}{y_0M} + \mathcal{O}\left(\frac{1}{M^2}\right), \\
  & \sum_i \delta w_{i, M} = 0.
\end{align}
In random sampling methods, $y^{(i)}$ takes the value independent of $i$, 
which allows us to apply the relationship $\overline{\delta y^{(i)} \delta y^{(j)}} = \delta _{ij} \overline{\delta y^2}$. 
Accordingly, we find 
\begin{eqnarray}
   \overline{\delta w_{i, M} \delta w_{j, M}}
   &=& \frac{1}{y^2_0 M^2}\overline{\delta y^{(i)} \delta y^{(j)}}
   +\mathcal{O}\left(\frac{1}{M^3}\right) \nonumber\\
   &=& \frac{\delta_{ij}}{y^2_0 M^2} \overline{\delta y^2}
   +\mathcal{O}\left(\frac{1}{M^3}\right) \nonumber\\
   &=& \frac{\delta_{ij}}{M^2} \delta z^2+\mathcal{O}\left(\frac{1}{M^3}\right).
\label{eq:dwdw}
\end{eqnarray}
Here, we introduce a key quantity, $\delta z^2$, of the present framework defined as 
\begin{align}
   \delta z^2
   &= \frac{1}{y_0^2} \overline{\delta y^2} \notag \\
   &= \frac{\mathrm{Var}(\braket{\psi_\beta|\psi_\beta})}
   {\left(\overline{\braket{\psi_\beta|\psi_\beta}}\right)^2},
   \label{eq:def_nfpf}
\end{align}
where $\mathrm{Var}(\cdots)$ is the sample variance of a variable $(\cdots)$. 
We denote $\delta z^2$ as {\it normalized fluctuation of partition function} and abbreviate it as NFPF. 
\par
The entropy of weights is expanded up to the second order of $\delta w_{i, M}$ as 
\begin{align}
   S_M
   &= -\sum_i w_{i, M} \log w_{i, M} \notag \\
   &= -\sum_i \left(\frac{1}{M}+\delta w_{i, M}\right)
   \log\left(\frac{1}{M}+\delta w_{i, M}\right) \notag \\
   &= \log M - \frac{1}{M}\sum_i(1+M\delta w_{i, M})\log(1+M\delta w_{i, M}) \notag \\
   &= \log M - \frac{1}{2M} \sum_i (M\delta w_{i, M})^2 + \mathcal{O}(\delta^3), 
\end{align}
and using this, the efficiency $\eta$ in Eq.(\ref{eq:eta}) is given as 
\begin{align}
   \frac{e^{S_M}}{M}
   &= \exp\left[- \frac{1}{2M} \sum_i (M\delta w_{i, M})^2 + \mathcal{O}(\delta^3)\right] \notag \\
   &= 1 - \frac{1}{2M} \sum_i (M\delta w_{i, M})^2 + \mathcal{O}(\delta^3), 
\end{align}
whose random average becomes 
\begin{align}
   \frac{\overline{e^{S_M}}}{M}
   &= 1 - \frac{\delta z^2}{2} + \mathcal{O}\left(\frac{1}{M}\right)
   + \mathcal{O}\left(\overline{\delta ^3}\right).
\end{align}
where we applied the relationship in Eq.(\ref{eq:dwdw}). 
By taking the limit $M \to \infty$, we finally obtain 
\begin{align}
   \eta 
   &= 1 - \frac{\delta z^2}{2} + \mathcal{O}\left(\overline{\delta ^3}\right) \notag \\
   &= \exp \left( -\frac{\delta z^2}{2} \right)+\mathcal{O}\left(\overline{\delta ^3}\right)
   \label{eq:efficiency}
\end{align}
This equation shows that {\it the efficiency of the random sampling method 
is an inverse exponential of NFPF}. 
Intuitively, a larger fluctuation of the norm means a larger variance in the weights of random sampling, 
since NFPF represents the degree of fluctuation of the norm of 
the finite temperature wave function.

\subsection{Random fluctuation}
\label{sec:variance_phys}
We now analytically evaluate the variance of physical quantities in the random sampling method against the value obtained by the Gibbs ensemble average and relate it to NFPF. 
In the previous section, 
we assumed $M\rightarrow\infty$ and expanded the efficiency up to several leading orders of $1/M$. 
However, the following formula does not require $M$ to be infinitely large, 
and the equations are satisfied under the random average $\overline{\sum_{i=1}^M}$ 
with the finite $M$-samples generated from the random distribution. 
Similarly to the process given in the previous section, 
we expand the variance in terms of $\delta x$ and $\delta y$ 
up to second-order as
\begin{align}
   &\overline{\left( \frac{\braket{\psi_\beta|\Hat{O}|\psi_\beta}}{\braket{\psi_\beta|\psi_\beta}}
   - \langle \Hat{O} \rangle_\beta \right)^2} \notag \\
   &= \overline{\left( \frac{x}{y} - \frac{x_0}{y_0} \right) ^2} \notag \\
   &= \overline{\left( \frac{1}{y_0} \delta x - \frac{x_0}{y_0^2} \delta y
   + \mathcal{O}(\delta ^2) \right) ^2} \notag \\
   &= \frac{1}{y_0^2} \overline{\delta x^2} - \frac{2x_0}{y_0^2} \overline{\delta x \delta y}
   + \frac{x_0^2}{y_0^2} \overline{\delta y^2} + \mathcal{O}\left(\overline{\delta ^3}\right) \notag \\
   &\lesssim \frac{1}{y_0^2} \overline{\delta x^2} + \frac{2 \| \Hat{O} \|}{y_0} \left| \overline{\delta x \delta y} \right|
   + \frac{\| \Hat{O} \| ^2}{y_0^2} \overline{\delta y^2}.
\end{align}
Here, $\|\cdot\|$ is an operator norm.
Then, using the Cauchy-Schwarz inequality we find,
\begin{align}
   \left| \overline{\delta x \delta y} \right|
   &= \left| \int \delta x \delta y d \mu \right| \notag \\
   &\le \int | \delta x \delta y | d \mu \notag \\
   &\le \left( \int | \delta x | ^2 d \mu \right) ^{1/2}
   \left( \int | \delta y | ^2 d \mu \right) ^{1/2} \notag \\
   &= \sqrt{\overline{\delta x^2}} \sqrt{\overline{\delta y^2}}.
\end{align}
The variance is rewritten as 
\begin{align}
   \overline{\left( \frac{x}{y} - \frac{x_0}{y_0} \right) ^2}
   \lesssim \frac{1}{y_0^2} \overline{\delta x^2} +
   \frac{2 \| \Hat{O} \|}{y_0} \sqrt{\overline{\delta x^2}} \sqrt{\overline{\delta y^2}}
   + \frac{\| \Hat{O} \| ^2}{y_0^2} \overline{\delta y^2} .
\label{eq:varxytmp}
\end{align}
Next, to further develop this inequality, we introduce the following important assumption; 
\begin{align}
   \mathrm{Var}(\braket{\psi_\beta|\Hat{O}|\psi_\beta}) \le \mathrm{(const.)} \times
   \|\Hat{O}\|^2 \mathrm{Var}(\braket{\psi_\beta|\psi_\beta}), 
   \label{eq:assumption}
\end{align}
which is rewritten using $\delta x$ and $\delta y$ as 
\begin{align}
   \overline{\delta x^2} \le \mathrm{(const.)} \times
   \|\Hat{O}\|^2 \overline{\delta y^2}.
\end{align}
Here, $\mathrm{(const.)}$ is a constant of order $\mathcal{O}(N^0)$.
\par
When we refer to the original random state as ``Haar random" in numerical calculations, 
it means the normalized state. 
In general, the normalization of the initial state is quite commonly adopted in numerical calculations. 
Such normalized initial states have the zero-variance of the norm by definition, 
and accordingly, Eq.(\ref{eq:assumption}) breaks down at the relevant high-temperature limit. 
Even in such a case, after the imaginary time evolution, 
we acquire thermal states that typically fulfill Eq.(\ref{eq:assumption}). 
Here, we notice that the unnormalized initial random states naturally apply to standard calculations, 
which could provide superior results, although its implication had not been examined so far. 
We notice in Appendix \ref{sec:validity} the two particular and exceptional cases 
where the assumption breaks down at all temperatures; they utilize the energy eigenstates to build random states. 
Such choice is rather unusual and for the general choices of the basis for constructing a random state, 
Eq.(\ref{eq:assumption}) naturally holds from low to extremely high temperatures. 
\par
Now, by utilizing Eq.(\ref{eq:assumption}), 
Eq.(\ref{eq:varxytmp}) is converted to 
\begin{eqnarray}
   \overline{\left( \frac{x}{y} - \frac{x_0}{y_0} \right) ^2}
   &\lesssim& (\mathrm{const.}) \times \frac{\| \Hat{O} \| ^2}{y_0^2} \overline{\delta y^2} \nonumber\\
   &=& (\mathrm{const.}) \times \| \Hat{O} \| ^2 \delta z^2, 
\end{eqnarray}
or equivalently to 
\begin{align}
   \overline{\left( \frac{\braket{\psi_\beta|\Hat{O}|\psi_\beta}}{\braket{\psi_\beta|\psi_\beta}}
   - \langle \Hat{O} \rangle_\beta \right)^2}
   \lesssim (\mathrm{const.}) \times \| \Hat{O} \|^2 \delta z^2 .
   \label{eq:fluctuation}
\end{align}
A similar evaluation can be given for the partition function $\overline{\braket{\psi_\beta|\psi_\beta}}$ as 
\begin{align}
   \overline{\left(\frac{\braket{\psi_\beta|\psi_\beta}}{cZ(\beta)}-1\right)^2} = \delta z^2.
\end{align}
These equations show that the variance of physical quantities is bounded by the NFPF. 
By using the relationship between the efficiency and the NFPF in Eq.(\ref{eq:efficiency}),
we obtain 
\begin{gather}
   \overline{\left( \frac{\braket{\psi_\beta|\Hat{O}|\psi_\beta}}{\braket{\psi_\beta|\psi_\beta}}
   - \langle \Hat{O} \rangle_\beta \right)^2}
   \lesssim (\mathrm{const.}) \times \| \Hat{O} \|^2 \log\left(\frac{1}{\eta}\right) ,
   \label{eq:qvarfinal} \\
   \overline{\left(\frac{\braket{\psi_\beta|\psi_\beta}}{cZ(\beta)}-1\right)^2}
   \simeq 2\log\left(\frac{1}{\eta}\right) .
\end{gather}
The relationship between the variance of physical quantities and the efficiency $\eta$ 
in random sampling methods is thus clarified using NFPF. 
\begin{figure}[tbp]
   \centering
   \includegraphics[width=0.8\hsize]{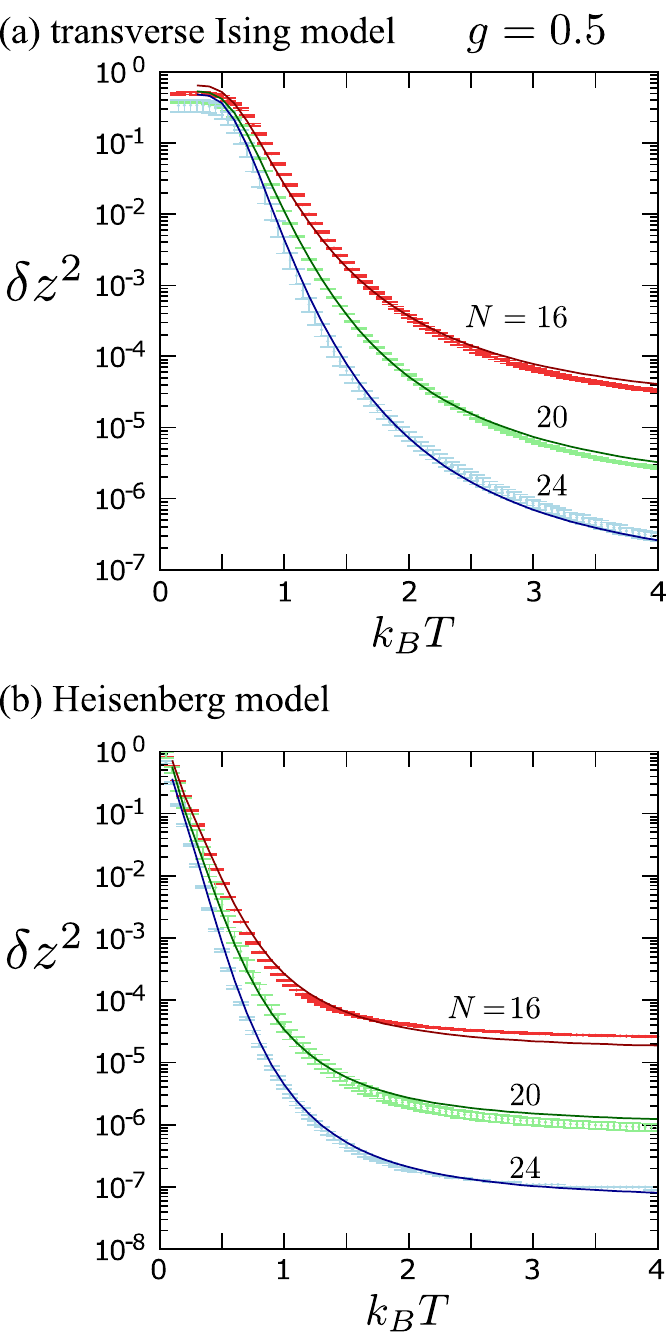}
   \caption{NFPF $\delta z^2$ of the TPQ states for 
      (a) transverse Ising model 
       in the Ne\'el state (we have critical points at $g=1.0$)
      and
      (b) Heisenberg model both in one-dimension. 
      We take $M=20, 15, 8$ samples for $N = 16, 20, 24$, respectively.
      Data points with error bars are directly calculated using Eq.(\ref{eq:def_nfpf}) and 
      solid lines are obtained by using Eq.(\ref{eq:nfpf_tpq}) and $s_\mathrm{th}$ 
      obtained in the same numerical calculations. 
   }
   \label{fig:nfpf_tpq}
\end{figure}
\par
In the TPQ state, the assumption (\ref{eq:assumption}) is satisfied with $(\mathrm{const.})=1$ and
the NFPF can be evaluated as 
\begin{align}
   \delta z^2_\mathrm{TPQ}
   &= \frac{Z(2\beta)}{Z(\beta)^2} = \mathcal{P}_\mathrm{Gibbs} \notag \\
   &= e^{-Ns_\mathrm{th} (\Tilde{\beta})}.
   \label{eq:nfpf_tpq}
\end{align}
Then, from Eq.(\ref{eq:fluctuation}), we immediately see that the random fluctuations of the TPQ method 
decreases exponentially with $N$ and is easily suppressed to negligibly small values. 
In this way, our formulation provides an alternative of Ref.[\onlinecite{sugiura2013}] to evaluate Eq.(\ref{error:tpq}). 
Accordingly, the efficiency is calculated as
\begin{align}
   \eta \simeq 1 - \frac{1}{2} e^{-N s_\mathrm{th}(\Tilde{\beta})} , 
\end{align}
where we find $\eta \rightarrow 1$ in the thermodynamical limit. 
The TPQ method thus has the highest efficiency, 
e.g. for entropy density of $s_\mathrm{th}\sim 0.1J$ at temperatures $0.1\lesssim k_BT/J \lesssim 1$, 
where $J$ is the typical energy scale of the model, it is roughly $\eta \gtrsim 0.9$ for $N \gtrsim 20$. 
\par
In Figs.~\ref{fig:nfpf_tpq}(a) and \ref{fig:nfpf_tpq}(b), we show the NFPF of TPQ states 
obtained for the two models with $N=16,20,24$ (for numerical details, see \S.\ref{sec:full-tpq}). 
We use the definition (\ref{eq:def_nfpf}) to obtain the data points 
and compare them with the solid line derived by the analytical form (\ref{eq:nfpf_tpq}) 
using the numerically obtained $s_{\rm th}(\tilde\beta)$. 
The two are nearly identical. 
In the low-temperature limit, we find $\delta z^2\rightarrow 1$, 
which is verified analytically.  
\par
In this way, the random fluctuation is exponentially small which yields $\mathcal{P} \sim 1$, 
meaning that a single TPQ state can represent the thermal equilibrium. 
In the following, we evaluate $\mathcal{P}_\mathrm{rand}$ 
of other methods by relying on the purity-1 of the TPQ state 
of the same system size $N$. 

\subsection{Definition of purity}
We start by introducing the density operator for 
a mixed state consisting of $M$ samples $\{\ket{\psi_\beta^{(i)}}\}_{i=1}^M$, 
which gives the expectation value in the form of Eq.(\ref{eq:osamp}), 
\begin{align}
   \rho(M) = \sum_{i=1}^M w_{i,M} \frac{\ket{\psi_\beta^{(i)}}\bra{\psi_\beta^{(i)}}}
   {\braket{\psi_\beta^{(i)}|\psi_\beta^{(i)}}}. 
\end{align}
Then, the purity of this mixed state is given as 
\begin{align}
   \Tr[\rho(M)^2]
   &= \sum_{ij}^M \frac{|\braket{\psi_\beta^{(i)}|\psi_\beta^{(j)}}|^2 w_{i,M}w_{j,M}}
   {\braket{\psi_\beta^{(i)}|\psi_\beta^{(i)}} \braket{\psi_\beta^{(j)}|\psi_\beta^{(j)}}}.
\label{eq:rhom2}
\end{align}
We now newly define random vaiables 
\begin{gather}
   Y^{(ij)} = |\braket{\psi_\beta^{(i)}|\psi_\beta^{(j)}}|^2
   \quad (i \neq j) , \\
   Y_0 = \overline{Y^{(ij)}}, \rule{30mm}{0mm}
\end{gather}
and rewrite Eq.(\ref{eq:rhom2}) as 
\begin{align}
   \Tr[\rho(M)^2]
   &= \sum_{i=1}^M w_{i,M}^2
   + \sum_{i \neq j}^M \frac{Y^{(ij)} w_{i,M}w_{j,M}}{y^{(i)} y^{(j)}}. 
   \label{eq:purity1}
\end{align}
The first term is evaluated in the same manner as in \S.\ref{sec:variance_phys} as 
\begin{align}
   \text{(first term)}
   &= \sum_{i=1}^M \left( \frac{1}{M} + \delta w_{i,M} \right)^2 \notag \\
   &= \frac{1}{M} + \sum_{i=1}^M \left( \frac{2\delta w_{i,M}}{M} + \delta w_{i,M}^2 \right),
\end{align}
and by further taking a random average, we obtain 
\begin{align}
   \overline{\text{(first term)}}
   &= \frac{1}{M} + M\overline{\delta w_{i,M}^2} \notag \\
   &= \frac{1}{M} \left[ 1+ \left( 1-\frac{1}{M} \right) \delta z^2 \right]
   + \mathcal{O}\left(\overline{\delta^3}\right). 
\end{align}
The second term of Eq.(\ref{eq:purity1}) is generally 
as small as the purity of the Gibbs state, although it is difficult to 
derive analytically its exact form. 
Still, to estimate its magnitude, the zeroth order of expansion is sufficient, 
and by replacing all the variables with their mean values we find, 
\begin{align}
   \text{(second term)}
   &= \sum_{i \neq j}^M \frac{Y_0}{M^2 y_0^2} + \mathcal{O}(\delta) \notag \\
   &= \left( 1-\frac{1}{M} \right) \frac{Z(2\beta)}{Z(\beta)^2} + \mathcal{O}(\delta) \notag \\
   &= \left( 1-\frac{1}{M} \right) e^{-Ns_{\mathrm{th}}(\Tilde{\beta})} + \mathcal{O}(\delta), 
\end{align}
which is indeed comparable to ${\cal P}_\mathrm{Gibbs}$ in Eq.(\ref{p_gibbs}). 
Since they serve as the lower bound of purity and approach zero exponentially with $N$, 
we consider the second term as an offset, 
and consider the purity ${\cal P}_\mathrm{rand}$ as those measured from this offset. 
By considering only the first term, 
the random average of Eq.(\ref{eq:purity1}) is reduced to 
\begin{align}
   &\overline{\Tr[\rho(M)^2]} \notag \\
   &= \frac{1}{M} \left[ 1+ \left( 1-\frac{1}{M} \right) \delta z^2 \right]
   + \mathcal{O}\left(\overline{\delta^3}\right)
   + e^{-\Theta(N)}.
\end{align}
\par
Finally, we choose a number of random sampling $M$ to a physically meaningful ``appropriate" value $N_\mathrm{samp}$ and rewrite the purity of the obtained mixed state as 
\begin{align}
\mathcal{P}_\mathrm{rand} =
   \frac{1}{N_\mathrm{samp}} \left[ 1+ \left( 1-\frac{1}{N_\mathrm{samp}} \right)
   \delta z^2 \right]
   \label{eq:purity2}
\end{align}

Next discussion is about how the ``appropriate" value $N_\mathrm{samp}$ is determined.
For simplicity, we consider the random state at finite temperature $\ket{\psi_\beta}$
which satisfies
\begin{align}
   \overline{\ket{\psi_\beta} \bra{\psi_\beta}} = \rho_\beta.
\end{align}
The fluctuation of physical quantities is decomposed into two terms.
\begin{align}
   &\langle(\Hat{O}-\langle\Hat{O}\rangle_\beta)^2\rangle_\beta \notag \\
   &= \overline{\braket{\psi_\beta|(\Hat{O}
   -\braket{\psi_\beta|\Hat{O}|\psi_\beta})^2|\psi_\beta}}
   + \overline{\left(\braket{\psi_\beta|\Hat{O}|\psi_\beta}
   -\langle\Hat{O}\rangle_\beta\right)^2}
\end{align}
The first term is the random average of quantum fluctuation and the second term is the random fluctuation. 
In random sampling methods, we try to decrease the random fluctuation. 
The TPQ state maximizes the quantum fluctuation for arbitrary operators and suppresses the random fluctuation, 
and in that sense, it is an ideal random state at finite temperature. 
As a result, one can set $N_\mathrm{samp}=1$ conceptually for TPQ states. 
Then, $N_\mathrm{samp}$ can be defined as the number of samples 
required to obtain physical quantities with the same degrees of accuracy as the TPQ state.  
Since the random fluctuation of a physical quantity is inversely proportional to $N_\mathrm{samp}$, 
and since the random fluctuation is bounded by NFPF, 
we reach the representation, 
\begin{align}
   N_\mathrm{samp} = \frac{\delta z^2}{\delta z^2_\mathrm{TPQ}},
   \label{eq:nsamp}
\end{align}
where $\delta z^2_\mathrm{TPQ}$ is the NFPF of the TPQ method.
\par
By substituting Eq.(\ref{eq:nsamp}) for Eq.(\ref{eq:purity2}),
an explicit representation of the purity is obtained as
\begin{align}
   \mathcal{P}_\mathrm{rand} =
   \frac{\delta z^2_\mathrm{TPQ}}{\delta z^2}
   \left(1+\delta z^2-\delta z^2_\mathrm{TPQ}\right).
   \label{eq:purity}
\end{align}
Since we find $\delta z^2_\mathrm{TPQ} \to 1$ for $\beta \to \infty$, 
$\mathcal{P}_\mathrm{rand} = 1$ is fulfilled at zero temperature for any random sampling methods. 
This property legitimates the definition of purity in Eq.(\ref{eq:purity}) 
because all sampled states generated by the imaginary time evolution approach the pure ground state. 
The form in Eq.(\ref{eq:purity}) allows us to evaluate the purity from the numerically measurable quantities. 
This form is self-contained since $\delta z^2_\mathrm{TPQ}$ is obtained using Eq.(\ref{eq:nfpf_tpq}) 
without knowing the TPQ state itself. 
We note that it does not necessarily exclude the possibility of other ways of describing purity.  
\par
So far, we have developed a set of formulas assuming that $\{|\psi_\beta^{(i)}\rangle\}$ are pure states. 
However, depending on numerical methods the sampled states can sometimes be mixed states. 
In such a case, it may be physically meaningful to consider the corrections to 
$\mathcal{P}_\mathrm{rand}$ by measuring 
the purity for each sampled state. 
The details will be discussed in Appendix \ref{sec:purity2} 
using TPQ-MPS as an example. 


\section{Numerical demonstration}
\label{sec:demonstration}
In this section, we demonstrate that purity can be obtained by evaluating the NFPF and 
using Eq.(\ref{eq:purity}) 
for two different random sampling methods, the RPMPS+T method and the TPQ-MPS method.  
\subsection{MPS based methods}
We consider two representative one-dimensional spin-1/2 models on a chain of length $N$; 
a transverse Ising model whose Hamiltonian reads
\begin{align}
   \Hat{H} = - 4 \sum_{i=1}^{N-1} \Hat{S}_i^z \Hat{S}_{i+1}^z
   -2g \sum_{i=1}^N \Hat{S}_i^x, 
\end{align}
and a Heisenberg model with
\begin{align}
   \Hat{H} = \sum_{i=1}^{N-1} \Hat{\bm{S}}_i \cdot \Hat{\bm{S}}_{i+1}, 
\end{align}
where $\Hat{\bm{S}}_i = (\Hat{S}_i^x, \Hat{S}_i^y, \Hat{S}_i^z)$ is the spin operator at site $i$. 
\par
The two methods we apply are based on the MPS representation of the wave functions. 
The MPS was first proposed \cite{fannes1992} 
and developed as a variational wave function \cite{ostlund1995, rommer1997, dukelsky1998} for the density matrix renormalization group (DMRG) method \cite{white1992, white1993}. 
The general form of MPS in one-dimensional quantum system with open boundary condition (OBC) is 
\begin{align}
   \ket{\psi} = &\sum_{\{\alpha\}} \sum_{\{i\}} 
   A_{\alpha_1}^{[1]i_1} A_{\alpha_1\alpha_2}^{[2]i_2} 
   \cdots A_{\alpha_{N-1}}^{[N]i_N} \notag \\
   &\times \ket{i_1, i_2, \dots, i_N}, 
   \label{eq:mps}
\end{align}
where the matrix $A^{[m]i_m}$ has $\chi \times \chi$ for each local degrees of freedom $i=1,\cdots,d$, 
with $d=2$ for the spin-1/2 models. 
For edge sites, $A^{[1]i_1}$ and $A^{[N]i_N}$ are $1\times \chi$ and $\chi \times 1$, respectively. 
\par
Since the computational memory for the description of MPS scales linearly with $Nd\chi^2$, 
the methods using MPS can afford a much larger system size compared to the diagonalization method using the full Hilbert space. 
However, because of the limited dimension of matrices $\chi$, 
the entanglement entropy of the subsystem 
is bounded as $S_A \le \log\chi \propto N_A^0$, which is the area law in one dimension. 
The ground state of a gapped one-dimensional quantum many-body systems \cite{calabrese2004, hastings2006-2, hastings2007, eisert2010} or 
the excited state with many-body localization \cite{basko2006, bauer2013, swingle2013, friesdorf2015, khemani2016, yu2017}, both having the area law entanglement, are efficiently described by a single MPS. 
Whereas it is evident that a single MPS in Eq.(\ref{eq:mps}) cannot cover the full entanglement 
in a finite temperature state with the volume law entanglement.  
Therefore one needs to mix large numbers of MPS, which is efficiently done in the RPMPS+T method. 
The TPQ-MPS, on the other hand, stores the volume law entanglement by a few sampled pure states; 
they can hold more entanglement by using a special type of MPS in which the auxiliaries are attached to both edges of the open boundary.  
\begin{figure}[tbp]
   \centering
   \includegraphics[width=\hsize]{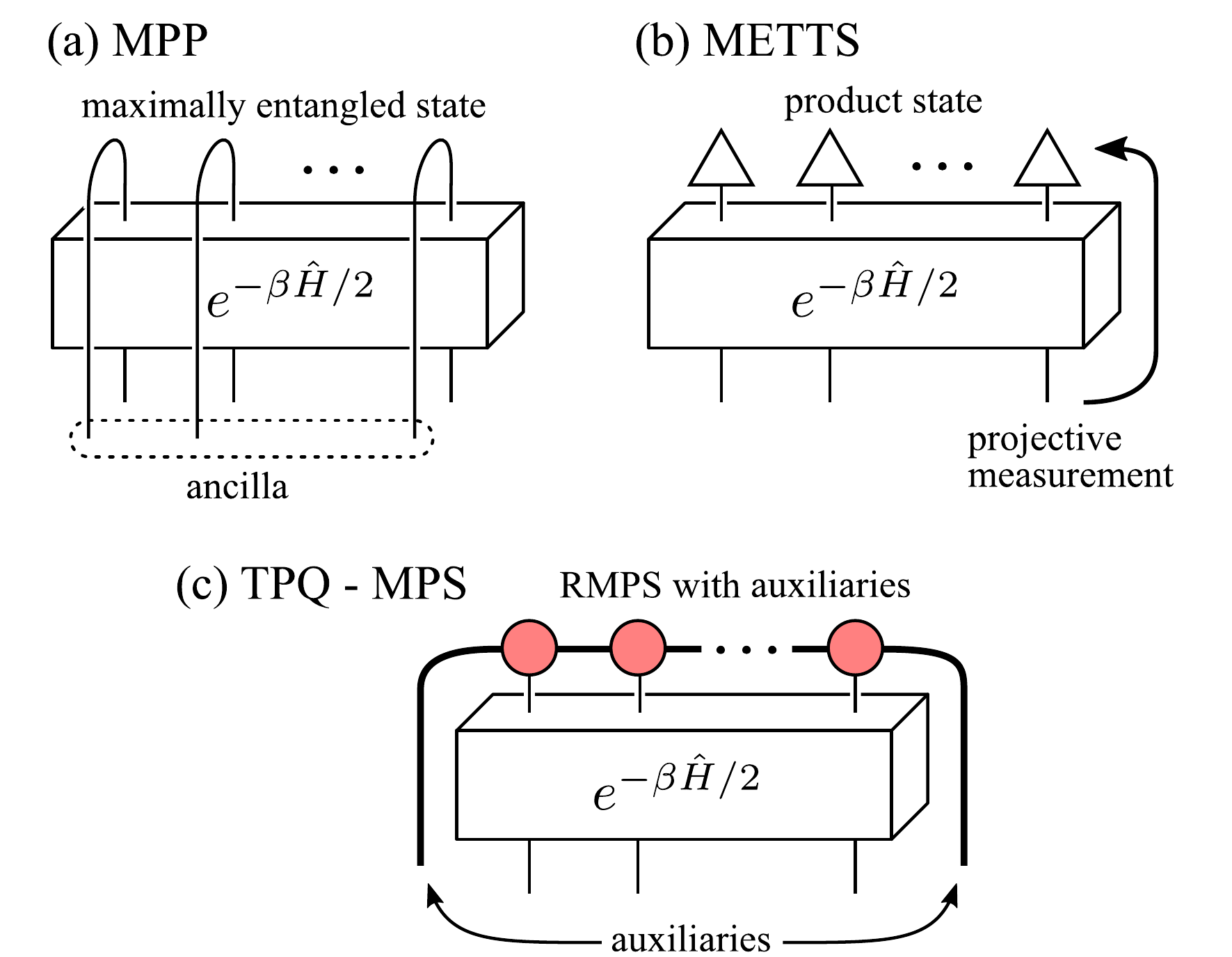}
   \caption{
      Schematic illustration of the way how the imaginary time evolutions in the MPS-based methods for finite temperature are performed. 
      (a) The initial state of MPP is maximally entangled with an ancilla system at each site. 
      The imaginary time evolution is performed only on the physical system which is expected to reach 
      the purified form of the Gibbs state, and finally, the ancilla is traced out. 
      (b) In METTS, for each step, a classical product state is prepared as an initial state 
      that undergoes an imaginary time evolution. Then, by performing an appropriate projective measurement, 
      another classical product state is generated, which is used as the initial state of the next step. 
      These processes form a Markov chain. 
      (c) In the TPQ-MPS method, the highly entangled random matrix product state connected to the auxiliaries 
       is chosen as the initial state and after the imaginary time evolution of the physical system, 
       the auxiliaries are finally traced out.    
   }
   \label{fig:mpsmethods2}
\end{figure}
\begin{figure}[tbp]
   \centering
   \includegraphics[width=\hsize]{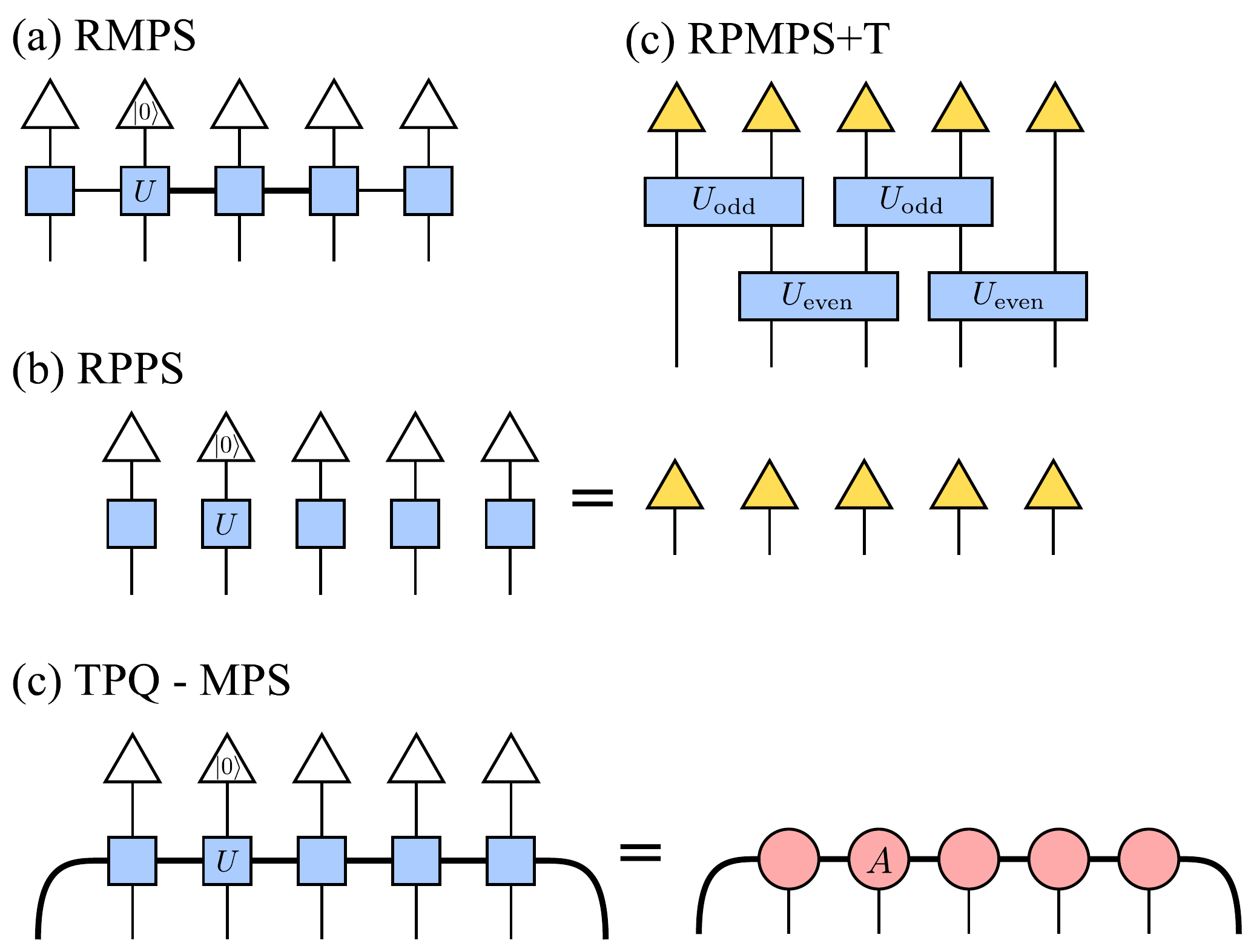}
   \caption{
      Schematic diagrams of initial random states. 
      (a) Standard RMPS with OBC given in the form of Eq.(\ref{eq:mps}) with $A^{[m]i}_{\alpha\beta}=U^{[m]}_{(1,\alpha)(i,\beta)}$. 
      (b) RPPS with each triangle being a single local state, where the system consists of their classical product state. 
      (c) RPMPS+T operating Trotter gate to the RPPS. 
      (d) TPQ-MPS constructed by replacing the $1\times\chi$ and $\chi\times1$ matrices $A^{[1]}$ and $A^{[N]}$ to 
       $\chi\times \chi$ matrices at left and right edges by 
       adding the auxiliaris of $\chi\times\chi$. 
   }
   \label{fig:mpsmethods1}
\end{figure}
\par
A time-evolving block decimation (TEBD) \cite{vidal2004} or time-dependent DMRG (tDMRG) \cite{white2004, daley2004} 
can be used for the imaginary time evolution of MPS, which is incorporated in the core framework 
of the random sampling methods for finite temperatures. 
Technically, the local Hamiltonian can be represented by a simple matrix product operator (MPO) \cite{mcculloch2007}  of bond dimension $\chi_\mathrm{op}$, which allows for continuous operation of the Hamiltonian and simplifies the procedure. The transverse Ising and the Heisenberg models have $\chi_\mathrm{op}=3$ and $5$, respectively.  
\par
Previously, major numerical approaches for finite temperature utilizing MPS or tensor network 
were designed to approximate the density matrix operator or a Gibbs state. 
The matrix product density operator (MPDO) \cite{verstraete2004-2, zwolak2004} approximates the density operator of Gibbs state by the MPO. 
The matrix product purification (MPP) \cite{feiguin2005} combines the physical system and the auxiliaries of the same size 
and describes that doubled system at finite temperature in an MPS form. 
The minimally entangled typical thermal state (METTS) \cite{white2009, stoudenmire2010} has a structure similar to the quantum Monte Carlo method, successively generating a Markov chain of MPS. At each step, the pure MPS state is obtained 
from the classical product state by imaginary time evolution. 
To further accelerate the Markov relaxation of METTS another algorithm 
that gives a better projection to initial state is proposed\cite{binder2017}. 
These methods are schematically shown in Fig.~\ref{fig:mpsmethods2}. 
The direct comparison of MPP and METTS is given in Ref.[\onlinecite{binder2015}], 
which showed that METTS is more efficient at low temperatures. 
\par
Here, we briefly mention the adequacy of using MPS based method for finite temperature calculation. 
It is established that mutual information of the Gibbs state follows an area law \cite{wolf2008}. 
Since the mutual information is a quantity that includes both the classical and quantum correlation \cite{groisman2005}, 
the quantum correlation are localized and shall also follow the area law. 
Consequently, the bond dimension of the MPO representation of the Gibbs state used in MPDO becomes 
moderately small. 
\par
Recently, the random sampling methods based on MPS have been actively studied 
\cite{garnerone2013-1, garnerone2013-2, iitaka2020, iwaki2021,goto2021}. 
Their standard initial state is the random matrix product state (RMPS) for OBC, 
which takes the form of Eq.(\ref{eq:mps}) by using the $d\chi \times d\chi$ random unitary matrix as 
$A^{[m]i}_{\alpha\beta}=U_{(i,\alpha)(1,\beta)}$ or $U_{(1,\alpha)(i,\beta)}$ \cite{garnerone2010-1, garnerone2010-2}. 
These matrices fulfill the left or right canonical form, respectively, and are shown schematically in Fig.~\ref{fig:mpsmethods1}(a). 
The random phase product state (RPPS) is the RMPS with $\chi=1$, 
described in other way round as 
$\ket{\psi_0^\mathrm{RPPS}}= \sum_{\{i\}} \exp(i\theta^{[1]i_1}) \exp(i\theta^{[1]i_2}) \dots  \exp(i\theta^{[1]i_N}) \ket{i_1, i_2, \dots, i_N}$, 
where $\theta^{[m]i_m}$ is chosen uniformly from $[0,2\pi)$ [see Fig.~\ref{fig:mpsmethods1}(b)]. 
\par
Unlike MPDO and MPP, these methods do not need to increase the size of the Hilbert space by orders of magnitude 
, 
and unlike the importance sampling used in METTS, the process of samplings is parallelized. 
However, the efficiency of the random sampling methods depends 
sensitively on how the initial random states are chosen. 
For example, in RPPS \cite{iitaka2020}, the efficiency of the Heisenberg model with $N=100$ is 
less than 0.05, which corresponds to $\delta z^2\sim 6$, 
and since $\delta z_\mathrm{TPQ}^2$ is less than $10^{-8}$ from our evaluation (see Fig.~\ref{fig:nfpf_tpq}), 
$N_\mathrm{samp}=\delta z^2/\delta z_\mathrm{TPQ}^2$ can be huge. 
In contrast, by constructing MPS wave functions with a relatively higher entanglement, 
RPMPS+T and TPQ-MPS methods are able to efficiently reduce the number of samplings. 
We demonstrate it quantitatively by using our framework in the following 
\S.\ref{sec:rpmpst} and \S.\ref{sec:tpqmps}. 
The schematic illustrations of these highly entangled MPS wave functions are shown 
in Figs.~\ref{fig:mpsmethods1}(c) and \ref{fig:mpsmethods1}(d) which will be explained shortly.

\subsection{Full-TPQ method}
\label{sec:full-tpq}
There are several methods to obtain the finite temperature states equivalent to full-TPQ states, 
such as the quantum transfer Monte Carlo method \cite{imada1986}, the finite temperature Lanczos methods \cite{jaklic1994}, 
and the TPQ methods \cite{hams2000,sugiura2012, sugiura2013}. 
The variations among them depend on how the initial random states are prepared, 
and the comparative studies are given in Ref.[\onlinecite{jin2021}]. 
Here, we adopt the {\it unnormalized} initial random states $\ket{\psi_0}$, 
slightly different from the Haar measure used in Ref.[\onlinecite{sugiura2012}], 
and apply the protocol called microcanonical TPQ (mTPQ) method \cite{sugiura2012}. 
Instead of directly calculating the time evolutions in Eq.(\ref{tpqstate}), 
this method successively applies $(l-\Hat{h})$ to the initial state where 
$l$ is a real number larger than the maximal eigenvalue of $\Hat{h}= \Hat{H}/N$. 
The $k$-th mTPQ state, $\ket{\psi_k} = (l-\Hat{h})^k \ket{\psi_0}$, is one of the 
TPQ states belonging to the energy shell of energy density $u_k = \braket{\psi_k|\Hat{h}|\psi_k}/\braket{\psi_k|\psi_k}$. 
The temperature of this energy shell is given within  the accuracy of $\mathcal{O}(1/N)$ \cite{sugiura2012, yoneta2019} as 
\begin{align}
   k_B T_k = \frac{1}{\beta_k} = \frac{N(l-u_k)}{2k}, 
   \label{eq:temperature}
\end{align}
which decreases roughly inversely proportional to $k$. 
The full-TPQ state is written using a set of $\ket{\psi_k}$ as
\begin{align}
   \ket{\psi_\beta} = e^{-N\beta l/2} \sum_{k=0}^\infty \frac{1}{k!}
   \left(\frac{N\beta}{2}\right)^k \ket{\psi_k} .
\end{align}
We prepare a set of mTPQ states $k=1,\cdots,k_\mathrm{max}$, to obtain the physical properties 
within the temperature range of $k_BT \gtrsim k_BT_{k_\mathrm{max}}$. 
The NFPF in Fig.~\ref{fig:nfpf_tpq} are obtained using this protocol.

\subsection{RPMPS+T method} 
\label{sec:rpmpst}
\begin{figure*}[tbp]
   \centering
   \includegraphics[width=\hsize]{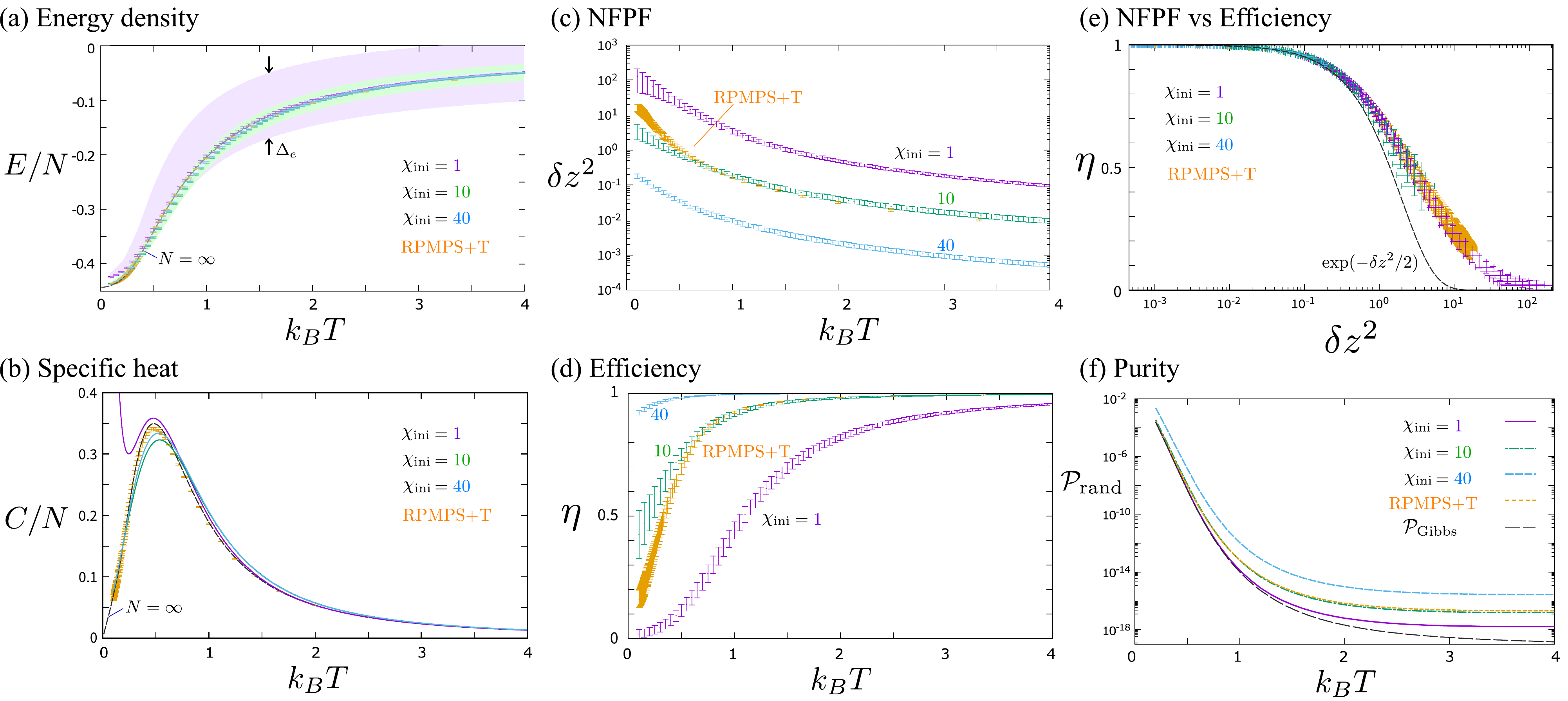}
   \caption{
      (a) Energy density $E/N$ and (b) specific heat $C/N$ 
      for the Heisenberg chain of $N=64$ calculated using RPMPS+T method and TPQ-MPS method. 
      We take $M=500$ samples for RPMPS+T method and TPQ-MPS method with $\chi_\mathrm{ini} =1$, and
      $M=100$ samples for TPQ-MPS method with $\chi_\mathrm{ini} =10, 40$.
      The exact solution of $N=\infty$ obtained by the quantum transfer matrix method \cite{klumper1998} 
      is shown for reference. 
      Regions highlighted in purple($\chi_\mathrm{ini} =1$), green(10) and blue(40) are 
      the range of distribution of the sampled data.
     (c) NFPF $\delta z^2$ and (d) efficiency $\eta$ evaluated from Eq.(\ref{eq:eta})
     for the Heisenberg model with $N=64$.
     (e) Relationships between $\eta$ and $\delta z^2$ for the data points of all temperatures. 
     (f) The purity of RPMPS+T and TPQ-MPS $\mathcal{P}_\mathrm{rand}$, 
     forming a TMQ state from $N_\mathrm{samp}\times \chi_\mathrm{ini}^2$ effective samples. 
}
   \label{fig:hei_purity}
\end{figure*}

The RPMPS+T approach chooses RPPS as initial random states \cite{iitaka2020}. 
If one straightforwardly performs an imaginary time evolution to RPPS, the lack of important sampling 
makes the result inefficient by several orders of magnitude than the other approaches. 
The RPMPS+T overcomes this issue by operating a Trotter gate to RPPS, 
which is the unitary transformation making the state entangled \cite{goto2021}. 
The schematic diagram of the RPMPS+T is shown in Fig.~\ref{fig:mpsmethods1}(c). 
The Trotter gate is described as a unitary operator
\begin{align}
   \Hat{U} = e^{-i\tau\Hat{H}_\mathrm{even}'}
   e^{-i\tau\Hat{H}_\mathrm{odd}'},
\end{align}
where $\tau$ is a real number which we choose as 0.5 and $\Hat{H}_\mathrm{even(odd)}'$ is a sum of even(odd)-bond interactions of Trotter Hamiltonian. 
Following Ref.[\onlinecite{goto2021}], for the Heisenberg model we apply a spin-1/2 XXZ chain Hamiltonian as Trotter gate, 
\begin{align}
   \Hat{H}' = \sum_{i=1}^{N-1} \left( \Hat{S}_i^x \Hat{S}_{i+1}^x
      +\Hat{S}_i^y \Hat{S}_{i+1}^y +J_z \Hat{S}_i^z \Hat{S}_{i+1}^z
   \right), 
\end{align}
with $J_z = 9.0$ and take the number of gates to operate as $n=1$. 
For the calculation of the Heisenberg model with $N=64$ we take $M=500$, 
which reproduced the results of purification in Ref.[\onlinecite{goto2021}]. 
We plot the energy and specific heat of RPMPS+T down to $k_BT=0.1J$ in Fig.~\ref{fig:hei_purity} together with the results of 
TPQ-MPS which we see shortly. 
These RPMPS+T data reproduces the $N=\infty$ result of the exact solution \cite{klumper1993, klumper1998}. 
For the same RPMPS+T states, we calculate $\delta z^2$, $\eta$ and $\mathcal{P}_\mathrm{rand}$ 
as shown in Fig.~\ref{fig:hei_purity}. 
We find a high enough efficiency $\eta \ge 0.6$ that reproduces Ref.\cite{goto2021}. 
At $k_BT \sim 0.5$, 
we find $\delta z^2 \sim 1$, while since $\delta z^2_\mathrm{TPQ} \sim 10^{-6}$ for the corresponding full-TPQ state, 
the purity is suppressed to $\mathcal{P}\lesssim 10^{-6}$. 

\subsection{TPQ-MPS method}
\label{sec:tpqmps}
The TPQ-MPS shown in Fig.~\ref{fig:mpsmethods1}(d) is proposed by the authors [\onlinecite{iwaki2021}]. 
There, we confirmed that TPQ-MPS fulfills the entanglement volume law in Eq.(\ref{eq:sA}) 
throughout the whole system at low temperature 
when the entropy density is $s_\mathrm{th}\lesssim (2\log\chi) /N$, 
e.g. $k_BT \lesssim 0.5$ with $s_\mathrm{th}\lesssim 0.1J$ for $N=64$, 
while the volume law is satisfied for shorter lengthscale for higher temperatures. 
Since typicality and the entanglement volume law are two sides of the same coin, 
TPQ-MPS is regarded as a good approximation of the TPQ state. 
To overcome the small area-law bound of the entanglement, 
the TPQ-MPS abandoned the standard expression of MPS in Eq.(\ref{eq:mps}) 
and attaches the auxiliary systems at both edges, 
having the same degree of freedom as $\chi$. 
These auxiliaries are the order-1 physical degrees of freedom 
that are not interacting with the physical system, 
and thus can be practically neglected when $N$ is large enough. 
However, this small modification prevents the entanglement entropy 
from going to zero at both edges of the system: 
the entanglement entropy can grow with subsystem size up to $N_A \rightarrow N$ 
without ``feeling" the bounds of entanglement, $S_A\lesssim 2\log\chi$. 
This allows us to successfully extract the thermal entropy from the entanglement entropy \cite{iwaki2021}. 
By contrast, the Page curve of the standard MPS decreases to zero near the boundary, 
significantly deteriorating the volume law behavior. 
\par
In the TPQ-MPS method, we start from RMPS with auxiliaries at both edges, 
\begin{align}
   \ket{\psi_0}
   = &\sum_{\{\alpha\}} \sum_{\{i\}} A_{\alpha_0\alpha_1}^{[1]i_1}
   A_{\alpha_1\alpha_2}^{[2]i_2} \cdots A_{\alpha_{N-1}\alpha_N}^{[N]i_N} \notag \\
   &\times \ket{\alpha_0, i_1, i_2, \dots, i_N, \alpha_N}, 
   \label{eq:rmps}
\end{align}
where we take the bond dimension $\chi=\chi_\mathrm{ini}$ for all matrices. 
We perform the same set of calculation as mTPQ in \S.\ref{sec:full-tpq}, 
generating $\ket{\psi_k} = (l-\Hat{h})^k \ket{\psi_0}$ successively. 
During the calculation, the bond dimension $\chi$ becomes larger than $\chi_\mathrm{ini}$, 
which is decided so as to keep the truncation error less than a given constant. 
Since $\chi$ varies depending on $k$ and $l$, 
but its denendence is ruled by the initial choice $\chi_\mathrm{ini}$, the plots we made are classified 
according to $\chi_\mathrm{ini}$. 
The $N$-dependence of the results is negligibly small as we demonstrated in Ref.~[\onlinecite{iwaki2021}]. 
Some details updated from Ref.[\onlinecite{iwaki2021}] is given in Appendix \ref{sec:detail_tpqmps}. 

\begin{figure*}[tbp]
   \centering
   \includegraphics[width=\hsize]{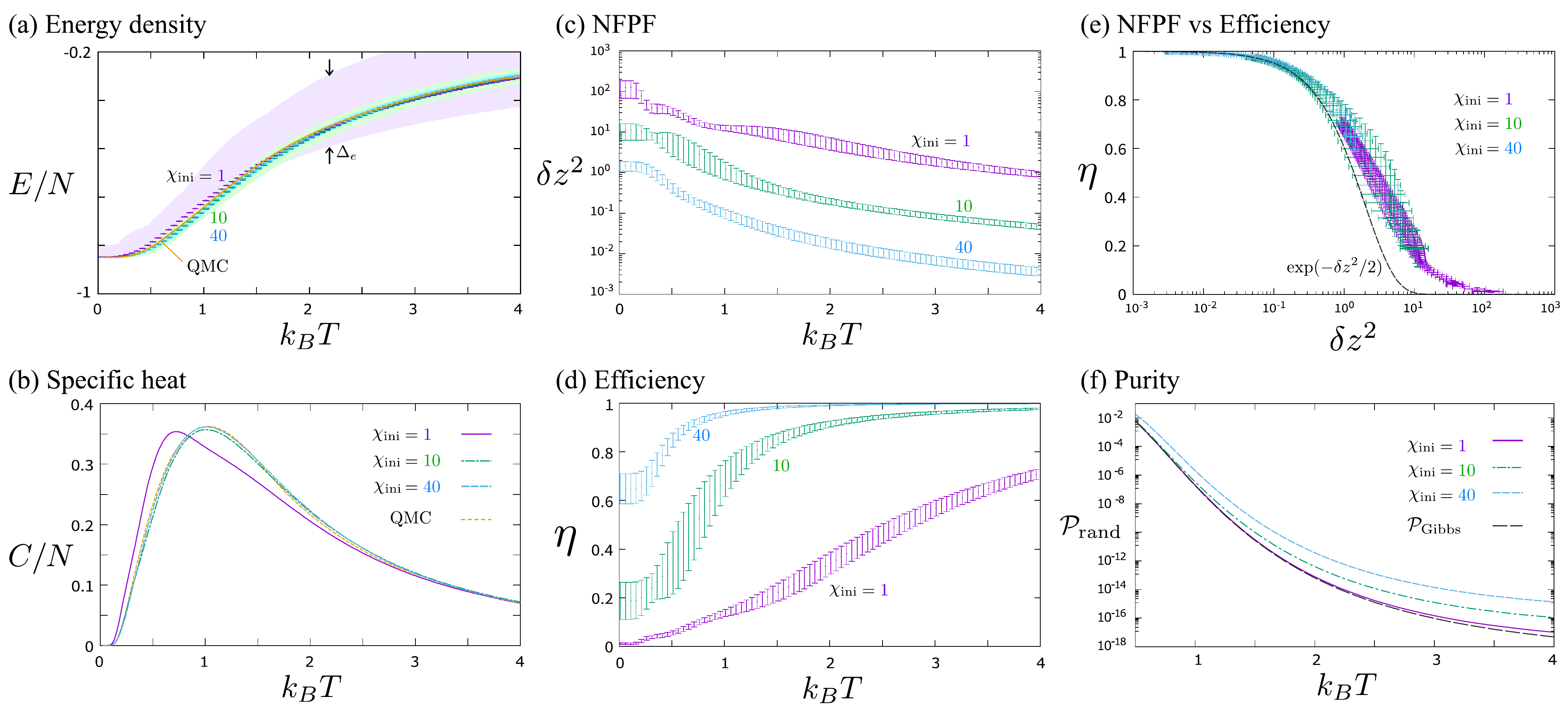}
   \caption{
      Results of the transverse Ising model with $g=0.5, N=64$ 
      using TPQ-MPS with $l=5.0$ and $\chi_\mathrm{ini} =1, 10, 40$ 
      (a) Energy density $E/N$, (b) specific heat $C/N$, 
      (c) NFPF $\delta z^2$ and (d) efficiency $\eta$ as functions of temperature. 
      The number of samples are taken as $M=500$ with $\chi_\mathrm{ini}=1$ and $M=100$ with $\chi_\mathrm{ini}=10, 40$. 
       In (a), the ranges of distribution of the sampled data are 
       highlighted in purple ($\chi_\mathrm{ini}=1$), green ($\chi_\mathrm{ini}=10$), and blue ($\chi_\mathrm{ini}=40$). 
      (e) Relationships between $\eta$ and $\delta z^2$. 
      The data points with $\chi_\mathrm{ini} =1, 10, 40$ and different temperatures form a unique line.
      The solid line is analytical prediction. 
      In the region in which the NFPF is small, 
      numerical results coincide with analytical prediction
      but in the region in which the NFPF is large, it does not.
      This can be attributed to the influence of the term $\mathcal{O}(\overline{\delta^3})$
      in Eq.(\ref{eq:efficiency}).
      (f) The purity from Eq.(\ref{eq:purity}) 
      of the TPQ-MPS method.
   }
   \label{fig:purity_tri}
\end{figure*}
%
%
\begin{figure*}[t]
   \centering
   \includegraphics[width=\hsize]{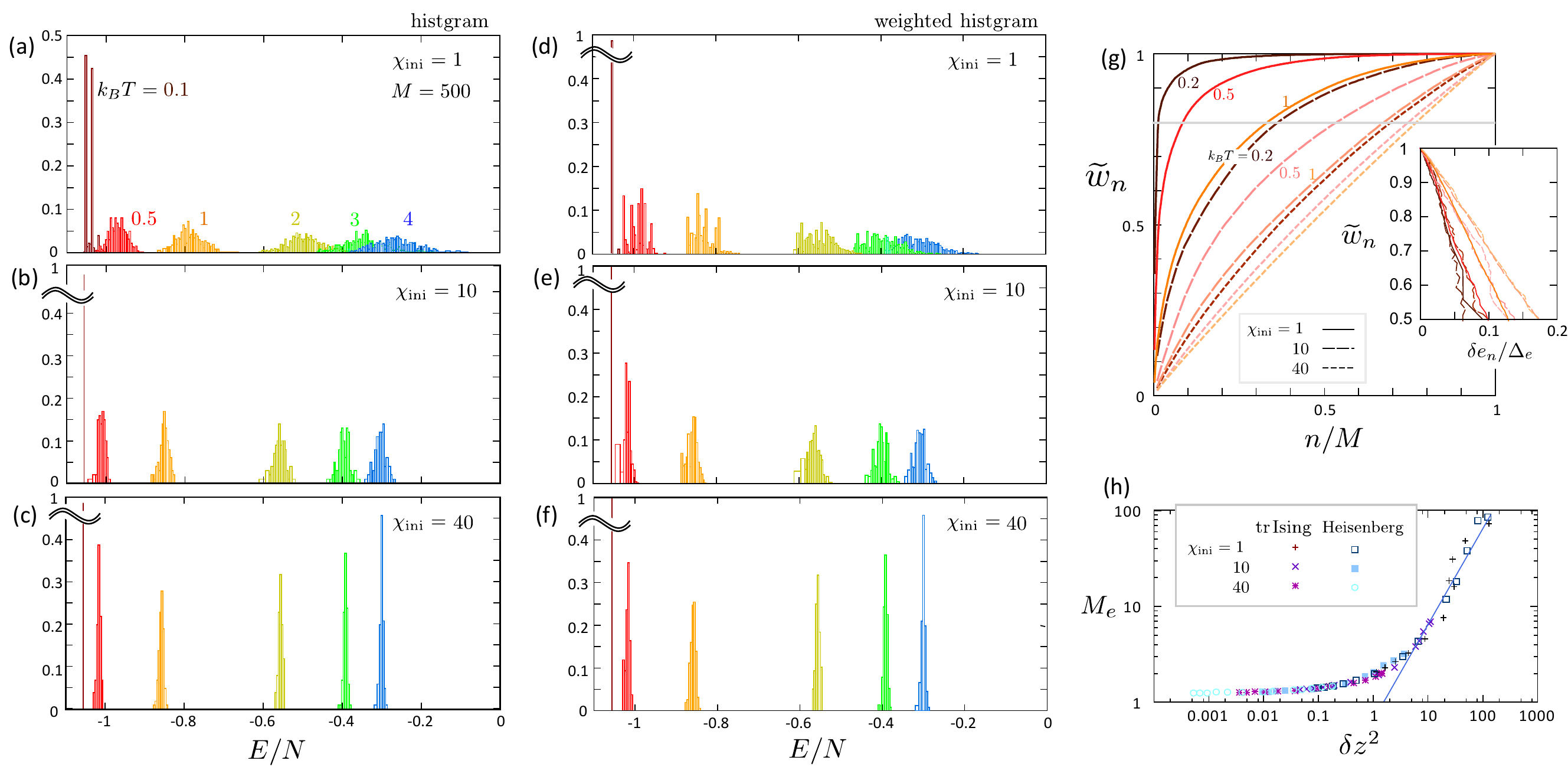}
   \caption{
      Histograms (a)-(c) and the weighted histograms (d)-(f) of the sampled energy density $E/N$ 
      of the transverse Ising model calculated by TPQ-MPS in Fig.~\ref{fig:purity_tri}(a). 
      We show the results for $\chi_{\rm ini}=1,10,40$ and $k_BT=0.1, 0.5, 1,2,3,4$. 
      (g) Integrated $n$-largest weights $\tilde w_n$ among $M$ samples of the transverse Ising model 
       as a function of $n/M$. The inset shows $\widetilde w_n$ as a function of $\delta e_n/\Delta_e$. 
       Here, $\delta e_n$ is the difference of energy evaluated for $n$ and $M$-samples from the largest weight. 
      (h) Number of samples $M_e=M/n$ giving $\widetilde w_n=0.8$ as a function of $\delta z^2$ 
      for the transverse Ising and Heisenberg models. Solid line $M_e \propto \delta z^2$ is the guide to the eye. 
   }
   \label{fig:samp}
\end{figure*}
To evaluate the quality of the TPQ-MPS, 
we first perform the same set of calculation of as RPMPS+T for the Heisenberg model with $N=64$. 
Here, we take $\chi_\mathrm{ini}=1,10,40$, $l=1.0$ and $k_\mathrm{max}=500$ 
which allows us to reach $k_BT_\mathrm{min} \sim 0.1$. 
The number of sample average is taken as $M=500$ for $\chi_\mathrm{ini} =1$ and $M=100$ for $\chi_\mathrm{ini}=10, 40$ 
to sufficiently suppress the sample variance of NFPF. 
However, for standard physical quantities such as energy, the reduced $M$ for $\chi_\mathrm{ini} \ge 10$ 
do not change the variance much. 
Here, notice that $M$ used for each calculation generally depends on the quantities 
and is different from $N_\mathrm{samp}$ defined to qualify the TMQ states. 
Here, TPQ-MPS with $\chi_\mathrm{ini} =1$ exactly corresponds to RPPS. 
\par
Figures~\ref{fig:hei_purity}(a) and \ref{fig:hei_purity}(b) show the energy and the specific heat. 
These results are shown to confirm that they overall agree with each other, 
where we judge that they converge to the values with visible but small differences 
due to different $\chi_{\rm ini}$ or to different methods. 
We have previously shown in Ref.[\onlinecite{iwaki2021}] that 
energy density, specific heat, and susceptibility calculated by TPQ-MPS for $N=16$ with $\chi=20$ 
agrees well with those obtained by the exact (full) diagonalization of the same system size. 
Here, our data for $N=64$ have no reference data to directly compare with, 
while the exact solution for $N=\infty$ takes a reasonably close value. 
The highlighted range indicates the distribution of the sampled data for different $\chi_{\rm ini}$'s. 
Its relationship with the sample number $M_e$ and $\chi_{\rm ini}$ 
will be discussed shortly in \S.\ref{sec:sample}. 
\par
Figures~\ref{fig:hei_purity}(c) and \ref{fig:hei_purity}(d) 
show NFPF and the efficiency, respectively, 
where the differences between different numerical conditions become visible. 
The ones obtained for $\chi_\mathrm{ini}=10$ agree well with the RPMPS+T results. 
From the numerical data, we find $\delta z^2 \propto \chi_\mathrm{ini}^{-2}$, 
which means that $N_\mathrm{samp}$ decreases as $\propto \chi_\mathrm{ini}^{-2}$. 
Intuitively, $\chi_\mathrm{ini}^2$ is the number of auxiliary degrees of freedom attached to the system, 
which replaces the samplings. 
Since the numerical cost of increasing $\chi_\mathrm{ini}$ by a few times is small, 
the TPQ-MPS can gain a high efficiency as 
\begin{equation}
   \eta = \exp[\Theta(\chi_\mathrm{ini}^{-2})], 
\end{equation}
which agrees with Eq.(\ref{eq:efficiency}). 
In our demonstration, by increasing $\chi_\mathrm{ini}$ from 10 to 40, we are able to increase the 
efficiency from $\eta \gtrsim 0.5$ to $\sim 1$ at the lowest temperature. 
\par
Meanwhile, focusing only on the bond dimension, the numerical cost of taking one sample is roughly estimated as 
$\mathcal{O}(\chi_\mathrm{ini}^3)$ considering the multiplications of the matrices, 
and if the practical number of samples required should increase as 
$\propto \delta z^2 \propto \chi_\mathrm{ini}^{-2}$, 
the overall cost will increase by $\propto \chi_\mathrm{ini}$. 
However, we further made analytical calculations assuming the RMPS state, 
and found that $\delta z^2$ can be expanded by $\chi_\mathrm{ini}$ to higher order as 
\begin{align}
   \delta z^2 = \frac{a}{\chi_\mathrm{ini}^2} 
   + \frac{b}{\chi_\mathrm{ini}^4} 
   + \mathcal{O} \left( \frac{1}{\chi_\mathrm{ini}^6} \right)
\end{align}
where $a$ and $b$ are constants \cite{iwaki-prep}. 
Accordingly, the estimated numerical cost is modified to 
\begin{align}
   a\chi_\mathrm{ini} + \frac{b}{\chi_\mathrm{ini}}, 
\end{align}
which no longer increases linearly with $\chi_{\rm ini}$, 
but takes a minimal value at finite $\chi_\mathrm{ini}$. 
Therefore considering both the numerical resources and 
the relationships between $\delta z^2$ and the accuracy of the physical quantity at focus, 
the optimal $\chi_\mathrm{ini}$ can be chosen. 
\par
In Fig.~\ref{fig:hei_purity}(e), the NFPF is plotted as a function of efficiency 
for all different choices of $\chi_\mathrm{ini}$ and $k_BT$. 
They collapse to a single curve, which follows the analytical form Eq.(\ref{eq:efficiency}) shown in solid line 
where the NFPF $\delta z^2$ is small. 
The formulation given in the previous section is thus numerically confirmed. 
Figure~\ref{fig:hei_purity}(f) is the purity $\mathcal{P}_\mathrm{rand}$. 
Again, we find good agreement between TPQ-MPS with $\chi_\mathrm{ini}=10$ and the RPMPS+T result. 
\par
To show that the tendencies found in the Heisenberg model hold for other models, 
we perform the same set of calculations for the transverse Ising model 
with $g=0.5$ for which the ground state is in the N\'eel ordered state. 
Figures \ref{fig:purity_tri}(a)-\ref{fig:purity_tri}(d) show the energy, specific heat, $\delta z^2$ and 
the efficiency $\eta$. 
We choose $N=64, l=5.0$, and $M=500$ with $\chi_\mathrm{ini}=1$, and $M=100$ with $\chi_\mathrm{ini}=10$ and $40$. 
The energy and the specific heat at $\chi_{\rm ini}=10$ and $40$ show very good agreement with 
the quantum Monte Carlo (QMC) results of the same $N=64$; 
the results of the TPQ-MPS and QMC should ideally coincide, 
and the slightly visible differences remain smaller than the errors of the QMC results 
(see also Ref.[\onlinecite{iwaki2021}] ). 
With increasing $\chi_\mathrm{ini}$, $\eta$ increases particularly at higher temperatures, 
indicating that choosing a large initial value of $\chi_\mathrm{ini}$ works effectively 
to obtain accurate numerical results in TPQ-MPS which is generally 
known for other variational methods like DMRG.  
In Fig.~\ref{fig:purity_tri}(e) we again find that $\eta$ as a function of $\delta z^2$ 
for various $\chi_\mathrm{ini}$ and temperature collapes to a single curve. 
Figure~\ref{fig:purity_tri}(f) shows the purity $\mathcal{P}_\mathrm{rand}$. 
The same tendency as those of the Heisenberg model holds, 
while the absolute values of purity becomes larger, because $\delta z^2_\mathrm{TPQ}$ differs 
as we saw in Fig.~\ref{fig:nfpf_tpq}. 
\subsection{Number of samples}
\label{sec:sample}
We want to estimate the number of required samples $M_O$ in measuring the physical quantity $\hat O$ 
to support its relationship between our analytical results Eq.(\ref{eq:nsamp}). 
The physical implication of Eq.(\ref{eq:nsamp}) was such that 
the number of required samples is proportional to $\delta z^2$, 
while in the actual calculation, $M_O$ differs depending on the choice of $\hat O$ 
even though we choose the same Hamiltonian and the model parameters. 
\par
For a series of calculations performed in Figs.~\ref{fig:hei_purity} and \ref{fig:purity_tri}, 
we try to make reasonable connection between $M_e$ of the energy density $e=E/N$ and $\delta z^2$. 
In Figs.~\ref{fig:hei_purity}(a) and \ref{fig:purity_tri}(a), we presented the range of distribution 
of sampled data when taking $M=500$ for $\chi_{\rm ini}=1$ and $M=100$ for $\chi_{\rm ini}=10, 40$, 
highlighted with different colors, which we denote as $\Delta_e$. 
The histograms of the distribution of the sampled data of the transverse Ising model 
are shown in Figs.~\ref{fig:samp}(a)-\ref{fig:samp}(c) for $\chi_{\rm ini}=1,10,40$, 
where the width of the histogram is $\Delta_e$. 
\par
In general, the histograms are expected to follow a Gaussian distribution and the width of the 
histogram is expected to be roughly proportional to $M_{e}^{1/2}$. 
If this standard applies, the numerical error of our data becomes smaller for lower temperatures. 
It, however, contradicts the general tendency 
that the numerical error is larger for lower temperatures. 
Therefore it is natural to consider that the accuracy of the data does not simply scale with $\Delta_e$. 
\par
To clarify this point, we present the modified histograms measured not by $1$ for each sample but by 
the sample-dependent weight $w_{i,M}$, in Eq.(\ref{eq:weight}) 
as shown in Figs.~\ref{fig:samp}(d)-\ref{fig:samp}(f). 
By comparing them with Figs.~\ref{fig:samp}(a)-\ref{fig:samp}(c), 
we find that for $\chi_{\rm ini}=1$, the weighted histogram has a larger weight on the lower energy side, 
and particularly at low temperature, $k_BT=0.1$, one of the two peaks dissapear, 
namely it has no influence on the averaged value. 
For $\chi_{\rm ini}=40$, the weight seems to distribute uniformly while the width of the histogram 
is already narrow and the obtained results should show enough accuracy, 
as has been shown in Figs.\ref{fig:purity_tri}(a) and \ref{fig:purity_tri}(b) in comparison with the QMC results. 
These results indicate that for small $\chi_{\rm ini}$ and low temperatures, the width of the histogram 
(or its variance) does not provide a measure for numerical accuracy. 
\par
To extract the reasonable estimate of the quality of the hisgram, 
we arrange the sampled data in the descending order of their weight, 
and integrate them up to $n$-largest weight, $\widetilde w_n=\sum_{i=1}^{n} w_{i,M}$, 
as shown in Fig.~\ref{fig:samp}(g). 
When $\chi_{\rm ini}=1$ and $k_BT=0.2$, only $n=6$ samples among $M=500$ contribute to 
80$\%$ of the whole weight, whereas for $\chi_{\rm ini}=40$ and $k_BT=1$, 
$\widetilde w_n$ increases almost linearly with $n/M$, namely the distribution is close to uniform and $\eta\sim 1$. 
Although we are not able to measure the accuray of $E/N$ on the same ground 
for these different numerical conditions, 
the diffence between the energy densities 
evaluated for these top $n$-samples and for $M$-samples, 
$\delta e_n$, will provide us a clue. 
We show in the inset the relationships between $\tilde w_n$ and the $\delta e_n/\Delta_e$, 
which behave linearly, having the slopes of the same order. 
Based on these results, we estimate $M_e$ as such that 
the value of $n$ that gives $\tilde w_n=0.8$ [gray line in Fig.~\ref{fig:samp}(g)], 
to be $1/M_e$; this means that we need to collect $M_e$-sample-data
in order to accumulate the weight of $\widetilde w_n=0.8$ of the histogram. 
Figure~\ref{fig:samp}(h) shows $M_e$ as a function of $\delta z^2$ 
obtained for $k_BT=0.1, 0.5, 1$ and $\chi_{\rm ini}=1,10,40$ for the two models. 
When $\delta z^2 \gtrsim 1$, the data scale linearly, whereas 
for smaller $\delta z^2$, we already find $M_e$ to be of order 1, which means 
that the system is relatively pure and we only need a few samples. 
We checked that $M_e$ obtained by setting, e.g., $\widetilde w_n=0.9$, differs only 
by few factors without changing the profile of Fig.~\ref{fig:samp}(g). 
\par
The practical usage of $M_e$ should be such that to compare the quality of data 
from different numerical conditions such as $\chi_{\rm ini}$, 
or the data from different methods. 
The difference in $\delta z^2$ tells us the difference in the order $M_e$ to find a similar quality of accuracy. 
In fact, the RPPS method corresponding to our $\chi_{\rm ini}=1$ 
was reported to require more than a few hundred samples, 
and in Fig.~\ref{fig:samp}(h) we indeed see almost two orders of magnitude different $M_e$ 
compared to our TPQ-MPS with $\chi_{\rm ini}=10$ and $40$. 
\section{summary and discussion}
\label{sec:summary}
We introduced a systematic way of viewing a series of thermal equilibrium states with purity $0< {\mathcal P} <1$, 
which we named ``thermal mixed quantum (TMQ) states". 
Our theory is based on the framework of random sampling method; 
starting from a set of $M$ independent states $\{\ket{\psi_0^{(i)}}\}_{i=1}^M$ 
generated from a random distribution, 
we performed an imaginary time evolution to obtain a set of states as, 
$\ket{\psi_\beta^{(i)}}=e^{-\beta\Hat{H}/2} \ket{\psi_0^{(i)}}$. 
The norm $\langle \psi_\beta^{(i)}| \psi_\beta^{(i)}\rangle$ 
averaged over the samples is the partition function $Z(\beta)$ 
of a TMQ state consisting of $\{\ket{\psi_\beta^{(i)}}\}_{i=1}^M$. 
At the same time, since this norm depends on $i$, 
the average of the physical quantities over $i=1- M$ becomes a weighted average, 
with its weight being proportional to the norm. 
The efficiency of the random sampling method is the largest when 
all the samples equally contribute, which means zero fluctuation of the norm. 
Whereas the efficiency is lowered when the norm largely varies from sample to sample. 
Based on this consideration, we introduced a quantity called 
``normalized fluctuation of the partition function (NFPF)," denoted as $\delta z^2$, 
and showed analytically that the efficiency is described as $\eta=e^{-\delta z^2/2}$.
The physical quantities evaluated by most of the TMQ states obtained by random sampling methods 
are bounded by the NFPF. 
This fact further endows the NFPF with the physical meaning that 
it provides a number of samples $M=N_\mathrm{samp}$ to form a TMQ state. 
However, $N_\mathrm{samp}$ itself remains conceptual, 
since in practice, the necessary and sufficient number of samples $M_O$ to properly evaluate the 
physical quantity $\hat O$ differs for different $\hat O$. 
We showed numerically that $M_e$ for the energy density is proportional to $\delta z^2$ 
when the TMQ state is enough far from pure. 
\par
The density matrix of the TMQ state is given as a weighted mixture of 
the density matrix of sampled pure states, 
and the purity, which is the trace of the square of this density matrix, is described using NFPF. 
Accordingly, the purity of a TMQ state is expressed solely by NFPF, 
and its form is roughly the ratio of NFPF of the TPQ state against that of the TMQ state. 
Therefore the purity evaluated using NFPF has a physical implication of 
an effective dimension of the TMQ state. 
Previously, purity had rather been a conceptual quantity that was not available unless the wave function of the mixed state 
was known {\it a priori}. 
Our theory provides the explicit form of purity and related quantities 
that are calculated in numerical experiments.
We successfully applied the MPS-based random sampling methods, TPQ-MPS and RPMPS+T, 
to our theory and confirmed our analytical results. 
\par
We note here that the present theory cannot be directly applied to Markov-chain-based random sampling methods, 
such as the quantum Monte Carlo method and METTS. 
This is because in our analytical calculations we have performed the random average by assuming that the sampled states are independent of each other. 
With a few modifications, this problem can be resolved; 
when employing Markov chains, we can guarantee the independence between sampled states by 
taking samples at time intervals that are sufficiently longer than the timescale of the Markov chain's autocorrelation. 
NFPF and purity are evaluated by carefully performing these calculations. 
However, NFPF is no longer a suitable signal for determining $N_\mathrm{samp}$. 
This is because $N_\mathrm{samp} \propto$ (NFPF) $\times$ (autocorrelation), 
where the second biggest eigenvalue of the transition matrix can be used to estimate the autocorrelation of a Markov chain. 
\par
While the present theory targets a TMQ state, 
our formulas are straightforwardly applied to any of the mixed states generated by random sampling methods, 
regardless of the construction of the wave function or the numerical details to calculate the TMQ states. 
Suppose that we want to obtain a mixed state that follows a distribution function represented by 
an operator $\Hat{F} = F(\Hat{H})$. 
The expectation values of operator $\Hat{O}$ for this distribution are given as
\begin{align}
   & \langle \Hat{O} \rangle_F = \frac{\Tr(\Hat{F}\Hat{O})}{Z_F}, 
\label{eq:extend}
\end{align}
with $Z_F = \Tr(\Hat{F})$. 
A mixed state $\{\ket{\psi_F^{(i)}}\}$ obtained by the random sampling method 
can approximate Eq.(\ref{eq:extend}) as 
\begin{align}
   & \langle \Hat{O} \rangle_{F, M}^\mathrm{samp} = \frac{\sum_{i=1}^M \braket{\psi_F^{(i)}|\Hat{O}|\psi_F^{(i)}}}
   {\sum_{j=1}^M \braket{\psi_F^{(j)}|\psi_F^{(j)}}}. 
\end{align}
Here, instead of performing an imaginary time evolution, 
we operate $\Hat{F}^{1/2}$ to a set of random states $\{\ket{\psi_0^{(i)}}\}$ and obtain 
\begin{align}
   \ket{\psi_F^{(i)}} = \Hat{F}^{1/2} \ket{\psi_0^{(i)}}.
\end{align}
The purity and the necessary and sufficient number of samples $N_\mathrm{samp}$ is 
given by the same definition as those of the TMQ state where we consider the NFPF 
of the partition function, $Z_F$. 
For the Boltzmann distribution, $\hat F= e^{-\beta \hat H}$ we find a TMQ state.
\par
There is a recently growing demand for acquiring a tool to evaluate purity. 
An algorithm called quantum imaginary time evolution (QITE) is developed for quantum computers\cite{motta2020}. 
Although QITE rather remains a toy protocol since the computing cost is no smaller than 
that of the classical tensor network calculations, 
it is applicable to ground states and TMQ state calculations.  
In Ref.[\onlinecite{sun2021}], the thermal properties of the four-site system were studied 
by performing QITE on IBM's quantum computer. 
They adopted a finite temperature sampling method similar to the RPPS method.  
However, with increasing system size, QITE rapidly loses sample efficiency. 
There, they pointed out the importance of comparing diverse sampling methods,  
which can be done using the norm of the state after QITE is performed 
by using our definition of NFPF and purity. 
In this way, the comparison of various random sampling methods to obtain a mixed quantum wave function 
will become increasingly important in several fields including 
statistical mechanics and condensed matter, and computer science.

\begin{acknowledgments}
We thank Shimpei Goto, Ryui Kaneko, Ippei Danshita, and Tsuyoshi Okubo for the discussions. 
We used Goto's C++ code available at the Github repository \cite{goto_code} 
in the RPMPS+T calculation. 
A. I. was supported by a Grant-in-Aid for JSPS Research Fellow (Grant No. 21J21992).
This work was supported by a Grant-in-Aid for Transformative Research Areas 
"The Natural Laws of Extreme Universe---A New Paradigm for Spacetime and Matter from Quantum Information" (Grant No. 21H05191)
and other JSPS KAKENHI (No. 21K03440, 17K05533) of Japan.  
\end{acknowledgments}

\appendix

\section{Counter examples of assumption (\ref{eq:assumption})} 
\label{sec:validity}
The inequality Eq.(\ref{eq:qvarfinal}) relies on the assumption given in Eq.(\ref{eq:assumption}). 
Although Eq.(\ref{eq:assumption}) is valid for most cases, 
there are two particular and exceptional counterexamples. 
\par
The first one is the method using random phase state (RPS) \cite{iitaka2004}; 
for a $D$-dimensional Hibert space, 
a full set of energy eigenstate $\{\ket{n}\}$ is used as a basis, 
and the RPS at infinite temperature is given as 
\begin{align}
   \ket{\psi_0^{\mathrm{RPS}}} = \sum_n e^{i\theta_n} \ket{n}, \quad \theta_n \in [0, 2\pi), 
 \label{eq:rps}
\end{align}
where $\theta_n$ are uniformly distributed random variables. 
By the imaginary time evolution, we obtain a pure state at finite temperature $\beta^{-1}$ as 
$\ket{\psi_\beta^{\mathrm{RPS}}}=e^{-\beta\Hat{H}/2} \ket{\psi_0^{\mathrm{RPS}}}$. 
Since the Hamiltonian is written in the form of a spectral decomposition using the chosen basis as 
$\Hat{H} = \sum_{n=0}^{D-1} E_n \ket{n} \bra{n}$, 
it commutes with the imaginary-time-evolution operator, and 
the right-hand side of Eq.(\ref{eq:assumption}) is exactly equal to zero. 
For operator $\Hat{O}$ that does not commute with the Hamiltonian, 
the left-hand side of Eq.(\ref{eq:assumption}) is larger than zero. 
Therefore the assumption is broken despite that 
$\ket{\psi_\beta^{\mathrm{RPS}}}$ is typical as follows; 
one can evaluate 
\begin{align}
   \mathrm{Var}\left(\braket{\psi_\beta^{\mathrm{RPS}}|\Hat{O}|\psi_\beta^{\mathrm{RPS}}}\right)
   = \sum_{m\neq n} e^{-\beta(E_m+E_n)} |\braket{m|\Hat{O}|n}|^2, 
\end{align}
which has $\mathcal{O}(D^2)$ terms. Therefore the variance of the expectation value of 
a physical quantity is bounded as
\begin{align}
   \overline{\left( \frac{\braket{\psi_\beta^{\mathrm{RPS}}|\Hat{O}|\psi_\beta^{\mathrm{RPS}}}}
   {\braket{\psi_\beta^{\mathrm{RPS}}|\psi_\beta^{\mathrm{RPS}}}}
   - \langle \Hat{O} \rangle_\beta \right)^2}
   \lesssim \langle \Hat{O}^2 \rangle _{2\beta} \:e^{-Ns_\mathrm{th}(\Tilde{\beta})}. 
\end{align}
meaning that RPS is typical. 
\par
From the above counterexample, one might expect that 
the thermal state is typical if $\mathrm{Var}(\braket{\psi_\beta|\psi_\beta})=0$. 
However, it is not necessarily the case as we see in the second counterexample. 
This time we use the random state which we call random Bloch state (RBS), 
\begin{align}
   \ket{\psi_0^{\mathrm{RBS}}} = \sum_{n=0}^{D-1} e^{in\theta} \ket{n}, \quad \theta \in [0,2\pi) ,
\end{align}
which has only one random variable. 
As in the RPS, RBS also show no fluctuation of
operators which commute with the Hamiltonian. 
Hence, $\mathrm{Var}(\braket{\psi_\beta|\psi_\beta})=0$, 
and the assumption Eq.(\ref{eq:qvarfinal}) is broken. 
In contrast to the RPS, the RBS is not typical; 
this time, we find 
\begin{align}
   &\mathrm{Var}\left(\braket{\psi_\beta^{\mathrm{RBS}}|\Hat{O}|\psi_\beta^{\mathrm{RBS}}}\right) \notag \\
   &= \sum_{\substack{m \neq n, k \neq l\\ m+k=n+l}} e^{-\beta (E_m+E_n+E_k+E_l)/2}
   \braket{m|\Hat{O}|n} \braket{k|\Hat{O}|l}, 
\end{align}
which includes $\mathcal{O}(D^3)$ parameters, and becomes much larger than that of the RPS. 
Therefore the variance of a physical quantity can not be bounded. 

The reason why the RPS and the RBS break the assumption (\ref{eq:assumption}) is 
that we use the information of energy eigenstates when we construct these random states. 
In the actual calculation, the full information of energy eigenstates is usually inaccessible, 
and Eq.(\ref{eq:assumption}) is naturally expected.

\section{Definition of purity for a randomly sampled mixed states}
\label{sec:purity2}
In deriving Eq.(\ref{eq:purity}), we assumed that the samples are pure states. 
However, some methods sample the purified mixed states; each sampled state $\ket{\psi^{(i)}_\beta}$ is pure 
but it consists of a physical system $A$ and the ancilla or auxiliaries $B$, 
as we saw in Fig.~\ref{fig:mpsmethods2}. 
One is then able to formally obtain the mixed states in the system $A$ 
by tracing out the degrees of freedom in $B$ as 
\begin{align}
   \sigma_\beta^{(i)} = \Tr_B \ket{\psi_\beta^{(i)}} \bra{\psi_\beta^{(i)}}.
\end{align}
Here, we reformulate the purity of system $A$ using the sampled mixed states. 
Random variables $y^{(i)}$ and $w_{i, M}$ in Eqs.(\ref{eq:yi}) and (\ref{eq:wi}) 
are rewritten as
\begin{align}
   y^{(i)}=\Tr\sigma_\beta^{(i)}, \quad w_{i, M} = \frac{y^{(i)}}{\sum_{j=1}^M y^{(j)}}. 
\end{align}
Because the expectation value is given as 
\begin{align}
   \langle\Hat{O}\rangle_{\beta, M}^\mathrm{samp} = \sum_{i=1}^M w_{i, M} \frac{\Tr(\sigma_\beta^{(i)}\Hat{O})}{y^{(i)}} , 
\end{align}
the density operator becomes
\begin{align}
   \rho(M) = \sum_{i=1}^M w_{i, M} \frac{\sigma_\beta^{(i)}}{y^{(i)}}. 
\end{align}
Then, the purity of this density operator is given as 
\begin{align}
   &\Tr[\rho(M)^2] \notag \\
   &= \sum_{ij}^M w_{i, M} w_{j, M} \frac{\Tr[\sigma_\beta^{(i)} \sigma_\beta^{(j)}]}{y^{(i)} y^{(j)}} \notag \\
   &= \sum_{i=1}^M w_{i, M}^2 \frac{\Tr[(\sigma_\beta^{(i)})^2]}{(y^{(i)})^2}
   + \sum_{i \neq j}^M w_{i, M} w_{j, M} \frac{\Tr[\sigma_\beta^{(i)} \sigma_\beta^{(i)}]}{y^{(i)} y^{(j)}}. 
   \label{eq:mixed_purity}
\end{align}
However, unlike the case given in the main text, 
the random average of Eq.(\ref{eq:mixed_purity}) is not analytically evaluated, 
since they are the combinations of three different quantities that follow different distributions, 
$w_{i, M}$, $y^{(i)}$ and $\sigma_\beta^{(i)}$. 
Instead, we make a crude approximation of factorizing the random average into 
the random averages of three constituents. 
This process corresponds to replacing the three parts with their mean numbers, 
and provide an order estimate of Eq.(\ref{eq:mixed_purity}). 
\par
The random average of the first term in Eq.(\ref{eq:mixed_purity}) is evaluated as 
\begin{align}
   \overline{\text{(first term)}}
   &= \sum_{i=1}^M \overline{\left( w_{i, M}^2 \frac{\Tr[(\sigma_\beta^{(i)})^2]}{(y^{(i)})^2} \right)} \notag \\ 
   &\simeq \sum_{i=1}^M \overline{w_{i, M}^2} \overline{\left(\frac{\Tr[(\sigma_\beta^{(i)})^2]}{(y^{(i)})^2} \right)} \notag \\
   &\simeq \mathcal{P}_\mathrm{ave} \frac{1}{M} \left[ 1+ \left( 1-\frac{1}{M} \right) \delta z^2 \right] 
\end{align}
where $\mathcal{P}_\mathrm{ave} = \overline{\Tr(\sigma_\beta^2)/\Tr(\sigma_\beta)^2}$ is the averaged purity of sampled mixed states.
The random average of the second term in Eq.(\ref{eq:mixed_purity}) is transformed as 
\begin{align}
   \overline{\text{(second term)}}
   &= \sum_{i \neq j}^M \overline{\left(w_{i, M} w_{j, M} \frac{\Tr[\sigma_\beta^{(i)} \sigma_\beta^{(i)}]}{y^{(i)} y^{(j)}}\right)} \notag \\
   &\simeq \sum_{i \neq j}^M \overline{w_{i, M} w_{j, M}} 
   \frac{\overline{\Tr[\sigma_\beta^{(i)} \sigma_\beta^{(i)}]}}{\overline{y^{(i)} y^{(j)}}} \notag \\
   &\simeq \delta z^2_\mathrm{TPQ} \left(1-\frac{1}{M}\right) \left(1-\frac{1}{M}\delta z^2\right)
\end{align}
By combining the above two and substituting the appropriate number of samples 
$M=N_\mathrm{samp}=\delta z^2/\delta z^2_\mathrm{TPQ}$, 
we obtain an expression of purity for mixed state sampling, 
\begin{align}
   \mathcal{P}_\mathrm{rand}' = 
   & \mathcal{P}_\mathrm{ave} \frac{\delta z^2_\mathrm{TPQ}}{\delta z^2} (1+\delta z^2 - \delta z^2_\mathrm{TPQ}) \notag \\
   &+ \delta z^2_\mathrm{TPQ} \left(1-\frac{\delta z^2_\mathrm{TPQ}}{\delta z^2}\right) (1-\delta z^2_\mathrm{TPQ}).
   \label{eq:purity_mixed}
\end{align}
This expression is reasonable from several perspectives. 
In the low temperature limit $\beta\to\infty$, we find $\mathcal{P}_\mathrm{rand}'=1$, 
since $\delta z^2_\mathrm{TPQ}=1$ and $\mathcal{P}_\mathrm{ave}=1$. 
This meets the fact that the ground state wave function has a purity-1. 
If we apply the TPQ state with $\delta z^2 = \delta z^2_\mathrm{TPQ}$ to this form, 
we find $\mathcal{P}_\mathrm{rand}'=1$. 
It is important to guarantee that $\mathcal{P}_\mathrm{rand}'$ is larger that the purity of the Gibbs state 
$\mathcal{P}_\mathrm{Gibbs}$, which is satisfied as 
\begin{align}
   &\mathcal{P}_\mathrm{rand}' - \mathcal{P}_\mathrm{Gibbs} \notag \\
   &= (\mathcal{P}_\mathrm{ave} - \mathcal{P}_\mathrm{Gibbs}) \frac{\delta z^2_\mathrm{TPQ}}{\delta z^2} (1+\delta z^2 - \delta z^2_\mathrm{TPQ}) \geq 0
\label{eq:pmixed}
\end{align}
This equation takes the form of the multiplication of $(\mathcal{P}_\mathrm{ave} - \mathcal{P}_\mathrm{Gibbs})$ and 
Eq.(\ref{eq:purity}). 
The classical mixture is attributed to the number of sampling, $N_\mathrm{samp}$, 
which appears in the latter. 
Whereas, $\mathcal{P}_\mathrm{ave}$ is responsible for the entanglement with the bath 
which is explicitly introduced as auxiliaries. 
The purity in Eq.(\ref{eq:purity}) is thus the purity of states including the auxiliaries, 
and Eq.(\ref{eq:purity_mixed}) is the purity of the physical system without auxiliaries. 
\par
\begin{figure}[t]
   \centering
   \includegraphics[width=\hsize]{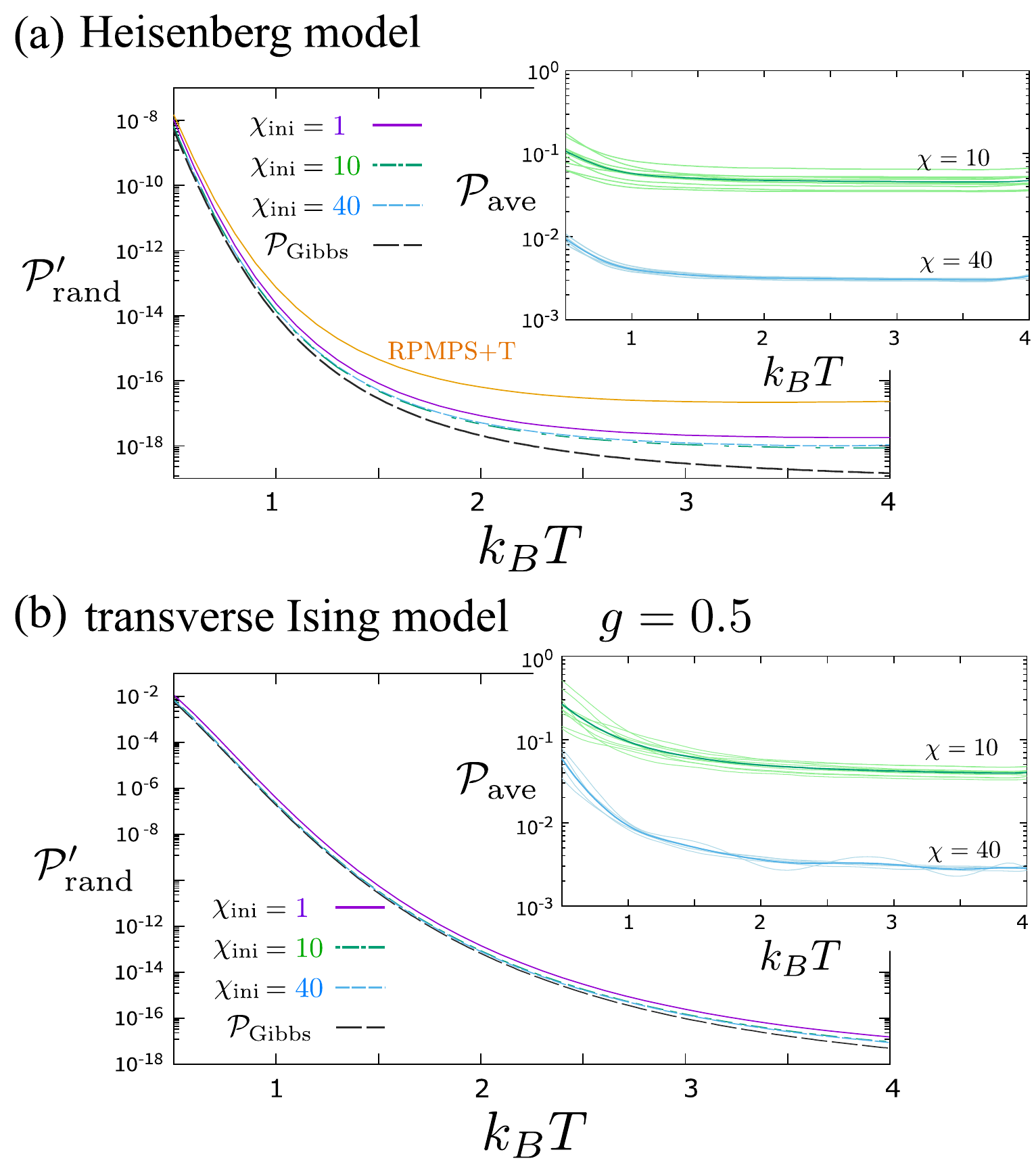}
   \caption{
      The purity defined in Eq.(\ref{eq:purity_mixed}) calculated for the physical system in TPQ-MPS method 
      for (a) spin-1/2 Heisenberg chain and (b) transverse Ising chain with $g=0.5$. 
      We demonstrate the TPQ-MPS method with $\chi_\mathrm{ini}=1, 10, 40$ and the RPMPS+T method. 
      Insets show the averaged purity $\mathcal{P}_\mathrm{ave}$ of each sampled TPQ-MPS. 
   }
   \label{fig:purity_mixed}
\end{figure}

Figure \ref{fig:purity_mixed} shows the purity in Eq.(\ref{eq:purity_mixed}) for TPQ-MPS which has auxiliaries. 
We demonstrate the results of the spin-1/2 Heisenberg chain and the transverse Ising chain with $g=0.5$  
corresponding to the ones in Figs.~\ref{fig:purity_tri} and \ref{fig:hei_purity}, respectively. 
In both models, $\mathcal{P}_\mathrm{rand}'$ of TPQ-MPS with $\chi_\mathrm{ini}=10, 40$ has almost the same curve; 
$\mathcal{P}_\mathrm{rand}'$ of TPQ-MPS is independent of $\chi_\mathrm{ini}$, 
because $\mathcal{P}_\mathrm{rand}$ scales as $\chi_\mathrm{ini}^2$ and $\mathcal{P}_\mathrm{ave}$ scales as $\chi_\mathrm{ini}^{-2}$. 
The RPMPS+T has relatively high purity when we focus on the physical system. 
The present benchmark result supports the high efficiency of the TPQ-MPS method; the dimension $\chi_\mathrm{ini}$ of the auxiliaries 
gives the degree of mixing, and the number of samples to take can be reduced very efficiently as $\chi_\mathrm{ini}^{-2}$. 

We notice that our formula (\ref{eq:purity_mixed}) can be applied to
subsystems of the TPQ state which we introduced 
in \S.\ref{sec:intro} as an example of TMQ states. 
In that case, random fluctuations are suppressed to 
the same degree as the TPQ state, 
and we obtain $\delta z^2 \simeq \delta z^2_\mathrm{TPQ}$ and then
$\mathcal{P}'_\mathrm{rand} \simeq \mathcal{P}_\mathrm{ave}$. 
Here, $\mathcal{P}_\mathrm{ave}$ coincides with the value 
calculated in Ref.[\onlinecite{nakagawa2018}]. 
Our discussion in \S.\ref{sec:intro} is justified
in the framework of our theory of random samplings.

\begin{figure}[t]
   \centering
   \includegraphics[width=0.8\hsize]{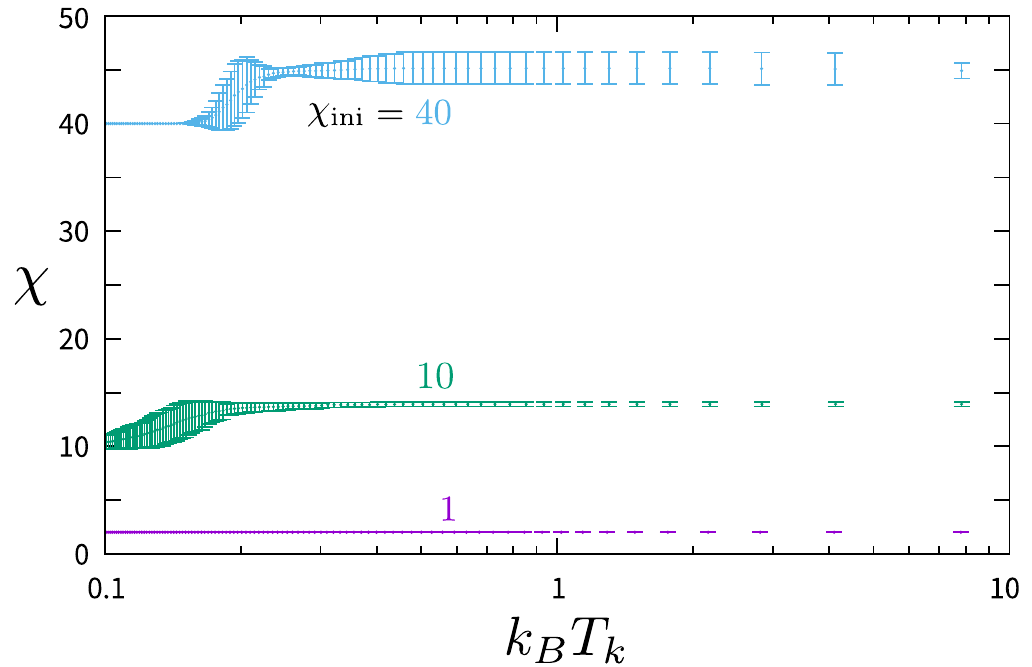}
   \caption{ Bond dimension $\chi$ of MPS of a $k$-th 
      TPQ-MPS state as a function of microcanonical temperature $k_BT_k$ in Eq.(\ref{eq:temperature}) 
      for a transverse Ising model with $g=0.5$ for the same calculation as Fig.~\ref{fig:purity_tri}. 
      We plot the cases with $\chi_\mathrm{ini}=1, 10, 40$. 
   }
   \label{fig:bonddim}
\end{figure}

\section{Details of the TPQ-MPS method}
\label{sec:detail_tpqmps}
In Ref.[\onlinecite{iwaki2021}], the authors have proposed the TPQ-MPS protocol 
which is applied in the present calculation. 
Here, we briefly explain the details updated from Ref.[\onlinecite{iwaki2021}]. 
Previously, we prepared the common bond dimension $\chi$ for all matrices and auxiliaries in the system, 
and kept the maximum bond dimension to this value throughout the calculation. 
In such a case, we found that the results depend much on $l$. 
In general, for larger $l$, operating $(l-\hat h)$ will not change the MPS wave function 
much and the truncation error becomes smaller. However, there is a trade-off that 
for larger $l$ it takes more steps $k$ to reach the low-temperature mTPQ state. 
In the previous Ref.[\onlinecite{iwaki2021}], these two tendencies needed to be optimized by varying $l$. 
In the present calculation, we set $\chi_\mathrm{ini}$ as an initial value of $\chi$. 
Each time we operate $(l-\hat h)$, the bond dimension increases from $\chi$ to $\chi\times \chi_\mathrm{op}$, 
which is truncated to a new $\chi$ by keeping the truncation error smaller than $10^{-7}$ 
independently for each bond after we transform the MPS to a canonical form. 
As we demonstrate in Fig.~\ref{fig:bonddim}, 
the bond dimension started at $\chi_\mathrm{ini}=1,10,40$ exceeds $\chi_\mathrm{ini}$ at $k\ge 1$, 
where we plot $\chi$ as a function of $k_BT_k$ for 
the $k$-th microcanonical state. 
However, the overall $\chi$ depends much on $\chi_\mathrm{ini}$, and accordingly, 
NFPF and purity depend systematically on $\chi_\mathrm{ini}$. 
This means that the initial quality of RMPS dominates the purity of the TPQ-MPS. 
We also confirmed that the $l$-dependence that influenced the quality of truncation in the 
previous protocol disappears by this update.

\nocite{apsrev41Control}
\bibliographystyle{apsrev4-1}
\bibliography{myref}

\begin{thebibliography}{71}%
\makeatletter
\providecommand \@ifxundefined [1]{%
 \@ifx{#1\undefined}
}%
\providecommand \@ifnum [1]{%
 \ifnum #1\expandafter \@firstoftwo
 \else \expandafter \@secondoftwo
 \fi
}%
\providecommand \@ifx [1]{%
 \ifx #1\expandafter \@firstoftwo
 \else \expandafter \@secondoftwo
 \fi
}%
\providecommand \natexlab [1]{#1}%
\providecommand \enquote  [1]{``#1''}%
\providecommand \bibnamefont  [1]{#1}%
\providecommand \bibfnamefont [1]{#1}%
\providecommand \citenamefont [1]{#1}%
\providecommand \href@noop [0]{\@secondoftwo}%
\providecommand \href [0]{\begingroup \@sanitize@url \@href}%
\providecommand \@href[1]{\@@startlink{#1}\@@href}%
\providecommand \@@href[1]{\endgroup#1\@@endlink}%
\providecommand \@sanitize@url [0]{\catcode `\\12\catcode `\$12\catcode
  `\&12\catcode `\#12\catcode `\^12\catcode `\_12\catcode `\%12\relax}%
\providecommand \@@startlink[1]{}%
\providecommand \@@endlink[0]{}%
\providecommand \url  [0]{\begingroup\@sanitize@url \@url }%
\providecommand \@url [1]{\endgroup\@href {#1}{\urlprefix }}%
\providecommand \urlprefix  [0]{URL }%
\providecommand \Eprint [0]{\href }%
\providecommand \doibase [0]{http://dx.doi.org/}%
\providecommand \selectlanguage [0]{\@gobble}%
\providecommand \bibinfo  [0]{\@secondoftwo}%
\providecommand \bibfield  [0]{\@secondoftwo}%
\providecommand \translation [1]{[#1]}%
\providecommand \BibitemOpen [0]{}%
\providecommand \bibitemStop [0]{}%
\providecommand \bibitemNoStop [0]{.\EOS\space}%
\providecommand \EOS [0]{\spacefactor3000\relax}%
\providecommand \BibitemShut  [1]{\csname bibitem#1\endcsname}%
\let\auto@bib@innerbib\@empty
\bibitem [{\citenamefont {von Neumann}(1929)}]{neumann1929}%
  \BibitemOpen
  \bibfield  {author} {\bibinfo {author} {\bibfnamefont {J.}~\bibnamefont {von
  Neumann}},\ }\bibfield  {title} {\enquote {\bibinfo {title} {Beweis des
  ergodensatzes und des {$H$}-theorems in der neuen mechanik},}\ }\href
  {https://link.springer.com/article/10.1007/BF01339852} {\bibfield  {journal}
  {\bibinfo  {journal} {Z. Phys.}\ }\textbf {\bibinfo {volume} {57}},\ \bibinfo
  {pages} {30} (\bibinfo {year} {1929})}\BibitemShut {NoStop}%
\bibitem [{\citenamefont {Popescu}\ \emph {et~al.}(2006)\citenamefont
  {Popescu}, \citenamefont {J.~Short},\ and\ \citenamefont
  {Winter}}]{popescu2006}%
  \BibitemOpen
  \bibfield  {author} {\bibinfo {author} {\bibfnamefont {S.}~\bibnamefont
  {Popescu}}, \bibinfo {author} {\bibfnamefont {A.}~\bibnamefont {J.~Short}}, \
  and\ \bibinfo {author} {\bibfnamefont {A.}~\bibnamefont {Winter}},\
  }\bibfield  {title} {\enquote {\bibinfo {title} {Entanglement and the
  foundations of statistical mechanics},}\ }\href {\doibase 10.1038/nphys444}
  {\bibfield  {journal} {\bibinfo  {journal} {Nature Phys.}\ }\textbf {\bibinfo
  {volume} {2}},\ \bibinfo {pages} {754} (\bibinfo {year} {2006})}\BibitemShut
  {NoStop}%
\bibitem [{\citenamefont {Goldstein}\ \emph {et~al.}(2006)\citenamefont
  {Goldstein}, \citenamefont {Lebowitz}, \citenamefont {Tumulka},\ and\
  \citenamefont {Zangh\`{\i}}}]{goldstein2006}%
  \BibitemOpen
  \bibfield  {author} {\bibinfo {author} {\bibfnamefont {S.}~\bibnamefont
  {Goldstein}}, \bibinfo {author} {\bibfnamefont {J.~L.}\ \bibnamefont
  {Lebowitz}}, \bibinfo {author} {\bibfnamefont {R.}~\bibnamefont {Tumulka}}, \
  and\ \bibinfo {author} {\bibfnamefont {N.}~\bibnamefont {Zangh\`{\i}}},\
  }\bibfield  {title} {\enquote {\bibinfo {title} {Canonical typicality},}\
  }\href {\doibase 10.1103/PhysRevLett.96.050403} {\bibfield  {journal}
  {\bibinfo  {journal} {Phys. Rev. Lett.}\ }\textbf {\bibinfo {volume} {96}},\
  \bibinfo {pages} {050403} (\bibinfo {year} {2006})}\BibitemShut {NoStop}%
\bibitem [{\citenamefont {Sugita}(2006)}]{sugita2006}%
  \BibitemOpen
  \bibfield  {author} {\bibinfo {author} {\bibfnamefont {A.}~\bibnamefont
  {Sugita}},\ }\href@noop {} {\bibfield  {journal} {\bibinfo  {journal} {RIMS
  kokyuroku}\ }\textbf {\bibinfo {volume} {1507}},\ \bibinfo {pages} {147}
  (\bibinfo {year} {2006})}\BibitemShut {NoStop}%
\bibitem [{\citenamefont {Reimann}(2007)}]{reimann2007}%
  \BibitemOpen
  \bibfield  {author} {\bibinfo {author} {\bibfnamefont {P.}~\bibnamefont
  {Reimann}},\ }\bibfield  {title} {\enquote {\bibinfo {title} {Typicality for
  generalized microcanonical ensembles},}\ }\href {\doibase
  10.1103/PhysRevLett.99.160404} {\bibfield  {journal} {\bibinfo  {journal}
  {Phys. Rev. Lett.}\ }\textbf {\bibinfo {volume} {99}},\ \bibinfo {pages}
  {160404} (\bibinfo {year} {2007})}\BibitemShut {NoStop}%
\bibitem [{\citenamefont {Sugiura}\ and\ \citenamefont
  {Shimizu}(2012)}]{sugiura2012}%
  \BibitemOpen
  \bibfield  {author} {\bibinfo {author} {\bibfnamefont {S.}~\bibnamefont
  {Sugiura}}\ and\ \bibinfo {author} {\bibfnamefont {A.}~\bibnamefont
  {Shimizu}},\ }\bibfield  {title} {\enquote {\bibinfo {title} {Thermal pure
  quantum states at finite temperature},}\ }\href {\doibase
  10.1103/PhysRevLett.108.240401} {\bibfield  {journal} {\bibinfo  {journal}
  {Phys. Rev. Lett.}\ }\textbf {\bibinfo {volume} {108}},\ \bibinfo {pages}
  {240401} (\bibinfo {year} {2012})}\BibitemShut {NoStop}%
\bibitem [{\citenamefont {Sugiura}\ and\ \citenamefont
  {Shimizu}(2013)}]{sugiura2013}%
  \BibitemOpen
  \bibfield  {author} {\bibinfo {author} {\bibfnamefont {S.}~\bibnamefont
  {Sugiura}}\ and\ \bibinfo {author} {\bibfnamefont {A.}~\bibnamefont
  {Shimizu}},\ }\bibfield  {title} {\enquote {\bibinfo {title} {Canonical
  thermal pure quantum state},}\ }\href {\doibase
  10.1103/PhysRevLett.111.010401} {\bibfield  {journal} {\bibinfo  {journal}
  {Phys. Rev. Lett.}\ }\textbf {\bibinfo {volume} {111}},\ \bibinfo {pages}
  {010401} (\bibinfo {year} {2013})}\BibitemShut {NoStop}%
\bibitem [{\citenamefont {Hyuga}\ \emph {et~al.}(2014)\citenamefont {Hyuga},
  \citenamefont {Sugiura}, \citenamefont {Sakai},\ and\ \citenamefont
  {Shimizu}}]{hyuga2014}%
  \BibitemOpen
  \bibfield  {author} {\bibinfo {author} {\bibfnamefont {M.}~\bibnamefont
  {Hyuga}}, \bibinfo {author} {\bibfnamefont {S.}~\bibnamefont {Sugiura}},
  \bibinfo {author} {\bibfnamefont {K.}~\bibnamefont {Sakai}}, \ and\ \bibinfo
  {author} {\bibfnamefont {A.}~\bibnamefont {Shimizu}},\ }\bibfield  {title}
  {\enquote {\bibinfo {title} {Thermal pure quantum states of many-particle
  systems},}\ }\href {\doibase 10.1103/PhysRevB.90.121110} {\bibfield
  {journal} {\bibinfo  {journal} {Phys. Rev. B}\ }\textbf {\bibinfo {volume}
  {90}},\ \bibinfo {pages} {121110(R)} (\bibinfo {year} {2014})}\BibitemShut
  {NoStop}%
\bibitem [{\citenamefont {Linden}\ \emph {et~al.}(2009)\citenamefont {Linden},
  \citenamefont {Popescu}, \citenamefont {Short},\ and\ \citenamefont
  {Winter}}]{linden2009}%
  \BibitemOpen
  \bibfield  {author} {\bibinfo {author} {\bibfnamefont {N.}~\bibnamefont
  {Linden}}, \bibinfo {author} {\bibfnamefont {S.}~\bibnamefont {Popescu}},
  \bibinfo {author} {\bibfnamefont {A.~J.}\ \bibnamefont {Short}}, \ and\
  \bibinfo {author} {\bibfnamefont {A.}~\bibnamefont {Winter}},\ }\bibfield
  {title} {\enquote {\bibinfo {title} {Quantum mechanical evolution towards
  thermal equilibrium},}\ }\href {\doibase 10.1103/PhysRevE.79.061103}
  {\bibfield  {journal} {\bibinfo  {journal} {Phys. Rev. E}\ }\textbf {\bibinfo
  {volume} {79}},\ \bibinfo {pages} {061103} (\bibinfo {year}
  {2009})}\BibitemShut {NoStop}%
\bibitem [{\citenamefont {Linden}\ \emph {et~al.}(2010)\citenamefont {Linden},
  \citenamefont {Popescu}, \citenamefont {Short},\ and\ \citenamefont
  {Winter}}]{linden2010}%
  \BibitemOpen
  \bibfield  {author} {\bibinfo {author} {\bibfnamefont {N.}~\bibnamefont
  {Linden}}, \bibinfo {author} {\bibfnamefont {S.}~\bibnamefont {Popescu}},
  \bibinfo {author} {\bibfnamefont {A.~J.}\ \bibnamefont {Short}}, \ and\
  \bibinfo {author} {\bibfnamefont {A.}~\bibnamefont {Winter}},\ }\bibfield
  {title} {\enquote {\bibinfo {title} {On the speed of fluctuations around
  thermodynamic equilibrium},}\ }\href {\doibase 10.1088/1367-2630/12/5/055021}
  {\bibfield  {journal} {\bibinfo  {journal} {New Journal of Physics}\ }\textbf
  {\bibinfo {volume} {12}},\ \bibinfo {pages} {055021} (\bibinfo {year}
  {2010})}\BibitemShut {NoStop}%
\bibitem [{\citenamefont {Page}(1993)}]{page1993}%
  \BibitemOpen
  \bibfield  {author} {\bibinfo {author} {\bibfnamefont {D.~N.}\ \bibnamefont
  {Page}},\ }\bibfield  {title} {\enquote {\bibinfo {title} {Average entropy of
  a subsystem},}\ }\href {\doibase 10.1103/PhysRevLett.71.1291} {\bibfield
  {journal} {\bibinfo  {journal} {Phys. Rev. Lett.}\ }\textbf {\bibinfo
  {volume} {71}},\ \bibinfo {pages} {1291} (\bibinfo {year}
  {1993})}\BibitemShut {NoStop}%
\bibitem [{\citenamefont {Garrison}\ and\ \citenamefont
  {Grover}(2018)}]{garrison2018}%
  \BibitemOpen
  \bibfield  {author} {\bibinfo {author} {\bibfnamefont {J.~R.}\ \bibnamefont
  {Garrison}}\ and\ \bibinfo {author} {\bibfnamefont {T.}~\bibnamefont
  {Grover}},\ }\bibfield  {title} {\enquote {\bibinfo {title} {Does a single
  eigenstate encode the full hamiltonian?}}\ }\href {\doibase
  10.1103/PhysRevX.8.021026} {\bibfield  {journal} {\bibinfo  {journal} {Phys.
  Rev. X}\ }\textbf {\bibinfo {volume} {8}},\ \bibinfo {pages} {021026}
  (\bibinfo {year} {2018})}\BibitemShut {NoStop}%
\bibitem [{\citenamefont {Nakagawa}\ \emph {et~al.}(2018)\citenamefont
  {Nakagawa}, \citenamefont {Watanabe}, \citenamefont {Fujita},\ and\
  \citenamefont {Sugiura}}]{nakagawa2018}%
  \BibitemOpen
  \bibfield  {author} {\bibinfo {author} {\bibfnamefont {Y.~O.}\ \bibnamefont
  {Nakagawa}}, \bibinfo {author} {\bibfnamefont {M.}~\bibnamefont {Watanabe}},
  \bibinfo {author} {\bibfnamefont {H.}~\bibnamefont {Fujita}}, \ and\ \bibinfo
  {author} {\bibfnamefont {S.}~\bibnamefont {Sugiura}},\ }\bibfield  {title}
  {\enquote {\bibinfo {title} {Universality in volume-law entanglement of
  scrambled pure quantum states},}\ }\href {\doibase
  10.1038/s41467-018-03883-9} {\bibfield  {journal} {\bibinfo  {journal}
  {Nature Communications}\ }\textbf {\bibinfo {volume} {9}},\ \bibinfo {pages}
  {1635} (\bibinfo {year} {2018})}\BibitemShut {NoStop}%
\bibitem [{\citenamefont {Iwaki}\ \emph {et~al.}(2021)\citenamefont {Iwaki},
  \citenamefont {Shimizu},\ and\ \citenamefont {Hotta}}]{iwaki2021}%
  \BibitemOpen
  \bibfield  {author} {\bibinfo {author} {\bibfnamefont {A.}~\bibnamefont
  {Iwaki}}, \bibinfo {author} {\bibfnamefont {A.}~\bibnamefont {Shimizu}}, \
  and\ \bibinfo {author} {\bibfnamefont {C.}~\bibnamefont {Hotta}},\ }\bibfield
   {title} {\enquote {\bibinfo {title} {Thermal pure quantum matrix product
  states recovering a volume law entanglement},}\ }\href {\doibase
  10.1103/PhysRevResearch.3.L022015} {\bibfield  {journal} {\bibinfo  {journal}
  {Phys. Rev. Research}\ }\textbf {\bibinfo {volume} {3}},\ \bibinfo {pages}
  {L022015} (\bibinfo {year} {2021})}\BibitemShut {NoStop}%
\bibitem [{\citenamefont {Garnerone}\ and\ \citenamefont
  {de~Oliveira}(2013)}]{garnerone2013-1}%
  \BibitemOpen
  \bibfield  {author} {\bibinfo {author} {\bibfnamefont {S.}~\bibnamefont
  {Garnerone}}\ and\ \bibinfo {author} {\bibfnamefont {T.~R.}\ \bibnamefont
  {de~Oliveira}},\ }\bibfield  {title} {\enquote {\bibinfo {title} {Generalized
  quantum microcanonical ensemble from random matrix product states},}\ }\href
  {\doibase 10.1103/PhysRevB.87.214426} {\bibfield  {journal} {\bibinfo
  {journal} {Phys. Rev. B}\ }\textbf {\bibinfo {volume} {87}},\ \bibinfo
  {pages} {214426} (\bibinfo {year} {2013})}\BibitemShut {NoStop}%
\bibitem [{\citenamefont {Garnerone}(2013)}]{garnerone2013-2}%
  \BibitemOpen
  \bibfield  {author} {\bibinfo {author} {\bibfnamefont {S.}~\bibnamefont
  {Garnerone}},\ }\bibfield  {title} {\enquote {\bibinfo {title} {Pure state
  thermodynamics with matrix product states},}\ }\href {\doibase
  10.1103/PhysRevB.88.165140} {\bibfield  {journal} {\bibinfo  {journal} {Phys.
  Rev. B}\ }\textbf {\bibinfo {volume} {88}},\ \bibinfo {pages} {165140}
  (\bibinfo {year} {2013})}\BibitemShut {NoStop}%
\bibitem [{\citenamefont {Iitaka}(2020)}]{iitaka2020}%
  \BibitemOpen
  \bibfield  {author} {\bibinfo {author} {\bibfnamefont {T.}~\bibnamefont
  {Iitaka}},\ }\bibfield  {title} {\enquote {\bibinfo {title} {Random phase
  product sate for canonical ensemble},}\ }\href
  {https://arxiv.org/abs/2006.14459} {\bibfield  {journal} {\bibinfo  {journal}
  {arXiv:2006.14459}\ } (\bibinfo {year} {2020})}\BibitemShut {NoStop}%
\bibitem [{\citenamefont {Goto}\ \emph {et~al.}(2021)\citenamefont {Goto},
  \citenamefont {Kaneko},\ and\ \citenamefont {Danshita}}]{goto2021}%
  \BibitemOpen
  \bibfield  {author} {\bibinfo {author} {\bibfnamefont {S.}~\bibnamefont
  {Goto}}, \bibinfo {author} {\bibfnamefont {R.}~\bibnamefont {Kaneko}}, \ and\
  \bibinfo {author} {\bibfnamefont {I.}~\bibnamefont {Danshita}},\ }\bibfield
  {title} {\enquote {\bibinfo {title} {Matrix product state approach for a
  quantum system at finite temperatures using random phases and trotter
  gates},}\ }\href {\doibase 10.1103/PhysRevB.104.045133} {\bibfield  {journal}
  {\bibinfo  {journal} {Phys. Rev. B}\ }\textbf {\bibinfo {volume} {104}},\
  \bibinfo {pages} {045133} (\bibinfo {year} {2021})}\BibitemShut {NoStop}%
\bibitem [{\citenamefont {Kaufman}\ \emph {et~al.}(2016)\citenamefont
  {Kaufman}, \citenamefont {Tai}, \citenamefont {Lukin}, \citenamefont
  {Rispoli}, \citenamefont {Schittko}, \citenamefont {Preiss},\ and\
  \citenamefont {Greiner}}]{kaufman2016}%
  \BibitemOpen
  \bibfield  {author} {\bibinfo {author} {\bibfnamefont {A.~M.}\ \bibnamefont
  {Kaufman}}, \bibinfo {author} {\bibfnamefont {M.~E.}\ \bibnamefont {Tai}},
  \bibinfo {author} {\bibfnamefont {A.}~\bibnamefont {Lukin}}, \bibinfo
  {author} {\bibfnamefont {M.}~\bibnamefont {Rispoli}}, \bibinfo {author}
  {\bibfnamefont {R.}~\bibnamefont {Schittko}}, \bibinfo {author}
  {\bibfnamefont {P.~M.}\ \bibnamefont {Preiss}}, \ and\ \bibinfo {author}
  {\bibfnamefont {M.}~\bibnamefont {Greiner}},\ }\bibfield  {title} {\enquote
  {\bibinfo {title} {Quantum thermalization through entanglement in an isolated
  many-body system},}\ }\href {\doibase 10.1126/science.aaf6725} {\bibfield
  {journal} {\bibinfo  {journal} {Science}\ }\textbf {\bibinfo {volume}
  {353}},\ \bibinfo {pages} {794} (\bibinfo {year} {2016})}\BibitemShut
  {NoStop}%
\bibitem [{\citenamefont {Brown}\ \emph {et~al.}(2008)\citenamefont {Brown},
  \citenamefont {Weinstein},\ and\ \citenamefont {Viola}}]{brown2008}%
  \BibitemOpen
  \bibfield  {author} {\bibinfo {author} {\bibfnamefont {W.~G.}\ \bibnamefont
  {Brown}}, \bibinfo {author} {\bibfnamefont {Y.~S.}\ \bibnamefont
  {Weinstein}}, \ and\ \bibinfo {author} {\bibfnamefont {L.}~\bibnamefont
  {Viola}},\ }\bibfield  {title} {\enquote {\bibinfo {title} {Quantum
  pseudorandomness from cluster-state quantum computation},}\ }\href {\doibase
  10.1103/PhysRevA.77.040303} {\bibfield  {journal} {\bibinfo  {journal} {Phys.
  Rev. A}\ }\textbf {\bibinfo {volume} {77}},\ \bibinfo {pages} {040303(R)}
  (\bibinfo {year} {2008})}\BibitemShut {NoStop}%
\bibitem [{\citenamefont {Dankert}\ \emph {et~al.}(2009)\citenamefont
  {Dankert}, \citenamefont {Cleve}, \citenamefont {Emerson},\ and\
  \citenamefont {Livine}}]{dankert2009}%
  \BibitemOpen
  \bibfield  {author} {\bibinfo {author} {\bibfnamefont {C.}~\bibnamefont
  {Dankert}}, \bibinfo {author} {\bibfnamefont {R.}~\bibnamefont {Cleve}},
  \bibinfo {author} {\bibfnamefont {J.}~\bibnamefont {Emerson}}, \ and\
  \bibinfo {author} {\bibfnamefont {E.}~\bibnamefont {Livine}},\ }\bibfield
  {title} {\enquote {\bibinfo {title} {Exact and approximate unitary 2-designs
  and their application to fidelity estimation},}\ }\href {\doibase
  10.1103/PhysRevA.80.012304} {\bibfield  {journal} {\bibinfo  {journal} {Phys.
  Rev. A}\ }\textbf {\bibinfo {volume} {80}},\ \bibinfo {pages} {012304}
  (\bibinfo {year} {2009})}\BibitemShut {NoStop}%
\bibitem [{\citenamefont {Harrow}\ and\ \citenamefont
  {Low}(2009)}]{harrow2009}%
  \BibitemOpen
  \bibfield  {author} {\bibinfo {author} {\bibfnamefont {A.~W.}\ \bibnamefont
  {Harrow}}\ and\ \bibinfo {author} {\bibfnamefont {R.~A.}\ \bibnamefont
  {Low}},\ }\bibfield  {title} {\enquote {\bibinfo {title} {Random quantum
  circuits are approximate 2-designs},}\ }\href {\doibase
  10.1007/s00220-009-0873-6} {\bibfield  {journal} {\bibinfo  {journal}
  {Commun. Math. Phys.}\ }\textbf {\bibinfo {volume} {291}},\ \bibinfo {pages}
  {257} (\bibinfo {year} {2009})}\BibitemShut {NoStop}%
\bibitem [{\citenamefont {Diniz}\ and\ \citenamefont
  {Jonathan}(2011)}]{diniz2011}%
  \BibitemOpen
  \bibfield  {author} {\bibinfo {author} {\bibfnamefont {I.~T.}\ \bibnamefont
  {Diniz}}\ and\ \bibinfo {author} {\bibfnamefont {D.}~\bibnamefont
  {Jonathan}},\ }\bibfield  {title} {\enquote {\bibinfo {title} {Comment on
  "random quantum circuits are approximate 2-designs" by a.w. harrow and r.a.
  low (commun. math. phys. 291, 257^^e2^^80^^93302 (2009))},}\ }\href {\doibase
  10.1007/s00220-011-1217-x} {\bibfield  {journal} {\bibinfo  {journal}
  {Commun. Math. Phys.}\ }\textbf {\bibinfo {volume} {304}},\ \bibinfo {pages}
  {281} (\bibinfo {year} {2011})}\BibitemShut {NoStop}%
\bibitem [{\citenamefont {Brand\~ao}\ \emph {et~al.}(2016)\citenamefont
  {Brand\~ao}, \citenamefont {Harrow},\ and\ \citenamefont
  {Horodecki}}]{brandao2016-1}%
  \BibitemOpen
  \bibfield  {author} {\bibinfo {author} {\bibfnamefont {F.~G. S.~L.}\
  \bibnamefont {Brand\~ao}}, \bibinfo {author} {\bibfnamefont {A.~W.}\
  \bibnamefont {Harrow}}, \ and\ \bibinfo {author} {\bibfnamefont
  {M.}~\bibnamefont {Horodecki}},\ }\bibfield  {title} {\enquote {\bibinfo
  {title} {Efficient quantum pseudorandomness},}\ }\href {\doibase
  10.1103/PhysRevLett.116.170502} {\bibfield  {journal} {\bibinfo  {journal}
  {Phys. Rev. Lett.}\ }\textbf {\bibinfo {volume} {116}},\ \bibinfo {pages}
  {170502} (\bibinfo {year} {2016})}\BibitemShut {NoStop}%
\bibitem [{\citenamefont {Brand^^c3^^a3o}\ \emph {et~al.}(2016)\citenamefont
  {Brand^^c3^^a3o}, \citenamefont {Harrow},\ and\ \citenamefont
  {Horodecki}}]{brandao2016-2}%
  \BibitemOpen
  \bibfield  {author} {\bibinfo {author} {\bibfnamefont {F.~G. S.~L.}\
  \bibnamefont {Brand^^c3^^a3o}}, \bibinfo {author} {\bibfnamefont {A.~W.}\
  \bibnamefont {Harrow}}, \ and\ \bibinfo {author} {\bibfnamefont
  {M.}~\bibnamefont {Horodecki}},\ }\bibfield  {title} {\enquote {\bibinfo
  {title} {Local random quantum circuits are approximate polynomial-designs},}\
  }\href {\doibase 10.1007/s00220-016-2706-8} {\bibfield  {journal} {\bibinfo
  {journal} {Commun. Math. Phys.}\ }\textbf {\bibinfo {volume} {346}},\
  \bibinfo {pages} {397} (\bibinfo {year} {2016})}\BibitemShut {NoStop}%
\bibitem [{\citenamefont {Cleve}\ \emph {et~al.}(2016)\citenamefont {Cleve},
  \citenamefont {Leung}, \citenamefont {Liu},\ and\ \citenamefont
  {Wang}}]{cleve2016}%
  \BibitemOpen
  \bibfield  {author} {\bibinfo {author} {\bibfnamefont {R.}~\bibnamefont
  {Cleve}}, \bibinfo {author} {\bibfnamefont {D.}~\bibnamefont {Leung}},
  \bibinfo {author} {\bibfnamefont {L.}~\bibnamefont {Liu}}, \ and\ \bibinfo
  {author} {\bibfnamefont {C.}~\bibnamefont {Wang}},\ }\bibfield  {title}
  {\enquote {\bibinfo {title} {Near-linear constructions of exact unitary
  2-designs},}\ }\href@noop {} {\bibfield  {journal} {\bibinfo  {journal}
  {Quantum Information and Computation}\ }\textbf {\bibinfo {volume} {16}},\
  \bibinfo {pages} {721} (\bibinfo {year} {2016})}\BibitemShut {NoStop}%
\bibitem [{\citenamefont {Nakata}\ \emph
  {et~al.}(2017{\natexlab{a}})\citenamefont {Nakata}, \citenamefont {Hirche},
  \citenamefont {Morgan},\ and\ \citenamefont {Winter}}]{nakata2017-1}%
  \BibitemOpen
  \bibfield  {author} {\bibinfo {author} {\bibfnamefont {Y.}~\bibnamefont
  {Nakata}}, \bibinfo {author} {\bibfnamefont {C.}~\bibnamefont {Hirche}},
  \bibinfo {author} {\bibfnamefont {C.}~\bibnamefont {Morgan}}, \ and\ \bibinfo
  {author} {\bibfnamefont {A.}~\bibnamefont {Winter}},\ }\bibfield  {title}
  {\enquote {\bibinfo {title} {Unitary 2-designs from random x- and z-diagonal
  unitaries},}\ }\href {\doibase 10.1063/1.4983266} {\bibfield  {journal}
  {\bibinfo  {journal} {Journal of Mathematical Physics}\ }\textbf {\bibinfo
  {volume} {58}},\ \bibinfo {pages} {052203} (\bibinfo {year}
  {2017}{\natexlab{a}})}\BibitemShut {NoStop}%
\bibitem [{\citenamefont {Nakata}\ \emph
  {et~al.}(2017{\natexlab{b}})\citenamefont {Nakata}, \citenamefont {Hirche},
  \citenamefont {Koashi},\ and\ \citenamefont {Winter}}]{nakata2017-2}%
  \BibitemOpen
  \bibfield  {author} {\bibinfo {author} {\bibfnamefont {Y.}~\bibnamefont
  {Nakata}}, \bibinfo {author} {\bibfnamefont {C.}~\bibnamefont {Hirche}},
  \bibinfo {author} {\bibfnamefont {M.}~\bibnamefont {Koashi}}, \ and\ \bibinfo
  {author} {\bibfnamefont {A.}~\bibnamefont {Winter}},\ }\bibfield  {title}
  {\enquote {\bibinfo {title} {Efficient quantum pseudorandomness with nearly
  time-independent hamiltonian dynamics},}\ }\href {\doibase
  10.1103/PhysRevX.7.021006} {\bibfield  {journal} {\bibinfo  {journal} {Phys.
  Rev. X}\ }\textbf {\bibinfo {volume} {7}},\ \bibinfo {pages} {021006}
  (\bibinfo {year} {2017}{\natexlab{b}})}\BibitemShut {NoStop}%
\bibitem [{\citenamefont {Fannes}\ \emph {et~al.}(1992)\citenamefont {Fannes},
  \citenamefont {Nachtergaele},\ and\ \citenamefont {Werner}}]{fannes1992}%
  \BibitemOpen
  \bibfield  {author} {\bibinfo {author} {\bibfnamefont {M.}~\bibnamefont
  {Fannes}}, \bibinfo {author} {\bibfnamefont {B.}~\bibnamefont
  {Nachtergaele}}, \ and\ \bibinfo {author} {\bibfnamefont {R.~F.}\
  \bibnamefont {Werner}},\ }\bibfield  {title} {\enquote {\bibinfo {title}
  {Finitely correlated states on quantum spin chains},}\ }\href {\doibase
  10.1007/BF02099178} {\bibfield  {journal} {\bibinfo  {journal} {Commun. Math.
  Phys.}\ }\textbf {\bibinfo {volume} {144}},\ \bibinfo {pages} {443} (\bibinfo
  {year} {1992})}\BibitemShut {NoStop}%
\bibitem [{\citenamefont {\"Ostlund}\ and\ \citenamefont
  {Rommer}(1995)}]{ostlund1995}%
  \BibitemOpen
  \bibfield  {author} {\bibinfo {author} {\bibfnamefont {S.}~\bibnamefont
  {\"Ostlund}}\ and\ \bibinfo {author} {\bibfnamefont {S.}~\bibnamefont
  {Rommer}},\ }\bibfield  {title} {\enquote {\bibinfo {title} {Thermodynamic
  limit of density matrix renormalization},}\ }\href {\doibase
  10.1103/PhysRevLett.75.3537} {\bibfield  {journal} {\bibinfo  {journal}
  {Phys. Rev. Lett.}\ }\textbf {\bibinfo {volume} {75}},\ \bibinfo {pages}
  {3537} (\bibinfo {year} {1995})}\BibitemShut {NoStop}%
\bibitem [{\citenamefont {Rommer}\ and\ \citenamefont
  {\"Ostlund}(1997)}]{rommer1997}%
  \BibitemOpen
  \bibfield  {author} {\bibinfo {author} {\bibfnamefont {S.}~\bibnamefont
  {Rommer}}\ and\ \bibinfo {author} {\bibfnamefont {S.}~\bibnamefont
  {\"Ostlund}},\ }\bibfield  {title} {\enquote {\bibinfo {title} {Class of
  ansatz wave functions for one-dimensional spin systems and their relation to
  the density matrix renormalization group},}\ }\href {\doibase
  10.1103/PhysRevB.55.2164} {\bibfield  {journal} {\bibinfo  {journal} {Phys.
  Rev. B}\ }\textbf {\bibinfo {volume} {55}},\ \bibinfo {pages} {2164}
  (\bibinfo {year} {1997})}\BibitemShut {NoStop}%
\bibitem [{\citenamefont {Dukelsky}\ \emph {et~al.}(1998)\citenamefont
  {Dukelsky}, \citenamefont {Mart{\'{\i}}n-Delgado}, \citenamefont {Nishino},\
  and\ \citenamefont {Sierra}}]{dukelsky1998}%
  \BibitemOpen
  \bibfield  {author} {\bibinfo {author} {\bibfnamefont {J.}~\bibnamefont
  {Dukelsky}}, \bibinfo {author} {\bibfnamefont {M.~A.}\ \bibnamefont
  {Mart{\'{\i}}n-Delgado}}, \bibinfo {author} {\bibfnamefont {T.}~\bibnamefont
  {Nishino}}, \ and\ \bibinfo {author} {\bibfnamefont {G.}~\bibnamefont
  {Sierra}},\ }\bibfield  {title} {\enquote {\bibinfo {title} {Equivalence of
  the variational matrix product method and the density matrix renormalization
  group applied to spin chains},}\ }\href {\doibase 10.1209/epl/i1998-00381-x}
  {\bibfield  {journal} {\bibinfo  {journal} {Europhysics Letters ({EPL})}\
  }\textbf {\bibinfo {volume} {43}},\ \bibinfo {pages} {457} (\bibinfo {year}
  {1998})}\BibitemShut {NoStop}%
\bibitem [{\citenamefont {White}(1992)}]{white1992}%
  \BibitemOpen
  \bibfield  {author} {\bibinfo {author} {\bibfnamefont {S.~R.}\ \bibnamefont
  {White}},\ }\bibfield  {title} {\enquote {\bibinfo {title} {Density matrix
  formulation for quantum renormalization groups},}\ }\href {\doibase
  10.1103/PhysRevLett.69.2863} {\bibfield  {journal} {\bibinfo  {journal}
  {Phys. Rev. Lett.}\ }\textbf {\bibinfo {volume} {69}},\ \bibinfo {pages}
  {2863} (\bibinfo {year} {1992})}\BibitemShut {NoStop}%
\bibitem [{\citenamefont {White}(1993)}]{white1993}%
  \BibitemOpen
  \bibfield  {author} {\bibinfo {author} {\bibfnamefont {S.~R.}\ \bibnamefont
  {White}},\ }\bibfield  {title} {\enquote {\bibinfo {title} {Density-matrix
  algorithms for quantum renormalization groups},}\ }\href {\doibase
  10.1103/PhysRevB.48.10345} {\bibfield  {journal} {\bibinfo  {journal} {Phys.
  Rev. B}\ }\textbf {\bibinfo {volume} {48}},\ \bibinfo {pages} {10345}
  (\bibinfo {year} {1993})}\BibitemShut {NoStop}%
\bibitem [{\citenamefont {Calabrese}\ and\ \citenamefont
  {Cardy}(2004)}]{calabrese2004}%
  \BibitemOpen
  \bibfield  {author} {\bibinfo {author} {\bibfnamefont {P.}~\bibnamefont
  {Calabrese}}\ and\ \bibinfo {author} {\bibfnamefont {J.}~\bibnamefont
  {Cardy}},\ }\bibfield  {title} {\enquote {\bibinfo {title} {Entanglement
  entropy and quantum field theory},}\ }\href {\doibase
  10.1088/1742-5468/2004/06/p06002} {\bibfield  {journal} {\bibinfo  {journal}
  {Journal of Statistical Mechanics: Theory and Experiment}\ }\textbf {\bibinfo
  {volume} {2004}},\ \bibinfo {pages} {P06002} (\bibinfo {year}
  {2004})}\BibitemShut {NoStop}%
\bibitem [{\citenamefont {Hastings}\ and\ \citenamefont
  {Koma}(2006)}]{hastings2006-2}%
  \BibitemOpen
  \bibfield  {author} {\bibinfo {author} {\bibfnamefont {M.~B.}\ \bibnamefont
  {Hastings}}\ and\ \bibinfo {author} {\bibfnamefont {T.}~\bibnamefont
  {Koma}},\ }\bibfield  {title} {\enquote {\bibinfo {title} {Spectral gap and
  exponential decay of correlations},}\ }\href {\doibase
  10.1007/s00220-006-0030-4} {\bibfield  {journal} {\bibinfo  {journal}
  {Commun. Math. Phys.}\ }\textbf {\bibinfo {volume} {265}},\ \bibinfo {pages}
  {781} (\bibinfo {year} {2006})}\BibitemShut {NoStop}%
\bibitem [{\citenamefont {Hastings}(2007)}]{hastings2007}%
  \BibitemOpen
  \bibfield  {author} {\bibinfo {author} {\bibfnamefont {M.~B.}\ \bibnamefont
  {Hastings}},\ }\bibfield  {title} {\enquote {\bibinfo {title} {An area law
  for one-dimensional quantum systems},}\ }\href {\doibase
  10.1088/1742-5468/2007/08/p08024} {\bibfield  {journal} {\bibinfo  {journal}
  {Journal of Statistical Mechanics: Theory and Experiment}\ }\textbf {\bibinfo
  {volume} {2007}},\ \bibinfo {pages} {P08024} (\bibinfo {year}
  {2007})}\BibitemShut {NoStop}%
\bibitem [{\citenamefont {Eisert}\ \emph {et~al.}(2010)\citenamefont {Eisert},
  \citenamefont {Cramer},\ and\ \citenamefont {Plenio}}]{eisert2010}%
  \BibitemOpen
  \bibfield  {author} {\bibinfo {author} {\bibfnamefont {J.}~\bibnamefont
  {Eisert}}, \bibinfo {author} {\bibfnamefont {M.}~\bibnamefont {Cramer}}, \
  and\ \bibinfo {author} {\bibfnamefont {M.~B.}\ \bibnamefont {Plenio}},\
  }\bibfield  {title} {\enquote {\bibinfo {title} {Colloquium: Area laws for
  the entanglement entropy},}\ }\href {\doibase 10.1103/RevModPhys.82.277}
  {\bibfield  {journal} {\bibinfo  {journal} {Rev. Mod. Phys.}\ }\textbf
  {\bibinfo {volume} {82}},\ \bibinfo {pages} {277} (\bibinfo {year}
  {2010})}\BibitemShut {NoStop}%
\bibitem [{\citenamefont {Basko}\ \emph {et~al.}(2006)\citenamefont {Basko},
  \citenamefont {Aleiner},\ and\ \citenamefont {Altshuler}}]{basko2006}%
  \BibitemOpen
  \bibfield  {author} {\bibinfo {author} {\bibfnamefont {D.}~\bibnamefont
  {Basko}}, \bibinfo {author} {\bibfnamefont {I.}~\bibnamefont {Aleiner}}, \
  and\ \bibinfo {author} {\bibfnamefont {B.}~\bibnamefont {Altshuler}},\
  }\bibfield  {title} {\enquote {\bibinfo {title} {Metal-insulator transition
  in a weakly interacting many-electron system with localized single-particle
  states},}\ }\href {\doibase https://doi.org/10.1016/j.aop.2005.11.014}
  {\bibfield  {journal} {\bibinfo  {journal} {Annals of Physics}\ }\textbf
  {\bibinfo {volume} {321}},\ \bibinfo {pages} {1126} (\bibinfo {year}
  {2006})}\BibitemShut {NoStop}%
\bibitem [{\citenamefont {Bauer}\ and\ \citenamefont
  {Nayak}(2013)}]{bauer2013}%
  \BibitemOpen
  \bibfield  {author} {\bibinfo {author} {\bibfnamefont {B.}~\bibnamefont
  {Bauer}}\ and\ \bibinfo {author} {\bibfnamefont {C.}~\bibnamefont {Nayak}},\
  }\bibfield  {title} {\enquote {\bibinfo {title} {Area laws in a many-body
  localized state and its implications for topological order},}\ }\href
  {\doibase 10.1088/1742-5468/2013/09/p09005} {\bibfield  {journal} {\bibinfo
  {journal} {Journal of Statistical Mechanics: Theory and Experiment}\ }\textbf
  {\bibinfo {volume} {2013}},\ \bibinfo {pages} {P09005} (\bibinfo {year}
  {2013})}\BibitemShut {NoStop}%
\bibitem [{\citenamefont {Swingle}(2013)}]{swingle2013}%
  \BibitemOpen
  \bibfield  {author} {\bibinfo {author} {\bibfnamefont {B.}~\bibnamefont
  {Swingle}},\ }\bibfield  {title} {\enquote {\bibinfo {title} {A simple model
  of many-body localization},}\ }\href {https://arxiv.org/abs/1307.0507}
  {\bibfield  {journal} {\bibinfo  {journal} {arXiv:cond-mat.dis-nn/1307.0507}\
  } (\bibinfo {year} {2013})}\BibitemShut {NoStop}%
\bibitem [{\citenamefont {Friesdorf}\ \emph {et~al.}(2015)\citenamefont
  {Friesdorf}, \citenamefont {Werner}, \citenamefont {Brown}, \citenamefont
  {Scholz},\ and\ \citenamefont {Eisert}}]{friesdorf2015}%
  \BibitemOpen
  \bibfield  {author} {\bibinfo {author} {\bibfnamefont {M.}~\bibnamefont
  {Friesdorf}}, \bibinfo {author} {\bibfnamefont {A.~H.}\ \bibnamefont
  {Werner}}, \bibinfo {author} {\bibfnamefont {W.}~\bibnamefont {Brown}},
  \bibinfo {author} {\bibfnamefont {V.~B.}\ \bibnamefont {Scholz}}, \ and\
  \bibinfo {author} {\bibfnamefont {J.}~\bibnamefont {Eisert}},\ }\bibfield
  {title} {\enquote {\bibinfo {title} {Many-body localization implies that
  eigenvectors are matrix-product states},}\ }\href {\doibase
  10.1103/PhysRevLett.114.170505} {\bibfield  {journal} {\bibinfo  {journal}
  {Phys. Rev. Lett.}\ }\textbf {\bibinfo {volume} {114}},\ \bibinfo {pages}
  {170505} (\bibinfo {year} {2015})}\BibitemShut {NoStop}%
\bibitem [{\citenamefont {Khemani}\ \emph {et~al.}(2016)\citenamefont
  {Khemani}, \citenamefont {Pollmann},\ and\ \citenamefont
  {Sondhi}}]{khemani2016}%
  \BibitemOpen
  \bibfield  {author} {\bibinfo {author} {\bibfnamefont {V.}~\bibnamefont
  {Khemani}}, \bibinfo {author} {\bibfnamefont {F.}~\bibnamefont {Pollmann}}, \
  and\ \bibinfo {author} {\bibfnamefont {S.~L.}\ \bibnamefont {Sondhi}},\
  }\bibfield  {title} {\enquote {\bibinfo {title} {Obtaining highly excited
  eigenstates of many-body localized hamiltonians by the density matrix
  renormalization group approach},}\ }\href {\doibase
  10.1103/PhysRevLett.116.247204} {\bibfield  {journal} {\bibinfo  {journal}
  {Phys. Rev. Lett.}\ }\textbf {\bibinfo {volume} {116}},\ \bibinfo {pages}
  {247204} (\bibinfo {year} {2016})}\BibitemShut {NoStop}%
\bibitem [{\citenamefont {Yu}\ \emph {et~al.}(2017)\citenamefont {Yu},
  \citenamefont {Pekker},\ and\ \citenamefont {Clark}}]{yu2017}%
  \BibitemOpen
  \bibfield  {author} {\bibinfo {author} {\bibfnamefont {X.}~\bibnamefont
  {Yu}}, \bibinfo {author} {\bibfnamefont {D.}~\bibnamefont {Pekker}}, \ and\
  \bibinfo {author} {\bibfnamefont {B.~K.}\ \bibnamefont {Clark}},\ }\bibfield
  {title} {\enquote {\bibinfo {title} {Finding matrix product state
  representations of highly excited eigenstates of many-body localized
  hamiltonians},}\ }\href {\doibase 10.1103/PhysRevLett.118.017201} {\bibfield
  {journal} {\bibinfo  {journal} {Phys. Rev. Lett.}\ }\textbf {\bibinfo
  {volume} {118}},\ \bibinfo {pages} {017201} (\bibinfo {year}
  {2017})}\BibitemShut {NoStop}%
\bibitem [{\citenamefont {Vidal}(2004)}]{vidal2004}%
  \BibitemOpen
  \bibfield  {author} {\bibinfo {author} {\bibfnamefont {G.}~\bibnamefont
  {Vidal}},\ }\bibfield  {title} {\enquote {\bibinfo {title} {Efficient
  simulation of one-dimensional quantum many-body systems},}\ }\href {\doibase
  10.1103/PhysRevLett.93.040502} {\bibfield  {journal} {\bibinfo  {journal}
  {Phys. Rev. Lett.}\ }\textbf {\bibinfo {volume} {93}},\ \bibinfo {pages}
  {040502} (\bibinfo {year} {2004})}\BibitemShut {NoStop}%
\bibitem [{\citenamefont {White}\ and\ \citenamefont
  {Feiguin}(2004)}]{white2004}%
  \BibitemOpen
  \bibfield  {author} {\bibinfo {author} {\bibfnamefont {S.~R.}\ \bibnamefont
  {White}}\ and\ \bibinfo {author} {\bibfnamefont {A.~E.}\ \bibnamefont
  {Feiguin}},\ }\bibfield  {title} {\enquote {\bibinfo {title} {Real-time
  evolution using the density matrix renormalization group},}\ }\href {\doibase
  10.1103/PhysRevLett.93.076401} {\bibfield  {journal} {\bibinfo  {journal}
  {Phys. Rev. Lett.}\ }\textbf {\bibinfo {volume} {93}},\ \bibinfo {pages}
  {076401} (\bibinfo {year} {2004})}\BibitemShut {NoStop}%
\bibitem [{\citenamefont {Daley}\ \emph {et~al.}(2004)\citenamefont {Daley},
  \citenamefont {Kollath}, \citenamefont {Schollw^^c3^^b6ck},\ and\
  \citenamefont {Vidal}}]{daley2004}%
  \BibitemOpen
  \bibfield  {author} {\bibinfo {author} {\bibfnamefont {A.~J.}\ \bibnamefont
  {Daley}}, \bibinfo {author} {\bibfnamefont {C.}~\bibnamefont {Kollath}},
  \bibinfo {author} {\bibfnamefont {U.}~\bibnamefont {Schollw^^c3^^b6ck}}, \
  and\ \bibinfo {author} {\bibfnamefont {G.}~\bibnamefont {Vidal}},\ }\bibfield
   {title} {\enquote {\bibinfo {title} {Time-dependent density-matrix
  renormalization-group using adaptive effective hilbert spaces},}\ }\href
  {\doibase 10.1088/1742-5468/2004/04/p04005} {\bibfield  {journal} {\bibinfo
  {journal} {Journal of Statistical Mechanics: Theory and Experiment}\ }\textbf
  {\bibinfo {volume} {2004}},\ \bibinfo {pages} {P04005} (\bibinfo {year}
  {2004})}\BibitemShut {NoStop}%
\bibitem [{\citenamefont {McCulloch}(2007)}]{mcculloch2007}%
  \BibitemOpen
  \bibfield  {author} {\bibinfo {author} {\bibfnamefont {I.~P.}\ \bibnamefont
  {McCulloch}},\ }\bibfield  {title} {\enquote {\bibinfo {title} {From
  density-matrix renormalization group to matrix product states},}\ }\href
  {\doibase 10.1088/1742-5468/2007/10/p10014} {\bibfield  {journal} {\bibinfo
  {journal} {Journal of Statistical Mechanics: Theory and Experiment}\ }\textbf
  {\bibinfo {volume} {2007}},\ \bibinfo {pages} {P10014} (\bibinfo {year}
  {2007})}\BibitemShut {NoStop}%
\bibitem [{\citenamefont {Verstraete}\ \emph {et~al.}(2004)\citenamefont
  {Verstraete}, \citenamefont {Garc\'{\i}a-Ripoll},\ and\ \citenamefont
  {Cirac}}]{verstraete2004-2}%
  \BibitemOpen
  \bibfield  {author} {\bibinfo {author} {\bibfnamefont {F.}~\bibnamefont
  {Verstraete}}, \bibinfo {author} {\bibfnamefont {J.~J.}\ \bibnamefont
  {Garc\'{\i}a-Ripoll}}, \ and\ \bibinfo {author} {\bibfnamefont {J.~I.}\
  \bibnamefont {Cirac}},\ }\bibfield  {title} {\enquote {\bibinfo {title}
  {Matrix product density operators: Simulation of finite-temperature and
  dissipative systems},}\ }\href {\doibase 10.1103/PhysRevLett.93.207204}
  {\bibfield  {journal} {\bibinfo  {journal} {Phys. Rev. Lett.}\ }\textbf
  {\bibinfo {volume} {93}},\ \bibinfo {pages} {207204} (\bibinfo {year}
  {2004})}\BibitemShut {NoStop}%
\bibitem [{\citenamefont {Zwolak}\ and\ \citenamefont
  {Vidal}(2004)}]{zwolak2004}%
  \BibitemOpen
  \bibfield  {author} {\bibinfo {author} {\bibfnamefont {M.}~\bibnamefont
  {Zwolak}}\ and\ \bibinfo {author} {\bibfnamefont {G.}~\bibnamefont {Vidal}},\
  }\bibfield  {title} {\enquote {\bibinfo {title} {Mixed-state dynamics in
  one-dimensional quantum lattice systems: A time-dependent superoperator
  renormalization algorithm},}\ }\href {\doibase 10.1103/PhysRevLett.93.207205}
  {\bibfield  {journal} {\bibinfo  {journal} {Phys. Rev. Lett.}\ }\textbf
  {\bibinfo {volume} {93}},\ \bibinfo {pages} {207205} (\bibinfo {year}
  {2004})}\BibitemShut {NoStop}%
\bibitem [{\citenamefont {Feiguin}\ and\ \citenamefont
  {White}(2005)}]{feiguin2005}%
  \BibitemOpen
  \bibfield  {author} {\bibinfo {author} {\bibfnamefont {A.~E.}\ \bibnamefont
  {Feiguin}}\ and\ \bibinfo {author} {\bibfnamefont {S.~R.}\ \bibnamefont
  {White}},\ }\bibfield  {title} {\enquote {\bibinfo {title}
  {Finite-temperature density matrix renormalization using an enlarged hilbert
  space},}\ }\href {\doibase 10.1103/PhysRevB.72.220401} {\bibfield  {journal}
  {\bibinfo  {journal} {Phys. Rev. B}\ }\textbf {\bibinfo {volume} {72}},\
  \bibinfo {pages} {220401(R)} (\bibinfo {year} {2005})}\BibitemShut {NoStop}%
\bibitem [{\citenamefont {White}(2009)}]{white2009}%
  \BibitemOpen
  \bibfield  {author} {\bibinfo {author} {\bibfnamefont {S.~R.}\ \bibnamefont
  {White}},\ }\bibfield  {title} {\enquote {\bibinfo {title} {Minimally
  entangled typical quantum states at finite temperature},}\ }\href {\doibase
  10.1103/PhysRevLett.102.190601} {\bibfield  {journal} {\bibinfo  {journal}
  {Phys. Rev. Lett.}\ }\textbf {\bibinfo {volume} {102}},\ \bibinfo {pages}
  {190601} (\bibinfo {year} {2009})}\BibitemShut {NoStop}%
\bibitem [{\citenamefont {Stoudenmire}\ and\ \citenamefont
  {White}(2010)}]{stoudenmire2010}%
  \BibitemOpen
  \bibfield  {author} {\bibinfo {author} {\bibfnamefont {E.~M.}\ \bibnamefont
  {Stoudenmire}}\ and\ \bibinfo {author} {\bibfnamefont {S.~R.}\ \bibnamefont
  {White}},\ }\bibfield  {title} {\enquote {\bibinfo {title} {Minimally
  entangled typical thermal state algorithms},}\ }\href {\doibase
  10.1088/1367-2630/12/5/055026} {\bibfield  {journal} {\bibinfo  {journal}
  {New Journal of Physics}\ }\textbf {\bibinfo {volume} {12}},\ \bibinfo
  {pages} {055026} (\bibinfo {year} {2010})}\BibitemShut {NoStop}%
\bibitem [{\citenamefont {Binder}\ and\ \citenamefont
  {Barthel}(2017)}]{binder2017}%
  \BibitemOpen
  \bibfield  {author} {\bibinfo {author} {\bibfnamefont {M.}~\bibnamefont
  {Binder}}\ and\ \bibinfo {author} {\bibfnamefont {T.}~\bibnamefont
  {Barthel}},\ }\bibfield  {title} {\enquote {\bibinfo {title} {Symmetric
  minimally entangled typical thermal states for canonical and grand-canonical
  ensembles},}\ }\href {\doibase 10.1103/PhysRevB.95.195148} {\bibfield
  {journal} {\bibinfo  {journal} {Phys. Rev. B}\ }\textbf {\bibinfo {volume}
  {95}},\ \bibinfo {pages} {195148} (\bibinfo {year} {2017})}\BibitemShut
  {NoStop}%
\bibitem [{\citenamefont {Binder}\ and\ \citenamefont
  {Barthel}(2015)}]{binder2015}%
  \BibitemOpen
  \bibfield  {author} {\bibinfo {author} {\bibfnamefont {M.}~\bibnamefont
  {Binder}}\ and\ \bibinfo {author} {\bibfnamefont {T.}~\bibnamefont
  {Barthel}},\ }\bibfield  {title} {\enquote {\bibinfo {title} {Minimally
  entangled typical thermal states versus matrix product purifications for the
  simulation of equilibrium states and time evolution},}\ }\href {\doibase
  10.1103/PhysRevB.92.125119} {\bibfield  {journal} {\bibinfo  {journal} {Phys.
  Rev. B}\ }\textbf {\bibinfo {volume} {92}},\ \bibinfo {pages} {125119}
  (\bibinfo {year} {2015})}\BibitemShut {NoStop}%
\bibitem [{\citenamefont {Wolf}\ \emph {et~al.}(2008)\citenamefont {Wolf},
  \citenamefont {Verstraete}, \citenamefont {Hastings},\ and\ \citenamefont
  {Cirac}}]{wolf2008}%
  \BibitemOpen
  \bibfield  {author} {\bibinfo {author} {\bibfnamefont {M.~M.}\ \bibnamefont
  {Wolf}}, \bibinfo {author} {\bibfnamefont {F.}~\bibnamefont {Verstraete}},
  \bibinfo {author} {\bibfnamefont {M.~B.}\ \bibnamefont {Hastings}}, \ and\
  \bibinfo {author} {\bibfnamefont {J.~I.}\ \bibnamefont {Cirac}},\ }\bibfield
  {title} {\enquote {\bibinfo {title} {Area laws in quantum systems: Mutual
  information and correlations},}\ }\href {\doibase
  10.1103/PhysRevLett.100.070502} {\bibfield  {journal} {\bibinfo  {journal}
  {Phys. Rev. Lett.}\ }\textbf {\bibinfo {volume} {100}},\ \bibinfo {pages}
  {070502} (\bibinfo {year} {2008})}\BibitemShut {NoStop}%
\bibitem [{\citenamefont {Groisman}\ \emph {et~al.}(2005)\citenamefont
  {Groisman}, \citenamefont {Popescu},\ and\ \citenamefont
  {Winter}}]{groisman2005}%
  \BibitemOpen
  \bibfield  {author} {\bibinfo {author} {\bibfnamefont {B.}~\bibnamefont
  {Groisman}}, \bibinfo {author} {\bibfnamefont {S.}~\bibnamefont {Popescu}}, \
  and\ \bibinfo {author} {\bibfnamefont {A.}~\bibnamefont {Winter}},\
  }\bibfield  {title} {\enquote {\bibinfo {title} {Quantum, classical, and
  total amount of correlations in a quantum state},}\ }\href {\doibase
  10.1103/PhysRevA.72.032317} {\bibfield  {journal} {\bibinfo  {journal} {Phys.
  Rev. A}\ }\textbf {\bibinfo {volume} {72}},\ \bibinfo {pages} {032317}
  (\bibinfo {year} {2005})}\BibitemShut {NoStop}%
\bibitem [{\citenamefont {Garnerone}\ \emph
  {et~al.}(2010{\natexlab{a}})\citenamefont {Garnerone}, \citenamefont
  {de~Oliveira},\ and\ \citenamefont {Zanardi}}]{garnerone2010-1}%
  \BibitemOpen
  \bibfield  {author} {\bibinfo {author} {\bibfnamefont {S.}~\bibnamefont
  {Garnerone}}, \bibinfo {author} {\bibfnamefont {T.~R.}\ \bibnamefont
  {de~Oliveira}}, \ and\ \bibinfo {author} {\bibfnamefont {P.}~\bibnamefont
  {Zanardi}},\ }\bibfield  {title} {\enquote {\bibinfo {title} {Typicality in
  random matrix product states},}\ }\href {\doibase 10.1103/PhysRevA.81.032336}
  {\bibfield  {journal} {\bibinfo  {journal} {Phys. Rev. A}\ }\textbf {\bibinfo
  {volume} {81}},\ \bibinfo {pages} {032336} (\bibinfo {year}
  {2010}{\natexlab{a}})}\BibitemShut {NoStop}%
\bibitem [{\citenamefont {Garnerone}\ \emph
  {et~al.}(2010{\natexlab{b}})\citenamefont {Garnerone}, \citenamefont
  {de~Oliveira}, \citenamefont {Haas},\ and\ \citenamefont
  {Zanardi}}]{garnerone2010-2}%
  \BibitemOpen
  \bibfield  {author} {\bibinfo {author} {\bibfnamefont {S.}~\bibnamefont
  {Garnerone}}, \bibinfo {author} {\bibfnamefont {T.~R.}\ \bibnamefont
  {de~Oliveira}}, \bibinfo {author} {\bibfnamefont {S.}~\bibnamefont {Haas}}, \
  and\ \bibinfo {author} {\bibfnamefont {P.}~\bibnamefont {Zanardi}},\
  }\bibfield  {title} {\enquote {\bibinfo {title} {Statistical properties of
  random matrix product states},}\ }\href {\doibase 10.1103/PhysRevA.82.052312}
  {\bibfield  {journal} {\bibinfo  {journal} {Phys. Rev. A}\ }\textbf {\bibinfo
  {volume} {82}},\ \bibinfo {pages} {052312} (\bibinfo {year}
  {2010}{\natexlab{b}})}\BibitemShut {NoStop}%
\bibitem [{\citenamefont {Imada}\ and\ \citenamefont
  {Takahashi}(1986)}]{imada1986}%
  \BibitemOpen
  \bibfield  {author} {\bibinfo {author} {\bibfnamefont {M.}~\bibnamefont
  {Imada}}\ and\ \bibinfo {author} {\bibfnamefont {M.}~\bibnamefont
  {Takahashi}},\ }\bibfield  {title} {\enquote {\bibinfo {title} {Quantum
  transfer monte carlo method for finite temperature properties and quantum
  molecular dynamics method for dynamical correlation functions},}\ }\href
  {\doibase 10.1143/JPSJ.55.3354} {\bibfield  {journal} {\bibinfo  {journal}
  {Journal of the Physical Society of Japan}\ }\textbf {\bibinfo {volume}
  {55}},\ \bibinfo {pages} {3354} (\bibinfo {year} {1986})}\BibitemShut
  {NoStop}%
\bibitem [{\citenamefont {Jakli\ifmmode~\check{c}\else \v{c}\fi{}}\ and\
  \citenamefont {Prelov\ifmmode~\check{s}\else
  \v{s}\fi{}ek}(1994)}]{jaklic1994}%
  \BibitemOpen
  \bibfield  {author} {\bibinfo {author} {\bibfnamefont {J.}~\bibnamefont
  {Jakli\ifmmode~\check{c}\else \v{c}\fi{}}}\ and\ \bibinfo {author}
  {\bibfnamefont {P.}~\bibnamefont {Prelov\ifmmode~\check{s}\else
  \v{s}\fi{}ek}},\ }\bibfield  {title} {\enquote {\bibinfo {title} {Lanczos
  method for the calculation of finite-temperature quantities in correlated
  systems},}\ }\href {\doibase 10.1103/PhysRevB.49.5065} {\bibfield  {journal}
  {\bibinfo  {journal} {Phys. Rev. B}\ }\textbf {\bibinfo {volume} {49}},\
  \bibinfo {pages} {5065} (\bibinfo {year} {1994})}\BibitemShut {NoStop}%
\bibitem [{\citenamefont {Hams}\ and\ \citenamefont
  {De~Raedt}(2000)}]{hams2000}%
  \BibitemOpen
  \bibfield  {author} {\bibinfo {author} {\bibfnamefont {A.}~\bibnamefont
  {Hams}}\ and\ \bibinfo {author} {\bibfnamefont {H.}~\bibnamefont
  {De~Raedt}},\ }\bibfield  {title} {\enquote {\bibinfo {title} {Fast algorithm
  for finding the eigenvalue distribution of very large matrices},}\ }\href
  {\doibase 10.1103/PhysRevE.62.4365} {\bibfield  {journal} {\bibinfo
  {journal} {Phys. Rev. E}\ }\textbf {\bibinfo {volume} {62}},\ \bibinfo
  {pages} {4365} (\bibinfo {year} {2000})}\BibitemShut {NoStop}%
\bibitem [{\citenamefont {Jin}\ \emph {et~al.}(2021)\citenamefont {Jin},
  \citenamefont {Willsch}, \citenamefont {Willsch}, \citenamefont {Lagemann},
  \citenamefont {Michielsen},\ and\ \citenamefont {De~Raedt}}]{jin2021}%
  \BibitemOpen
  \bibfield  {author} {\bibinfo {author} {\bibfnamefont {F.}~\bibnamefont
  {Jin}}, \bibinfo {author} {\bibfnamefont {D.}~\bibnamefont {Willsch}},
  \bibinfo {author} {\bibfnamefont {M.}~\bibnamefont {Willsch}}, \bibinfo
  {author} {\bibfnamefont {H.}~\bibnamefont {Lagemann}}, \bibinfo {author}
  {\bibfnamefont {K.}~\bibnamefont {Michielsen}}, \ and\ \bibinfo {author}
  {\bibfnamefont {H.}~\bibnamefont {De~Raedt}},\ }\bibfield  {title} {\enquote
  {\bibinfo {title} {Random state technology},}\ }\href {\doibase
  10.7566/JPSJ.90.012001} {\bibfield  {journal} {\bibinfo  {journal} {Journal
  of the Physical Society of Japan}\ }\textbf {\bibinfo {volume} {90}},\
  \bibinfo {pages} {012001} (\bibinfo {year} {2021})}\BibitemShut {NoStop}%
\bibitem [{\citenamefont {Yoneta}\ and\ \citenamefont
  {Shimizu}(2019)}]{yoneta2019}%
  \BibitemOpen
  \bibfield  {author} {\bibinfo {author} {\bibfnamefont {Y.}~\bibnamefont
  {Yoneta}}\ and\ \bibinfo {author} {\bibfnamefont {A.}~\bibnamefont
  {Shimizu}},\ }\bibfield  {title} {\enquote {\bibinfo {title} {Squeezed
  ensemble for systems with first-order phase transitions},}\ }\href {\doibase
  10.1103/PhysRevB.99.144105} {\bibfield  {journal} {\bibinfo  {journal} {Phys.
  Rev. B}\ }\textbf {\bibinfo {volume} {99}},\ \bibinfo {pages} {144105}
  (\bibinfo {year} {2019})}\BibitemShut {NoStop}%
\bibitem [{\citenamefont {Kl\"{u}mper}(1993)}]{klumper1993}%
  \BibitemOpen
  \bibfield  {author} {\bibinfo {author} {\bibfnamefont {A.}~\bibnamefont
  {Kl\"{u}mper}},\ }\bibfield  {title} {\enquote {\bibinfo {title}
  {Thermodynamics of the anisotropic spin-1/2 heisenberg chain and related
  quantum chains},}\ }\href {\doibase 10.1007/BF01316831} {\bibfield  {journal}
  {\bibinfo  {journal} {Z. Phys. B}\ }\textbf {\bibinfo {volume} {91}},\
  \bibinfo {pages} {507} (\bibinfo {year} {1993})}\BibitemShut {NoStop}%
\bibitem [{\citenamefont {Kl\"{u}mper}(1998)}]{klumper1998}%
  \BibitemOpen
  \bibfield  {author} {\bibinfo {author} {\bibfnamefont {A.}~\bibnamefont
  {Kl\"{u}mper}},\ }\bibfield  {title} {\enquote {\bibinfo {title} {The
  spin-1/2 heisenberg chain: thermodynamics, quantum criticality and
  spin-peierls exponents},}\ }\href {\doibase 10.1007/s100510050491} {\bibfield
   {journal} {\bibinfo  {journal} {Eur. Phys. J. B}\ }\textbf {\bibinfo
  {volume} {5}},\ \bibinfo {pages} {677} (\bibinfo {year} {1998})}\BibitemShut
  {NoStop}%
\bibitem [{\citenamefont {Iwaki}\ and\ \citenamefont {Hotta}()}]{iwaki-prep}%
  \BibitemOpen
  \bibfield  {author} {\bibinfo {author} {\bibfnamefont {A.}~\bibnamefont
  {Iwaki}}\ and\ \bibinfo {author} {\bibfnamefont {C.}~\bibnamefont {Hotta}},\
  }\href@noop {} {\bibinfo  {journal} {in preparation}\ }\BibitemShut {NoStop}%
\bibitem [{\citenamefont {Motta}\ \emph {et~al.}(2020)\citenamefont {Motta},
  \citenamefont {Sun}, \citenamefont {Tan}, \citenamefont {O'Rourke},
  \citenamefont {Ye}, \citenamefont {Minnich}, \citenamefont {Brand^^c3^^a3o},\
  and\ \citenamefont {Chan}}]{motta2020}%
  \BibitemOpen
\bibfield  {journal} {  }\bibfield  {author} {\bibinfo {author} {\bibfnamefont
  {M.}~\bibnamefont {Motta}}, \bibinfo {author} {\bibfnamefont
  {C.}~\bibnamefont {Sun}}, \bibinfo {author} {\bibfnamefont {A.~T.~K.}\
  \bibnamefont {Tan}}, \bibinfo {author} {\bibfnamefont {M.~J.}\ \bibnamefont
  {O'Rourke}}, \bibinfo {author} {\bibfnamefont {E.}~\bibnamefont {Ye}},
  \bibinfo {author} {\bibfnamefont {A.~J.}\ \bibnamefont {Minnich}}, \bibinfo
  {author} {\bibfnamefont {F.~G. S.~L.}\ \bibnamefont {Brand^^c3^^a3o}}, \ and\
  \bibinfo {author} {\bibfnamefont {G.~K.-L.}\ \bibnamefont {Chan}},\
  }\bibfield  {title} {\enquote {\bibinfo {title} {Determining eigenstates and
  thermal states on a quantum computer using quantum imaginary time
  evolution},}\ }\href {\doibase 10.1038/s41567-019-0704-4} {\bibfield
  {journal} {\bibinfo  {journal} {Nature Phys.}\ }\textbf {\bibinfo {volume}
  {16}},\ \bibinfo {pages} {205} (\bibinfo {year} {2020})}\BibitemShut
  {NoStop}%
\bibitem [{\citenamefont {Sun}\ \emph {et~al.}(2021)\citenamefont {Sun},
  \citenamefont {Motta}, \citenamefont {Tazhigulov}, \citenamefont {Tan},
  \citenamefont {Chan},\ and\ \citenamefont {Minnich}}]{sun2021}%
  \BibitemOpen
  \bibfield  {author} {\bibinfo {author} {\bibfnamefont {S.-N.}\ \bibnamefont
  {Sun}}, \bibinfo {author} {\bibfnamefont {M.}~\bibnamefont {Motta}}, \bibinfo
  {author} {\bibfnamefont {R.~N.}\ \bibnamefont {Tazhigulov}}, \bibinfo
  {author} {\bibfnamefont {A.~T.}\ \bibnamefont {Tan}}, \bibinfo {author}
  {\bibfnamefont {G.~K.-L.}\ \bibnamefont {Chan}}, \ and\ \bibinfo {author}
  {\bibfnamefont {A.~J.}\ \bibnamefont {Minnich}},\ }\bibfield  {title}
  {\enquote {\bibinfo {title} {Quantum computation of finite-temperature static
  and dynamical properties of spin systems using quantum imaginary time
  evolution},}\ }\href {\doibase 10.1103/PRXQuantum.2.010317} {\bibfield
  {journal} {\bibinfo  {journal} {PRX Quantum}\ }\textbf {\bibinfo {volume}
  {2}},\ \bibinfo {pages} {010317} (\bibinfo {year} {2021})}\BibitemShut
  {NoStop}%
\bibitem [{got()}]{goto_code}%
  \BibitemOpen
  \href {https://github.com/ShimpeiGoto/RPMPS-T} {\bibinfo  {journal}
  {https://github.com/ShimpeiGoto/RPMPS-T}\ }\BibitemShut {NoStop}%
\bibitem [{\citenamefont {Iitaka}\ and\ \citenamefont
  {Ebisuzaki}(2004)}]{iitaka2004}%
  \BibitemOpen
\bibfield  {journal} {  }\bibfield  {author} {\bibinfo {author} {\bibfnamefont
  {T.}~\bibnamefont {Iitaka}}\ and\ \bibinfo {author} {\bibfnamefont
  {T.}~\bibnamefont {Ebisuzaki}},\ }\bibfield  {title} {\enquote {\bibinfo
  {title} {Random phase vector for calculating the trace of a large matrix},}\
  }\href {\doibase 10.1103/PhysRevE.69.057701} {\bibfield  {journal} {\bibinfo
  {journal} {Phys. Rev. E}\ }\textbf {\bibinfo {volume} {69}},\ \bibinfo
  {pages} {057701} (\bibinfo {year} {2004})}\BibitemShut {NoStop}%
\end{thebibliography}%

\end{document}